\documentclass[reprint,aps,prx, amsmath, amssymb, tightenlines,natbib, superscriptaddress]{revtex4-2}

\usepackage{graphicx}
\usepackage{dcolumn}
\usepackage{bm}
\usepackage{float}
\usepackage{setspace}
\usepackage{multirow}   
\usepackage{amsmath}    
\usepackage{array}  
\usepackage{amssymb} 
\usepackage{xcolor} 
\usepackage{amsfonts} 
\usepackage{microtype}
\usepackage{wrapfig}

\usepackage{xr-hyper}  
\begin{document}
\title{Resonances in Overdamped Odd Materials}

\author{Julius Kiln}
\affiliation{Rudolf Peierls Centre for Theoretical Physics, Department of Physics, University of Oxford, Parks Road, Oxford OX1 3PU, United Kingdom}
\author{Alexander Mietke}
\email{alexander.mietke@physics.ox.ac.uk}
\affiliation{Rudolf Peierls Centre for Theoretical Physics, Department of Physics, University of Oxford, Parks Road, Oxford OX1 3PU, United Kingdom}

\begin{abstract}
Odd viscoelasticity arises in parity-violating nonequilibrium materials, where it leads to unconventional mechanical responses and oscillatory relaxation even in overdamped systems. While many living and active chiral materials present promising candidates to exhibit odd viscoelasticity, there is currently no approach that allows for a rheological inference of the large number of elastic and viscous moduli that even a minimal isotropic odd viscoelastic material can depend on. Generalizing the century-old Papkovich-Neuber ansatz to active materials, our work introduces an odd Papkovich-Neuber (OPN) solution -- an analytic solution for any isotropic linear odd fluid or solid, each described by up to 6 independent moduli -- that enable us to study the boundary-driven response in geometries that mimic common rheology methods. OPN solutions reveal three physically distinct resonances in odd viscoelastic solids that are characteristic of the underlying material moduli and can all be interpreted within a single geometric framework. Underlying this unification is an equivalent description of overdamped odd viscoelastic materials in terms of damped harmonic oscillators. Resonances appear as the effective damping coefficients of these oscillators vanish, which is facilitated by the activity that powers odd material properties. 

\end{abstract}


\maketitle

\setlength{\parindent}{0pt}

\section{Introduction}
Odd material properties emerge in systems with broken time-reversal and parity symmetry~\cite{fruchart2023odd}. They have been experimentally observed, or proposed to play a key role, in both synthetic~\cite{soni2019odd,mecke23,veenstra2025adaptive} and living systems~\cite{tan2022,shankar2024active,gu2025emergence}. Biological materials are particularly promising candidates to exhibit odd properties: They are inherently out of thermodynamic equilibrium and often contain chiral components, which manifest themselves most prominently during developmental left-right symmetry breaking processes~\cite{essner2005kupffer,schweickert2007cilia,naganathan2014active,pfanzelter2025active}. Many theoretical predictions have been made about unconventional phenomena that odd systems give rise to, including the possibility to extract work from deformation cycles~\cite{Scheibner:2020, Sousl2021}, lift forces acting on moving particles~\cite{hosaka2021nonreciprocal,lier2023lift}, pattern formation mechanisms~\cite{floyd2024pattern, de2024pattern}, and collision enhanced diffusion~\cite{kalz2022collisions}. Understanding if such phenomena are realized in biological systems is crucial to gain insight into the role of odd properties for biological function and to exploit biological matter for the design of actively responding materials~\cite{chao2026selective}.

Biological materials exhibit both fluid-like and solid-like behavior, depending largely on the time-scale at which they are actuated~\cite{Forgacs1998viscoelastic, serwane2017}. This generally viscoelastic behavior suggests living matter could harbor odd viscoelastic material properties. From a rheological inference perspective, this presents a major challenge as even a minimal isotropic odd viscoelastic material can in principle depend on up to~12 phenomenological couplings that connect deformations, flows and stresses. Current measurements of odd material parameters in viscous and viscoelastic systems typically exclude a large number of moduli \textit{a priori} and mostly rely on comparing indirect observables, such as dispersion relations~\cite{soni2019odd}, density variations~\cite{mecke23}, local strains near topological defects~\cite{tan2022}, and properties of coarse-grained cycles~\cite{shankar2024active}, with theoretical predictions. Overcoming these limitations requires tractable analytical methods to solve the general equations of motion of spatially extended viscoelastic materials in geometries that mimic common rheological approaches. Owing to the linearity of the force balance equations of odd elastic solids and odd viscous fluids, numerous partial analytic solutions already exist. These include singularity solutions~\cite{braverman2020topological, veenstra2025adaptive, hosaka2021nonreciprocal,khain2022stokes}, dispersion relations~\cite{Scheibner:2020, chen2021realization, floyd2024pattern, de2024pattern, soni2019odd, banerjee2021active, bililign2021chiral,lee2025disordered,lee2026nonhermitian}, solutions with preimposed spatial symmetry~\cite{lier2023lift, khain2022stokes, duclut2024probe, hosaka2021hydrodynamic, hosaka2021nonreciprocal, soni2019odd, bililign2021chiral}, or solutions in which only a subset of all possible moduli is included~\cite{hosaka2023lorentz, hosaka2021nonreciprocal, hosaka2021hydrodynamic, ganeshan2017, banerjee2017odd, avron1998odd}.

Despite these analyses and the general experimental need, there exists so far no method to solve general boundary-value problems for isotropic odd materials in different geometries and without any constraints on elastic and viscous moduli. Consequently, it is also not known if each and every modulus affects the material response in unique ways, or if there are degeneracies and unifying principles of how specific groupings of moduli impact the response. This is in contrast to passive linearly elastic systems and overdamped (Stokes) fluids, for which many general boundary-value problem solutions exist, such as Lamb's solution~\cite{lamb1932}, and methods including the Galerkin vector~\cite{GalerkinB1930Cttg} and the Papkovich-Neuber ansatz~\cite{papkovich1932,neuber1934}. In this work, we introduce the odd Papkovich-Neuber ansatz (OPN), which generalizes the Papkovich-Neuber ansatz for linear elastic equilibrium solids and Stokes fluids to analytically solve boundary-value problems for any linear isotropic odd material in two dimensions. We use the OPN solutions to quantify hallmark phenomena of spatially extended odd solids and to study resonances in odd viscoelastic materials under different boundary excitations. Our approach reveals three physically distinct types of odd resonances in general viscoelastic solids that can however all be interpreted within a single geometric framework. Underlying this unification is an equivalent description of \textit{overdamped} odd viscoelastic materials in terms of \textit{damped} harmonic oscillators. The activity that powers odd material properties can make the effective damping coefficient of these oscillators arbitrarily small, which gives rise to resonances that are characteristic of the underlying material properties. 

\section{Papkovich-Neuber solutions} \label{sec:PN Sols}
The classical Hooke's law of passive linear elasticity in two dimensions connects the stress $\boldsymbol{\sigma}$ in an isotropic two-dimensional solid to the displacement field $\boldsymbol{u}$ via a constitutive law~\cite{land70}
\begin{equation}\label{eq:constlawlinel}
\boldsymbol{\sigma}=B\nabla\cdot\boldsymbol{u}\,\mathbb{I} + \mu\left(\nabla\mathbf{u}+(\nabla\mathbf{u})^T-\nabla\cdot\boldsymbol{u}\,\mathbb{I}\right),
\end{equation}
where $B$ and $\mu$ are the bulk and shear moduli, respectively. In the absence of external forces, the force balance $\nabla\cdot\boldsymbol{\sigma}=0$ can be written as
\begin{equation}\label{eq:PassiveLinearElasticForceBalance}
a_0\nabla^2 \boldsymbol{u} + \nabla (\nabla \cdot \boldsymbol{u}) = 0,
\end{equation}
with an effective material parameter $a_0=\mu/B=(1 - \nu_0)/(1 + \nu_0)$, where $\nu_0$ is the two-dimensional Poisson ratio of an isotropic equilibrium solid~\cite{Scheibner:2020}. Papkovich and Neuber realized almost a hundred years ago \hbox{\cite{papkovich1932, neuber1934}} that Eq.~(\ref{eq:PassiveLinearElasticForceBalance}) is solved by an ansatz of the form~\cite{bower2025applied}
\begin{equation}\label{eq:ElasticPassivePNAnsatz}
\boldsymbol{u}(\boldsymbol{r}) = 2\left(a_0 + 1\right) \boldsymbol{B} - \nabla (\boldsymbol{r} \cdot \boldsymbol{B} + B_0),
\end{equation}
as long as $\boldsymbol{B}$ and $B_0$ are harmonic vector and scalar fields respectively, i.e. they must satisfy $\nabla^2 \boldsymbol{B} = 0, \nabla^2 B_0 = 0$. In separable coordinates, harmonic functions and vector fields can be easily found analytically, which makes the Papkovich-Neuber ansatz Eq.~(\ref{eq:ElasticPassivePNAnsatz}) a powerful tool to find exact solutions to boundary values problems in many different geometries. 

A similar ansatz can also be used for overdamped viscous fluid flows. This is because the stress in compressible Newtonian viscous fluids takes the same form as Eq.~(\ref{eq:constlawlinel}), only with the strain tensor replaced by the strain rate tensor and elastic moduli replaced by bulk and shear viscosities, $\eta_b$ and~$\eta_s$, respectively. This leaves the structure of the force balance equation unchanged and therefore allows for solutions of the form Eq.~(\ref{eq:ElasticGeneralisedPNAnsatz}) with $a_0=\eta_s/\eta_b$. With minor modifications, a Papkovich-Neuber solution can also be constructed for incompressible viscous fluids (Stokes flow) \cite{cong1982}.

In the following, we will show how this method can be generalized further to find deformation and flow solutions for isotropic nonequilibrium materials, in particular odd elastic and viscoelastic solids. We will use the resulting solutions to study resonances in overdamped odd materials.

\section{Odd Papkovich-Neuber Solution}\label{sec:OPN}
The most general constitutive law for a linear elastic material reads with Einstein notation
\begin{equation}\label{eq:ElasticInfinitessimalStrain}
\sigma_{ij} = C_{ijkl}u_{kl},
\end{equation}
where Latin indices denote Cartesian coordinates $(x,y)$, $C_{ijkl}$ is the elastic modulus tensor, and $u_{ij} = \partial_i u_j$ denotes the deformation gradient. For brevity, we neglect in Eq.~(\ref{eq:ElasticInfinitessimalStrain}) any displacement-independent prestress. Finite prestress can be incorporated into OPN solutions as an external forcing~\cite{SItext}, which does not affect our conclusions. The constitutive laws of odd elasticity can be introduced most conveniently using a tensor basis~\cite{Scheibner:2020} 
\begin{equation}\label{eq:ScheibnerBasis}
\boldsymbol{s}^0 = \big(\begin{smallmatrix}
    1 & 0 \\ 0 & 1
\end{smallmatrix} \big), \boldsymbol{s}^1 = \big(\begin{smallmatrix}
    0 & -1 \\ 1 & 0
\end{smallmatrix} \big), \boldsymbol{s}^2 = \big(\begin{smallmatrix}
    1 & 0 \\ 0 & -1
\end{smallmatrix} \big), \boldsymbol{s}^3 = \big(\begin{smallmatrix}
    0 & 1 \\ 1 & 0
\end{smallmatrix} \big),
\end{equation}
where the components of each $\boldsymbol{s}^\alpha$ are given with respect to the Cartesian basis. Displacement, stress and modulus tensor components in this basis are
$u^{\alpha} = s^{\alpha}_{ij} u_{ij}$, $\sigma^{\alpha} = s^{\alpha}_{ij} \sigma_{ij}$ and \smash{$C_{\alpha \beta} = \frac{1}{2} s^{\alpha}_{ij} s^{\beta}_{kl} C_{ijkl}$}, respectively, and Eq.~(\ref{eq:ElasticInfinitessimalStrain}) becomes
\begin{equation}\label{eq:stresscheib}
\sigma^{\alpha} = C_{\alpha \beta} u^{\beta}.    
\end{equation}

Strain tensor components $u^0, u^1$ correspond to dilation and rotation of the material respectively, and $u^2, u^3$ correspond to the two independent shear components in two dimensions. Similarly, the components of $\sigma^\alpha$ correspond to isotropic contractile and extensile stress ($\alpha=0$), torques ($\alpha=1$), and shear stresses ($\alpha=2,3$). The most general modulus tensor $C_{\alpha \beta}$ for an isotropic material is then given by~\cite{Scheibner:2020}
\begin{equation}\label{eq:ElasticModulusTensor}
    C_{\alpha \beta} = 2\begin{pmatrix}
        B & \Lambda & 0 & 0 \\
        A & \Gamma & 0 & 0 \\
        0 & 0 & \mu & K^o \\
        0 & 0 & -K^o & \mu \\
    \end{pmatrix}
\end{equation}
and contains in addition to the passive moduli from Eq.~(\ref{eq:constlawlinel}) also odd moduli that couple isotropic deformations to torques (odd bulk modulus $A$) and antisymmetrically mix shear components (odd shear modulus $K^o$). These moduli cannot be derived from an elastic free energy and therefore can only exists in active nonequilibrium materials. For completeness, we also included moduli $\Lambda$ and $\Gamma$ that couple rotations to stress and therefore require interactions with a substrate~\cite{braverman2020topological}. 

The force balance $\partial_i\sigma_{ij}=0$ for a stress described by Eq.~(\ref{eq:stresscheib}) and modulus tensor given in Eq.~(\ref{eq:ElasticModulusTensor}) can be written as
\begin{equation}\label{eq:OddElasticForceBalanceModuliVersion}
    \nabla^2[(\mu' \mathbb{I} + {K^o}' \boldsymbol{\epsilon}] \cdot \boldsymbol{u} + [B' \mathbb{I} + A' \boldsymbol{\epsilon}] \cdot \nabla (\nabla \cdot \boldsymbol{u}) = 0,
\end{equation}

where $\mu' = \mu + \Gamma, {K^o}' = K^o - \Lambda, B' = B - \Gamma, A' = A + \Lambda$ and $\boldsymbol{\epsilon}$ is the 2D Levi-Civita pseudo-tensor. The substrate-dependent moduli $\Lambda$ and~$\Gamma$ are therefore absorbed into effective bulk and shear moduli and are only explicitly relevant at boundaries. We drop the $'$s in following for convenience. Matrices of the form $m_1\mathbb{I} + m_2 \boldsymbol{\epsilon}$ describe scaling and rotations, hence we can write Eq.~(\ref{eq:OddElasticForceBalanceModuliVersion}) as
\begin{equation}\label{eq:ElasticGeneralForceBalance}
    a \nabla^2 \boldsymbol{u} + \boldsymbol{R}(\phi) \cdot \nabla (\nabla \cdot \boldsymbol{u}) = 0,
\end{equation}
where
\begin{equation}\label{eq:aodd}
a = \sqrt{\frac{\mu^2 + (K^o)^2}{B^2 + A^2}},
\end{equation}
and $\boldsymbol{R}(\phi)=\cos\phi\,\mathbb{I}-\sin\phi\boldsymbol{\epsilon}$ describes rotations by an angle~$\phi$ defined by
\begin{equation}\label{eq:OddElasticParameterDefs}
\begin{gathered}
    \cos\phi = \frac{\mu B + A K^o}{\sqrt{(B^2 + A^2)(\mu^2 + (K^o)^2)}},
    \\
    \sin\phi = \frac{B K^o - \mu A}{\sqrt{(B^2 + A^2)(\mu^2 + (K^o)^2)}}.
\end{gathered}
\end{equation}
It is instructive to directly compare the generalized force balance Eq.~(\ref{eq:ElasticGeneralForceBalance}) with the force balance Eq.~(\ref{eq:PassiveLinearElasticForceBalance}) of an equilibrium solid. The latter is recovered for $\phi=0\,(\Leftrightarrow K^o=A=0$ or $K^o/\mu=A/B$), suggesting the angle $\phi$ is a natural measure for the effective oddness of the material. Indeed, the rotation matrix in Eq.~(\ref{eq:ElasticGeneralForceBalance}) not only modifies the conventional Poisson ratio to $\nu=(\cos\phi-a)/(\cos\phi+a)$~\cite{Scheibner:2020,braverman2020topological,SItext}, but it also introduces a new odd ratio $\nu^o$  \cite{Scheibner:2020,braverman2020topological,SItext} satisfying the relationship $\tan\phi = 2 \nu^o/(1 + \nu)$. In odd elastic materials, $\phi$ appears as twice the angle by which the shear strain axis is locally rotated everywhere (compared to a passive material), in the presence of a topological defect~\cite{braverman2020topological}.

A first key result that we introduce and use throughout this work is the fact that the force balance for a general isotropic linear active solid Eq.~(\ref{eq:ElasticGeneralForceBalance}) is solved by an \textit{odd Papkovich-Neuber (OPN) ansatz} (see~\cite{SItext}, Sec.~\ref{appsec:OPN})
\begin{equation}\label{eq:ElasticGeneralisedPNAnsatz}
    \boldsymbol{u}(\boldsymbol{r}) = 2(a + \cos\phi) \boldsymbol{B} - \boldsymbol{R}(\phi) \cdot \nabla (\boldsymbol{r} \cdot \boldsymbol{B} + B_0),
\end{equation}
for any harmonic scalar and vector fields $B_0$ and $\boldsymbol{B}$, respectively. As expected, the OPN ansatz becomes equivalent to the conventional ansatz Eq.~(\ref{eq:ElasticPassivePNAnsatz}) when the effective oddness of the material vanishes, i.e. if $\phi=0$ (\hbox{$a\rightarrow a_0$, $\boldsymbol{R}(0)=\mathbb{I}$)}.
\subsection{General force balance solution in polar coordinates}\label{subsection: Displacement BCs}
To demonstrate how the OPN solution can be used to gain insights into the unconventional response of odd materials, we consider first an odd elastic solid with axisymmetric reference state and evaluate Eq.~(\ref{eq:ElasticGeneralisedPNAnsatz}) in polar coordinates ($r$,$\theta$).
Writing harmonic scalar and vector fields $B_0$ and $\mathbf{B}$, respectively, in polar coordinates in terms of angular modes $\sin(n \theta), \cos(n \theta)$  ($n \in \mathbb{Z}$) and substituting them into the OPN ansatz Eq.~(\ref{eq:ElasticGeneralisedPNAnsatz}), one finds the general solution (\cite{SItext}, Sec.~\ref{Appendix Section: General Solution Form})
\begin{equation}\label{eq:ElasticSolutionSumOverModes}
\boldsymbol{u}(r, \theta) = \sum_{n=-\infty}^{\infty} \boldsymbol{u}^{(n)}(r, \theta),
\end{equation}
with
\begin{align}
    \boldsymbol{u}^{(n)}(r, \theta) &= \frac{1}{r^{n+1}} \boldsymbol{R}(n \theta)\cdot\boldsymbol{\alpha}^{(n)} \notag\\ 
    &+ \frac{1}{r^{n-1}} \left[(n-1) \boldsymbol{\Tilde{R}} + \boldsymbol{Z}\right]\cdot\boldsymbol{R}(n \theta)\cdot\boldsymbol{\beta}^{(n)}.\label{eq:nthModeElasticSolution}
\end{align}
In this expression, we have omitted logarithmic terms that are associated with the Stokes paradox~\cite{lamb1932} and multivalued terms, which can used to construct displacement solutions around dislocations~\cite{braverman2020topological} (\cite{SItext}, Sec.~\ref{Appendix Section: Topological Defects}), as well as a Green's tensor (\cite{SItext}, Sec.~\ref{appsec:OPN with forcing}). The auxiliary matrices $\boldsymbol{\Tilde{R}}$ and $\boldsymbol{Z}$ read
\begin{align}
\boldsymbol{\Tilde{R}} &= \frac{1}{4a^2 + 4a \cos\phi + 1} \left( 2a \boldsymbol{R}(\phi) + \mathbb{I} \right)\label{eq:RTildeDefinition}\\
\boldsymbol{Z} &= \boldsymbol{e}_r \boldsymbol{e}_r - \boldsymbol{e}_\theta \boldsymbol{e}_\theta. \label{eq:ZDefinition}
\end{align}

The matrix $\boldsymbol{R}(n \theta)$ in Eq.~(\ref{eq:nthModeElasticSolution}) describes rotations by angle~$n \theta$, and $\boldsymbol{e}_r, \boldsymbol{e}_\theta$ are the usual radial and azimuthal unit basis vectors, respectively. The vectors \smash{$\boldsymbol{\alpha}^{(n)}= \alpha_1^{(n)} \boldsymbol{e}_r + \alpha_2^{(n)} \boldsymbol{e}_{\theta}$ and $\boldsymbol{\beta}^{(n)}=\beta_1^{(n)} \boldsymbol{e}_r + \beta_2^{(n)} \boldsymbol{e}_{\theta}$} collect the integration constants $\alpha_{k}^{(n)}, \beta_{k}^{(n)}$. Displacements proportional to \smash{$\boldsymbol{\alpha}^{(n)}$} are pure shear deformations, whilst $\boldsymbol{\beta}^{(n)}$ terms contribute rotational, dilational and shear deformations. The sign of $n$ determines the regularity of solutions at the origin and at infinity. The terms $\sim\boldsymbol{\alpha}^{(-1)}$ and $\sim\boldsymbol{\beta}^{(1)}$ are linearly dependent, and contain translational modes. This degeneracy of the solution corresponds to requiring force-free boundary conditions, which naturally avoids the Stokes paradox~\cite{lamb1932}. In the exemplary applications of this solution that follow, we will focus on the non-trivial modes~$|n|\ge2$.

\begin{figure}[t]
 	\includegraphics[width = 0.95\columnwidth]{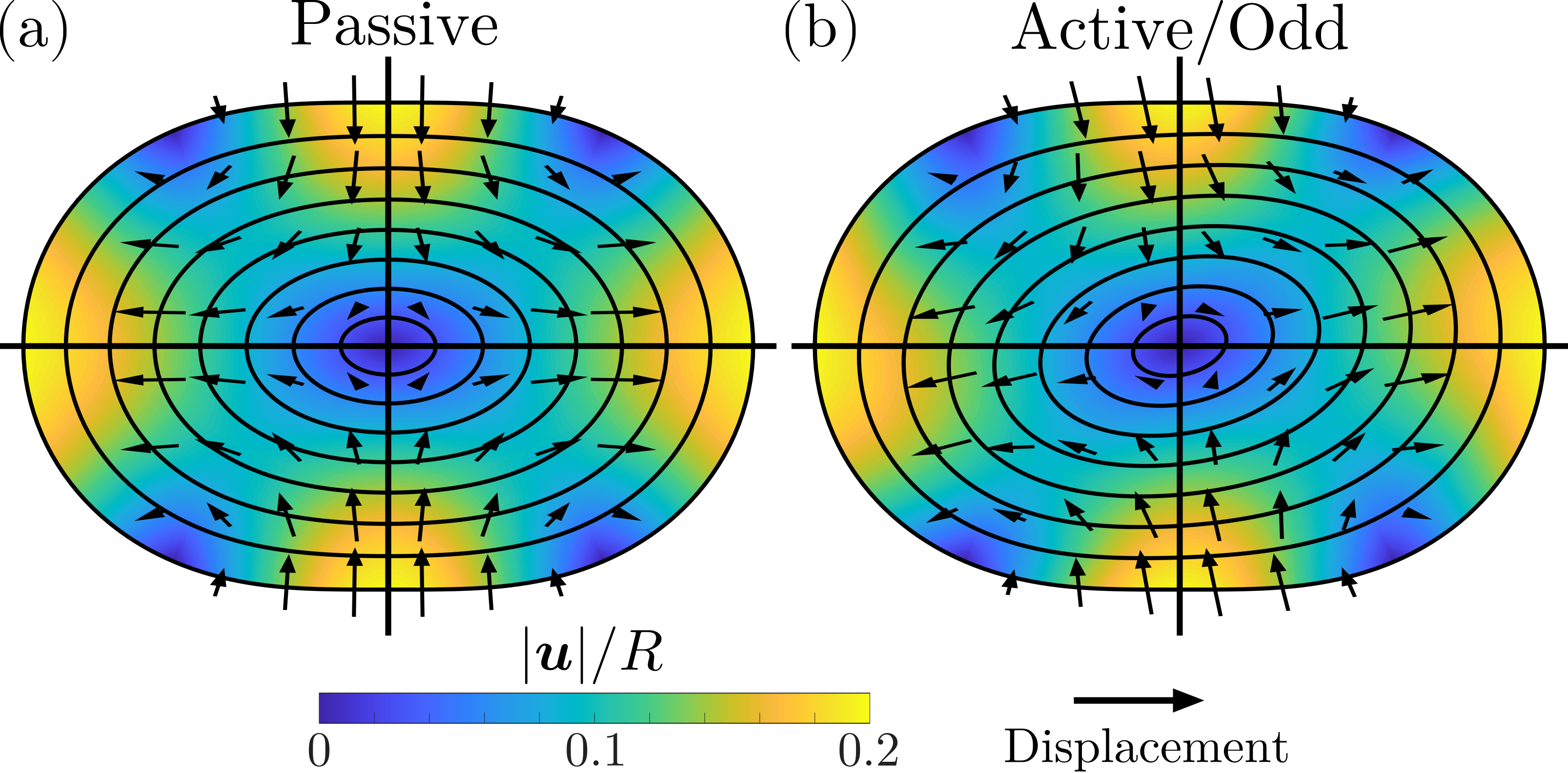}\vspace{-0.1cm}
	\caption{\textbf{Shear axis rotation of an odd disk.} \textbf{(a)}~Passive elastic material ($A = K^o = 0$) deformed under nematic DBCs [see Eq.~(\ref{eq:ElasticNematicBCs})]. Black lines show deformed equiradial lines, color represents magnitude of displacement, and arrows show displacement field. The symmetry of the applied shear is reflected in the symmetry of the displacement field. \textbf{(b)}~Analytic solution of an odd elastic material ($A = -1, K^o = 1$) deformed under same BCs as~(a). Chirality associated with odd elastic properties leads to a displacement field with a rotated shear axis. Other moduli: $B = 2, \mu =1, \Gamma = \Lambda = 0$.}
	\label{fig:ShearAxisRotation}
\end{figure}

\subsection{Shear axis rotation}
A hallmark feature of odd elastic materials is the emergence of a finite angle between the axes of the imposed shear and the shear response~\cite{Scheibner:2020, braverman2020topological}. To quantify this effect using OPN solutions, we consider a disk of radius $R$ and impose a displacement boundary condition (DBC) $\boldsymbol{\hat{u}}(\theta):=\boldsymbol{u}(R,\theta)$ with
\begin{equation}\label{eq:ElasticNematicBCs}
\boldsymbol{\hat{u}}(\theta) = u_0\cos(2 \theta) \boldsymbol{e}_r,
\end{equation}
where $u_0$ is the displacement magnitude. Such a nematic boundary deformation stretches the disk horizontally and compresses it vertically. Since any spatially constant strain tensor automatically satisfies the force balance Eq.~(\ref{eq:ElasticGeneralForceBalance}), Eq.~(\ref{eq:ElasticNematicBCs}) is the lowest mode of a DBC for which odd elasticity affects displacement in the bulk. Regularity at the origin and the symmetry of the boundary condition Eq.~(\ref{eq:ElasticNematicBCs}) suggest the solution of the force balance Eq.~(\ref{eq:ElasticGeneralForceBalance}) for this problem is given by the displacement mode Eq.~(\ref{eq:nthModeElasticSolution}) with $n=-2$. Indeed, direct comparison of $\boldsymbol{\hat{u}}^{(-2)}(R,\theta)$ with the boundary displacement Eq.~(\ref{eq:ElasticNematicBCs}) yields the solution
\begin{equation*}
    \boldsymbol{u}(r,\theta) =\frac{u_0r}{2R}\left[3 \boldsymbol{\Tilde{R}} + \mathbb{I} + \frac{r^2}{R^2} \left( -3 \boldsymbol{\Tilde{R}} + \boldsymbol{Z} \right) \right]\cdot\boldsymbol{R}^T(2 \theta)\cdot\boldsymbol{e}_r.
\end{equation*}
The strain tensor associated with this displacement at the origin can be expressed in terms of the tensor basis Eq.~(\ref{eq:ScheibnerBasis}) as
\begin{equation}\label{eq:nuorig}
    \nabla \boldsymbol{u} (0, \theta) = \frac{u_0}{2R} \left[ (3 \Tilde{R}_e + 1)\boldsymbol{s}^2 + 3 \Tilde{R}_{o} \boldsymbol{s}^3 \right],
\end{equation}
where the components of the moduli-dependent matrix \hbox{$\boldsymbol{\Tilde{R}} =  \Tilde{R}_e\mathbb{I} - \Tilde{R}_o \boldsymbol{\epsilon}$} given in Eq.~(\ref{eq:RTildeDefinition}) can be read off from
\begin{figure*}[t]
\includegraphics[width = 2.0\columnwidth]{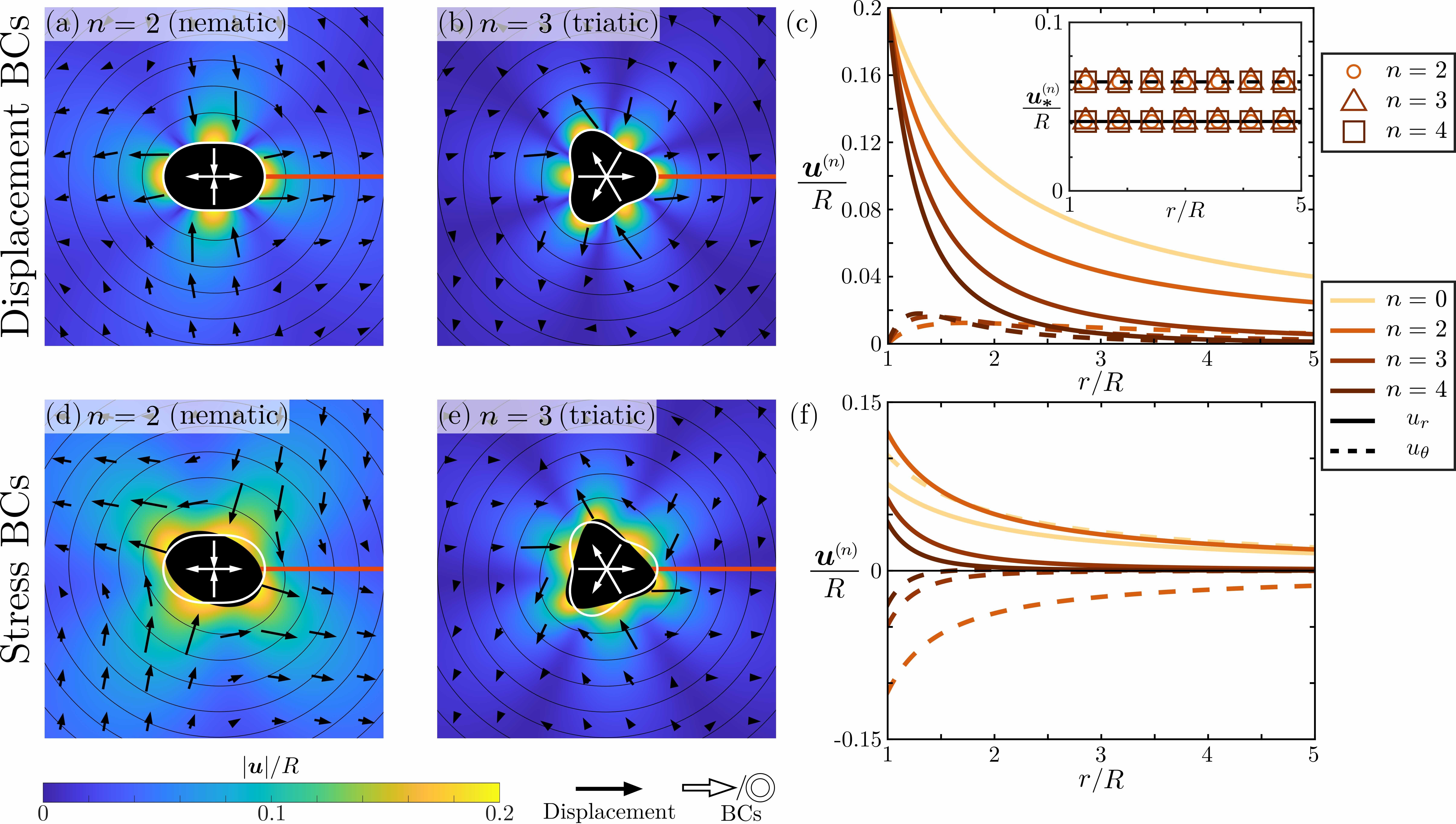}\vspace{-0.1cm}
    \caption{\textbf{Analytic displacement field solutions of odd elastic solids around cavities.} \textbf{(a)}~Displacement field around nematically deformed cavity [$P_2 \neq 0$ in Eq.~(\ref{eq:GeneralnthModeBcs})]. \textbf{(b)}~Same as (a) for a triatically deformed cavity ($P_3 \neq 0$). \textbf{(c)}~Radial (solid lines) and azimuthal (dashed lines) displacement field components for different symmetries of the cavity boundary displacement. Components are shown along the horizontal red lines in (a) and (b). Inset shows that suitably transformed displacement field solutions \hbox{\smash{$\mathbf{u}^{(n)}_*=\mathbf{R}^T(n\theta)\cdot\mathbf{u}_M/f_n(r)$}} [see Eq.~(\ref{eq:uM})] collapse for different cavity symmetries. \textbf{(d)}~~Displacement field around cavity deformed by stress boundary conditions (SBCs, see~\cite{SItext}) with nematic symmetry. \textbf{(e)}~Same as (d), but BCs have three-fold symmetry. SBCs are such that the boundary forces $\boldsymbol{\hat{f}}=-\mathbf{e}_r\cdot\boldsymbol{\sigma}|_R$ (white lines) are parallel to boundary displacements in (a) and (b). \textbf{(f)}~Same as (c), but displacement fields do not collapse. Elastic moduli: \hbox{$B = 4/3$}, \hbox{$\mu = 1$}, \hbox{$A = -2/3$}, \hbox{$K^o = 4/3$}, \hbox{$\Gamma = \Lambda = 0$}.}
    \label{fig:Elastic}
\end{figure*}
\begin{align}
\boldsymbol{\Tilde{R}}=\frac{2 a \cos\phi + 1}{4 a^2 + 4 a \cos\phi + 1}\mathbb{I}-\frac{2 a \sin\phi}{4 a^2 + 4 a \cos\phi +1}\boldsymbol{\epsilon}.
\end{align}

For an achiral material, parity invariance implies that the symmetry of the shear profile imposed along a boundary will be maintained in the bulk. For the DBC given in Eq.~(\ref{eq:ElasticNematicBCs}), whose principle shear axes are captured by $\boldsymbol{s}^2$ [see Eq.~(\ref{eq:ScheibnerBasis})], we indeed find the displacement at the origin Eq.~(\ref{eq:nuorig}) for $\phi=0$ contains only the shear strain component \hbox{$u^2=\text{tr}(\boldsymbol{s}^2\cdot\nabla\mathbf{u})\ne0$}, while $u^3=\text{tr}(\boldsymbol{s}^3\cdot\nabla\mathbf{u})=0$~(Fig.~\ref{fig:ShearAxisRotation}a). However, once there is a finite effective oddness in the material ($\phi\ne0$), the corresponding chirality leads to $\Tilde{R}_o\ne0$ in Eq.~(\ref{eq:nuorig}) and therefore $u^3 \neq 0$, i.e. we find a rotation of the shear axis of the deformed disk relative to the axis of the applied shear~(Fig.~\ref{fig:ShearAxisRotation}b). Denoting the angle of rotation of the shear axis by $\theta_s$, Eq.~(\ref{eq:nuorig}) implies
\begin{equation}\label{eq:dts}
\tan(2\theta_{\text{s}}) = \frac{3 a \sin\phi}{2a^2 + 5a \cos\phi + 2}.
\end{equation}
Interestingly, the angle of shear axis rotation $\theta_{\text{s}}$ is \textit{independent} of disk size and of the magnitude of the applied boundary strain. Instead, it is entirely determined by the material parameters of the odd solid and might therefore provide a robust experimental measure to probe effective material properties of such solids. 

\subsection{General solution for displacement boundary conditions}\label{subsection:General Elastic Solution DBCs}
We now consider arbitrary DBCs imposed at a fixed radial coordinate, $r=R$, and write the boundary displacement field~$\boldsymbol{\hat{u}}(\theta)$ in terms of a Fourier series
\begin{align}
    \hat{u}_r(\theta) &= P_0 + \sum_{n=2}^{\infty} \left[ P_n\cos(n \theta) + Q_n \sin(n \theta) \right], \notag \\
    \hat{u}_\theta(\theta) &= M_0 + \sum_{n=2}^{\infty} \left[ M_{n} \cos(n \theta) + N_{n} \sin(n \theta) \right]. \label{eq:GeneralnthModeBcs}
\end{align}
Adapting the general solution Eq.~(\ref{eq:nthModeElasticSolution}) to this boundary condition for solutions regular at infinity yields for $n\ge2$
\begin{align}
    \boldsymbol{u}^{(n)}(r, \theta) &= \notag\\
    &\hspace{-1.6cm}+\frac{R^n}{2r^n} \Bigg\{ \frac{R}{r} \Big[ (1-n)\boldsymbol{\Tilde{R}} + \mathbb{I} \Big]+\frac{r}{R} \Big[(n - 1) \boldsymbol{\Tilde{R}} + \boldsymbol{Z} \Big] \Bigg\}\notag\\
    &\hspace{-1.6cm}\cdot\boldsymbol{R}(n \theta)\cdot(P_{n} \boldsymbol{e}_r - Q_n \boldsymbol{e}_\theta) \notag\\
    &\hspace{-1.6cm}-\frac{R^n}{2r^n} \Bigg\{ \frac{R}{r} \Big[ (1-n) \boldsymbol{\Tilde{R}} - \mathbb{I} \Big]+\frac{r}{R} \Big[(n-1) \boldsymbol{\Tilde{R}} + \boldsymbol{Z} \Big] \Bigg\}\notag\\
    &\hspace{-1.6cm}\cdot\boldsymbol{R}(n \theta)\cdot(N_n \boldsymbol{e}_r + M_n \boldsymbol{e}_\theta),\label{eq:nthModeElasticDisplacementBCs}
\end{align}
and $\boldsymbol{u}^{(n)}(r, \theta) = 0$ for $n\le-2$. The solution which is regular at the origin can be found by taking \hbox{\smash{$n \mapsto -n$}} in Eq.~(\ref{eq:nthModeElasticDisplacementBCs}), as well as $P_{-n}\rightarrow P_n, M_{-n}\rightarrow M_n, Q_{-n} \rightarrow -Q_n$, \hbox{$N_{-n} \rightarrow -N_n$} to match the BCs in Eq.~(\ref{eq:GeneralnthModeBcs}). The mode $n=0$ contains a solution that is regular at the origin, \smash{$\boldsymbol{u}^{(0)}_{+}$}, or at infinity, \smash{$\boldsymbol{u}^{(0)}_{-}$}, given by \hbox{$\boldsymbol{u}^{(0)}_{\pm}(r, \theta) = (r/R)^{\pm1}(P_0 \boldsymbol{e}_r + M_0 \boldsymbol{e}_\theta)$}. For annulus geometries with both inner and outer DBCs, a linear combination of the inner and outer displacement field solution Eq.~(\ref{eq:nthModeElasticDisplacementBCs}) can be used.

A crucial insight from this general solution Eq.~(\ref{eq:nthModeElasticDisplacementBCs}) is that the moduli-dependence of displacement fields is -- up to a purely geometric rescaling -- the same for every $n$ and therefore independent of how the boundary is deformed. To see this, we first note that elastic moduli enter the displacement field only through $a$ and $\phi$ in $\boldsymbol{\Tilde{R}}$ [Eq.~(\ref{eq:RTildeDefinition})], which is in turn independent of the angular mode $n$. We can then write the moduli-dependent part of the full displacement field Eq.~(\ref{eq:nthModeElasticDisplacementBCs}) as
\begin{align}
\,\hspace{-0.15cm}\boldsymbol{u}_{\text{M}}^{(n)}(r, \theta) =f_n(r)\,\mathbf{R}(n \theta)\cdot\boldsymbol{u_*}^{(n)}(\theta)\label{eq:uM}
\end{align}
with $f_n(r)=(1-n)R^nr^{-n}(R/r - r/R )/2$, which corresponds to a factorization into the purely geometric, mode-dependent bulk behavior and an auxiliary displacement field
\begin{equation}
\,\hspace{-0.15cm}\boldsymbol{u_*}^{(n)}(\theta)=\boldsymbol{\Tilde{R}}\cdot\left[(P_n \boldsymbol{e}_r - Q_n \boldsymbol{e}_\theta) -  (N_n \boldsymbol{e}_r + M_n \boldsymbol{e}_\theta)\right]
\end{equation}
that encapsulates the full dependence on elastic moduli and on boundary conditions. Probing the material properties of an odd elastic solid by deforming its boundary and measuring bulk displacements is therefore futile: All one can infer from this measurement is the matrix $\boldsymbol{\Tilde{R}}$, which has two independent degrees of freedom, that are in turn parameterized by up to~6, possibly independent, elastic moduli. Therefore, an inference of material properties from a setup in which the material boundary undergoes prescribed deformations has to be constrained by additional priors or alternative measurement protocols have to be developed.

To illustrate the general solution for an experimentally relevant geometry, we use Eq.~(\ref{eq:nthModeElasticDisplacementBCs}) to determine solutions of cavity problems in which the undeformed material extends over $r\in[R,\infty]$ and deformations with different symmetries are imposed at the cavity boundary~(Fig.~\ref{fig:Elastic}a--c). Regularity at infinity allows for excitation only of modes with $n\geq2$ as well as $\boldsymbol{\alpha}^{(0)} \neq 0$, with all other constants of integration vanishing. The displacement fields surrounding radially deformed cavities via a nematic and a triatic profile are shown in Fig.~\ref{fig:Elastic}a~and~b, respectively. The emerging azimuthal displacements $\sim u_{\theta}$ in the material are a consequence of the chirality associated with odd elasticity. Even though the displacement magnitude decays differently for different symmetries of the cavity deformation (Fig.~\ref{fig:Elastic}c), their overall profiles all collapse for fixed material parameters after a geometric transformation implicitly defined by in Eq.~(\ref{eq:uM}) (see Fig.~\ref{fig:Elastic}c, inset). To overcome this degeneracy, one can alternatively use the OPN solution Eq.~(\ref{eq:nthModeElasticSolution}) as a starting point to determine displacement fields generated by different stress profiles imposed at the cavity boundary~\cite{SItext}. Two exemplary solutions, in which the boundary stresses $\boldsymbol{\sigma}|_R$ are such that $\boldsymbol{\hat{f}}=-\mathbf{e}_r\cdot\boldsymbol{\sigma}|_R\propto\cos(n\theta)\mathbf{e}_r$, are shown for nematic ($n=2$) and triatic ($n=3$) force density profiles in Fig.~\ref{fig:Elastic}d~and~e, respectively. In this case, resulting displacement profiles for different $n$ (Fig.~\ref{fig:Elastic}f) are not related anymore by a scaling and contain genuinely different information for different symmetries of the imposed force density. A rheological approach that combines stress boundary conditions (SBCs) with different symmetries is therefore more likely to achieve an unambiguous inference of all relevant elastic moduli if there is no other information available. 

\section{Odd Viscoelasticity}\label{sec: Odd Viscoelasticity}
The formal equivalence between the force balance equations of linear elastic solids and compressible viscous fluids suggests that the OPN ansatz, Eq.~(\ref{eq:ElasticGeneralisedPNAnsatz}), can be used to find force-balanced configurations of viscoelastic isotropic nonequilibrium materials. Such an OPN ansatz would enable us to investigate the impact of essentially arbitrary boundary conditions on finite domains, complementing many earlier works focusing typically on infinitely extended systems~\cite{Scheibner:2020, banerjee2021active, etien2021a, duclut2024probe, floyd2024pattern}. To demonstrate this idea, we consider in the following an odd Kelvin-Voigt (KV) model~\cite{banerjee2021active}, while noting that the same solution method and analysis can be applied to any other linear rheological model. In the tensor basis Eq.~(\ref{eq:ScheibnerBasis}), the constitutive equation for an odd KV model can be written as
\begin{equation}\label{eq:ViscoelasticConstitutiveEquation}
    \sigma^\alpha = (C_{\alpha \beta} + \eta_{\alpha \beta} \partial_t)u^\beta,
\end{equation}
where $C_{\alpha \beta}$ is the elastic modulus tensor Eq.~(\ref{eq:ElasticModulusTensor}), and $\eta_{\alpha \beta}$ is the viscous modulus tensor
\begin{equation}\label{eq:ViscousModuliTensor}
    \eta_{\alpha \beta} = 2\begin{pmatrix}
        \eta_b & \eta_\Lambda & 0 & 0 \\
        \eta_A & \eta_R & 0 & 0 \\
        0 & 0 & \eta_s & \eta^o \\
        0 & 0 & -\eta^o & \eta_s \\
    \end{pmatrix},
\end{equation}
where $\eta_b$ and $\eta_s$ are the conventional bulk and shear viscosities, respectively. Additionally, $\eta^o$ denotes the odd viscosity, or Hall viscosity~\cite{avron1998odd, fruchart2023odd}, which requires microscopic energy input and breaks parity invariance, and $\eta_R$ is the rotational viscosity~\cite{soni2019odd}. The viscous moduli $\eta_A$ and $\eta_\Lambda$ couple isotropic and rotary components of strain rate and stress tensor.

\begin{figure}[t]
 	\includegraphics[width = 0.99\columnwidth]{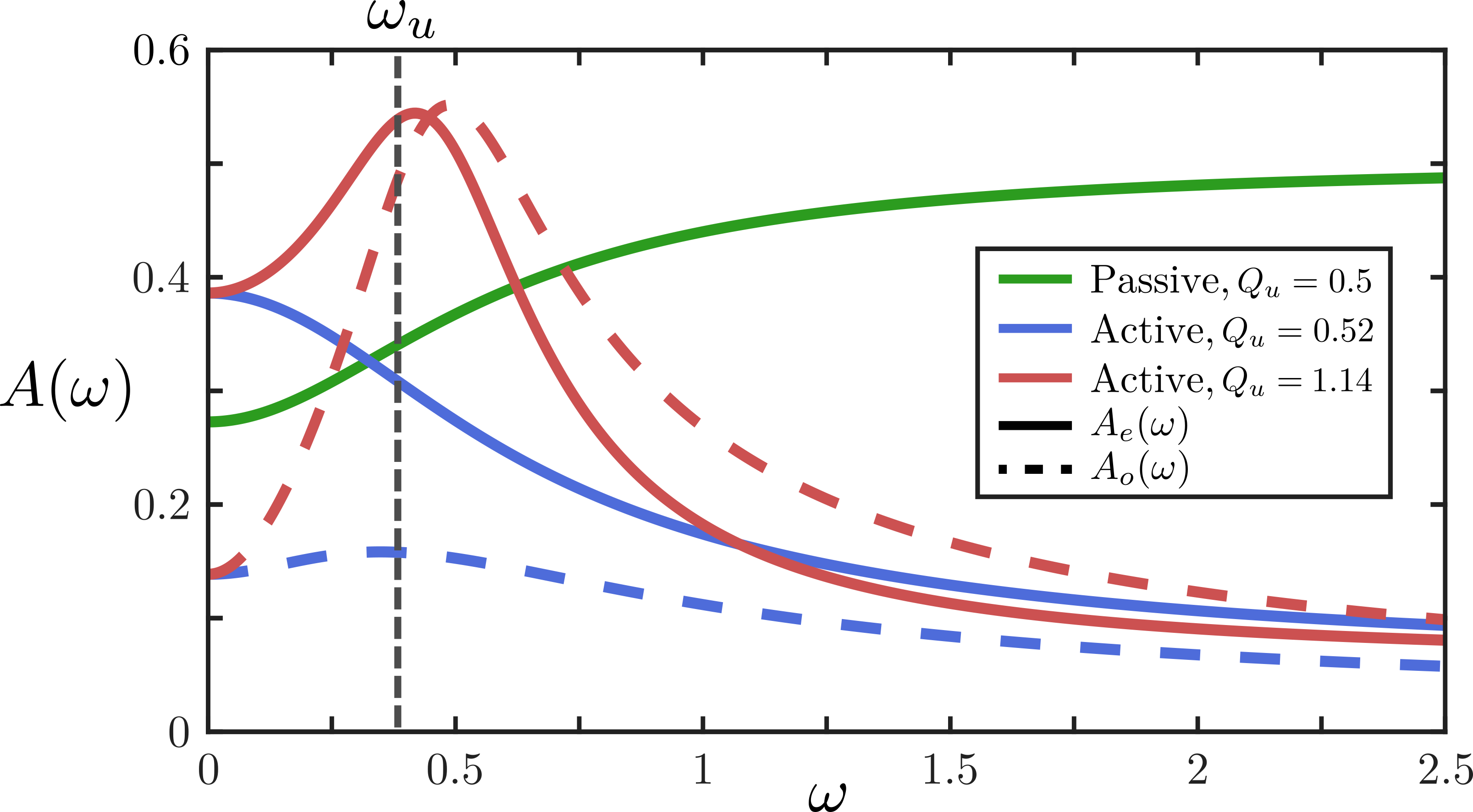}\vspace{-0.1cm}
	\caption{\textbf{Amplitude functions of odd viscoelastic solids driven by displacement boundary conditions.} Oscillation amplitudes $A_e$ (solid lines) and $A_o$ (dashed lines) [Eq.~(\ref{eq:VEgeneralSol})] for three different sets of material moduli and odd quality factors~$Q_{u}$~[Eq.~(\ref{eq:DispResonanceCondition})]. Vertical dashed line indicates intrinsic oscillatory frequency $\omega_u=\text{Im}(s_u^+)$ associated with the pole $s_u$ given in Eq.~(\ref{eq:sdisppm}), which provides a good approximation for the exact resonance frequency. Frequencies are in units $\omega_0=\mu/\eta_s$. Moduli used are given in~\cite{SItext} (Tab. S1).}
	\label{fig:DBCresonances}
\end{figure}

To solve the force balance $\partial_i \sigma_{ij} = 0$ for a stress tensor described by Eq.~(\ref{eq:ViscoelasticConstitutiveEquation}), it is convenient to work in Laplace space~\cite{duclut2024probe}. The force balance reads
\begin{equation}\label{eq:ViscoelasticForceBalanceLaplace}
    \partial_i C^{\text{eff}}_{ijkl}(s) \partial_k U_l(\boldsymbol{x},s) = \partial_i \eta_{ijkl} \partial_k [\boldsymbol{u}_0(\boldsymbol{x})]_l,
\end{equation}

where $s\in\mathbb{C}$ is the Laplace space variable, $\boldsymbol{U}(\boldsymbol{x},s) = \int_0^\infty{\boldsymbol{u}(\boldsymbol{x},t) e^{-s t}}d^2 \boldsymbol{x}$ is the Laplace transform of the displacement field $\boldsymbol{u}(\boldsymbol{x},t)$, $\boldsymbol{u}_0 (\boldsymbol{x}) = \boldsymbol{u}(\boldsymbol{x}, t = 0)$ is the initial displacement field, and $C^{\text{eff}}_{ijkl}$ is an effective, generally complex-valued, modulus tensor that reads in the tensor basis
\begin{equation}\label{eq:Ceff}
C^{\text{eff}}_{\alpha\beta}(s) = C_{\alpha\beta} + s \eta_{\alpha\beta}.
\end{equation}
Note that the left-hand side of Eq.~(\ref{eq:ViscoelasticForceBalanceLaplace}) is by construction formally equivalent to the force balance equation of an odd elastic solid, Eq.~(\ref{eq:ElasticGeneralForceBalance}), with moduli replaced according to Eq.~(\ref{eq:Ceff}). Consequently, the homogeneous solution of Eq.~(\ref{eq:ViscoelasticForceBalanceLaplace}) is given by the OPN ansatz Eq.~(\ref{eq:ElasticGeneralisedPNAnsatz}) using the effective moduli given in Eq.~(\ref{eq:Ceff}). A particular solution to Eq.~(\ref{eq:ViscoelasticForceBalanceLaplace}), which describes the relaxation of the material from some initial perturbed configuration to the reference configuration, can be found using a Helmholtz decomposition \cite{SItext}.

\subsection{Dynamics of an odd viscoelastic solid driven by boundary displacements}\label{sec:OVE DBCs}
We now characterize the long term steady-state dynamics of an odd viscoelastic solid that is periodically driven at the boundary of an enclosed cavity -- analog to the static elastic scenario shown in Fig.~\ref{fig:Elastic}a,b -- mimicking classical rheological approaches~\cite{squires2010microrheology,serwane2017}. The relevant solution is encoded by the homogeneous part of Eq.~(\ref{eq:ViscoelasticForceBalanceLaplace}) and can therefore be found from the OPN solution. We consider a time-periodic DBC with constant frequency $\omega$ and magnitude $u_0$ at the cavity boundary, i.e. we impose $\boldsymbol{\hat{u}}(\theta,t):=\boldsymbol{u}(R,\theta,t)$ with
\begin{equation}\label{eq:Viscoelastic DBCs}
\boldsymbol{\hat{u}}^{(n)}(\theta, t) = u_0\cos(\omega t) \cos(n \theta) \boldsymbol{e}_r.
\end{equation}
The OPN solution Eq.~(\ref{eq:nthModeElasticSolution}) then implies the emergence of an oscillatory steady state at late times, given by (\cite{SItext}, Sec.~\ref{appsec:osci})
\begin{align}\label{eq:VEgeneralSol}
\boldsymbol{u}(r, \theta, t) &= u_0\frac{\cos(\omega t)}{2}\frac{R^n}{r^n} \left( \frac{r}{R} \boldsymbol{Z} + \frac{R}{r} \mathbb{I} \right)\cdot\boldsymbol{R}(n \theta)\cdot\boldsymbol{e}_r\notag\\
&+u_0\frac{(n-1)}{2}\frac{R^n}{r^n}\left(\frac{r}{R} - \frac{R}{r} \right) \boldsymbol{R}(n \theta) \notag\\ 
&\hspace{-1.4cm}\cdot \left[A_e(\omega) \cos(\omega t - \varphi_e) \boldsymbol{e}_r + A_o(\omega) \cos(\omega t - \varphi_o) \boldsymbol{e}_\theta \right],
\end{align}
where $\varphi_e$ and $\varphi_o$ are phase differences between boundary driving and emerging  displacement oscillations. Key for the analysis of resonances are the mode-independent amplitude functions $A_e(\omega)$ and $A_o(\omega)$ that are closely related to the matrix~\smash{$\boldsymbol{\Tilde{R}}$} given in Eq.~(\ref{eq:RTildeDefinition})~\cite{SItext}. These functions capture the purely viscous response as $\omega \rightarrow \infty$ and the purely elastic response as $\omega \rightarrow 0$, and $A_o(\omega)$ vanishes in the absence of effective oddness. Representative examples of $A_e(\omega)$ and $A_o(\omega)$ are illustrated in Fig.~\ref{fig:DBCresonances}, where we find distinct resonance behavior in certain material parameter regimes. To understand these resonances quantitatively, we analyze the intrinsic timescales of the material, which are described mathematically by the poles of the displacement field, $\mathbf{U}(\boldsymbol{x},s)$, in Laplace space~\cite{duclut2024probe} [see Eq.~(\ref{eq:ViscoelasticForceBalanceLaplace})]. These timescales are in turn determined by the poles of the matrix \smash{$\boldsymbol{\Tilde{R}}$} [see Eq.~(\ref{eq:RTildeDefinition})], as this is the only moduli-dependent factor in the solution under DBCs. Poles of \smash{$\boldsymbol{\Tilde{R}}$} arise for $4 a^2 + 4 a \cos\phi + 1 = 0$, where the effective material parameters $a$ and $\phi$ defined in Eqs.~(\ref{eq:aodd})~and~(\ref{eq:OddElasticParameterDefs}), respectively, now become functions $a(s)$ and $\phi(s)$ through the complex moduli substitution Eq.~(\ref{eq:Ceff}). The roots of the resulting expression yield the poles
\begin{align}
    s_{u}^{\pm} &= -\frac{(B + 2 \mu)(\eta_b + 2 \eta_s) + (A + 2 {K^o})(\eta_A + 2 {\eta^o})}{(\eta_b + 2 \eta_s)^2 + (\eta_A + 2 {\eta^o})^2} \notag\\ 
    &\hspace{-0.3cm}\pm i \frac{(B + 2 \mu)(\eta_A + 2 {\eta^o}) - (A + 2 {K^o})(\eta_b + 2 \eta_s)}{(\eta_b + 2 \eta)^2 + (\eta_A + 2 {\eta^o})^2}.\label{eq:sdisppm}
\end{align}

The sign of the real part of $s_{u}^{\pm}$ determines whether deformations grow or decay in time. If \smash{$\text{Re}(s_{u}^\pm) > 0$}, then deformations grow exponentially in time, and the material is unstable. Reintroducing for completeness moduli coupling to rotations, this provides the stability condition
\begin{equation}\label{eq:sdispStabilityCondition}
(A - \Lambda + 2K^o)(\eta_A - \eta_\Lambda + 2 \eta^o) \geq - (B + \Gamma + 2 \mu)(\eta_b + \eta_R + 2\eta_s).
\end{equation}
The left-hand side of Eq.~(\ref{eq:sdispStabilityCondition}) depends on purely antisymmetric active components of the modulus tensors, whilst the right-hand side can be written as a product of traces of the elastic and viscous modulus tensors,  $- C_{\alpha \alpha} \eta_{\beta \beta}$. Bound energies and the second law of thermodynamics imply both $C_{\alpha \alpha}\ge0$ and $\eta_{\alpha \alpha}\ge0$~\cite{katch65}, such that Eq.~(\ref{eq:sdispStabilityCondition}) is trivially satisfied for passive materials. As such, we interpret Eq.~(\ref{eq:sdispStabilityCondition}) as a thermodynamic stability condition, generalizing energetic and entropic constraints for a passive material to an odd one. 

\begin{figure*}[t]
\includegraphics[width = 2.05\columnwidth]{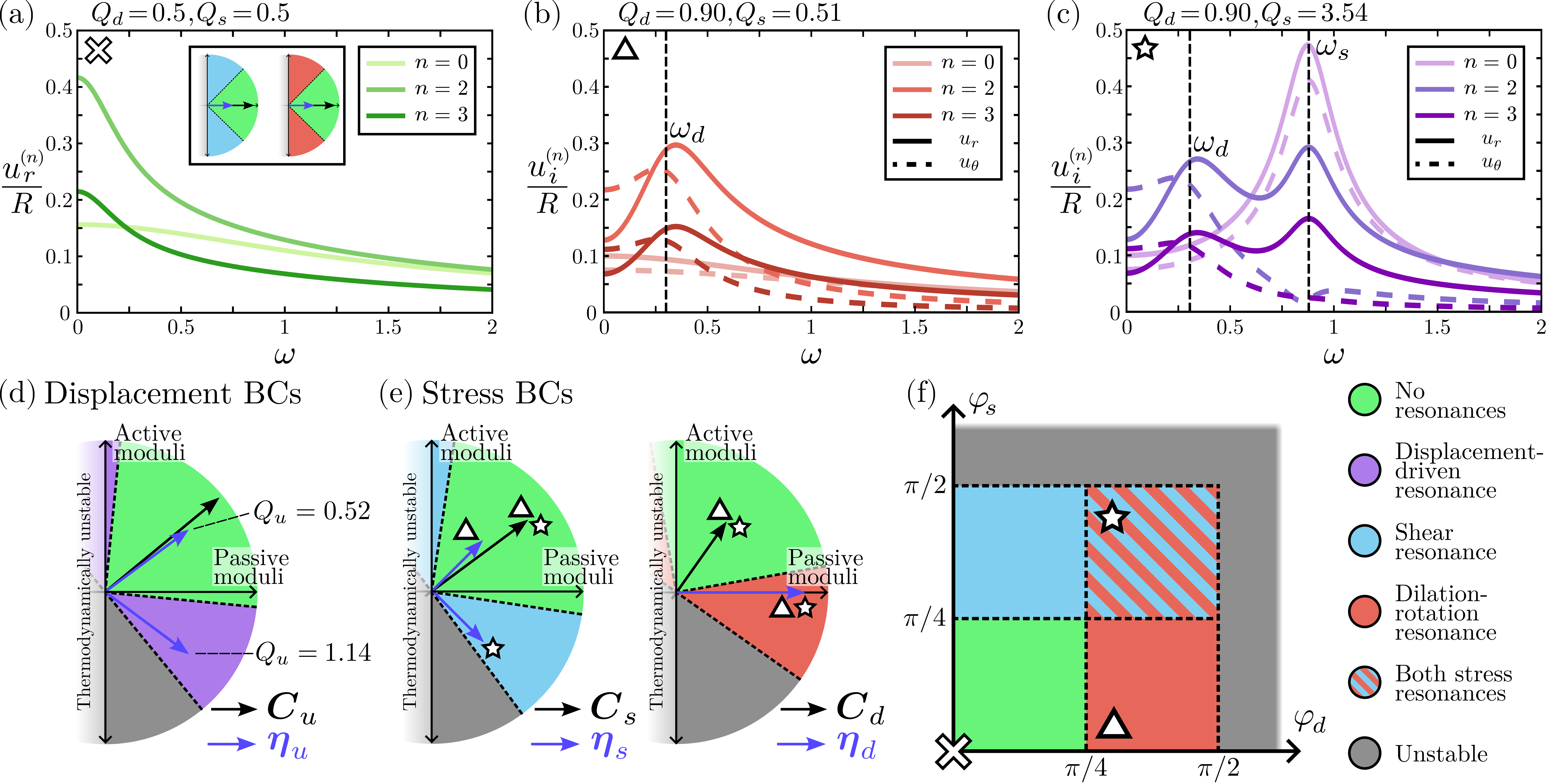}\vspace{-0.1cm}
    \caption{\textbf{Unified representation of resonances in odd viscoelastic solids.} (a)~Maximal radial displacement amplitude at steady state under SBCs [see Eq.~(\ref{eq:fdbc})] with forcing frequency $\omega$ for a passive viscoelastic material. Insets show geometric representations of the poles $s_d^{\pm}$ and $s_s^{\pm}$ given in Eqs.~(\ref{eq:sss}) and (\ref{eq:ssd}), which are trivial when only passive moduli are present. Quality factors $Q_s$ and $Q_d$ are defined in Eqs.~(\ref{eq:Qs}) and (\ref{eq:Qd}). (b)~Same data as in (a) but for an odd viscoelastic material exhibiting a dilation-rotation resonance associated with the pole $s_d^{\pm}$. Vertical dashed line indicates intrinsic oscillatory frequency $\omega_d=\text{Im}(s_d^+)$, which provides a good estimate for the exact resonance frequency. (c)~Same as (b), but for an odd viscoelastic material exhibiting two independently tunable resonances. In addition to the dilation-rotation resonance, there is now also a shear resonance associated with the pole $s_s^{\pm}$, whose imaginary part provides a good approximation $\omega_s$ for the exact resonance frequency. (d)~Geometric representation of resonances and instabilities, via effective elasticity and viscosity vectors within 2D space defined by symmetric and odd parts of the material modulus tensors, for a displacement-driven boundary. In the left half-plane passive moduli take negative values and materials are thermodynamically unstable. (e)~Same as (d), but for a stress-stress driven boundary, with symbols in reference to (e),(f). (f)~Phase diagram for the number of resonances and type of resonance under stress boundary conditions. Symbols indicate where the material moduli for graphs (a)--(c) lie. Frequencies in panels (a)--(c) are in units of $\omega_0=\mu/\eta_s$. Moduli used are given in~\cite{SItext} (Tab. S1).}
    \label{fig:ViscoelasticResonances}
\end{figure*}

The imaginary part, $\text{Im}(s_{u}^\pm)$, indicates oscillatory dynamics, facilitated by odd nonequilibrium properties: In the passive limit ($\eta_A,\eta^o,A,K^o=0$), we have $\text{Im}(s_{u}^\pm)=0$ and no oscillations occur, as expected for the overdamped scenario discussed here. For \smash{$\text{Im}(s_{u}^\pm)\ne0$}, on the other hand, oscillations may occur at various amplitudes. To quantify this, we use the quality factor~\cite{taylor2005classical}
\begin{equation}\label{eq:DispResonanceCondition}
Q_{u}:=-\frac{|s_{u}^\pm|}{2 \, \text{Re}(s_{u}^\pm)}=\frac{1}{2 \cos\varphi_u},
\end{equation}
that is related to the complex phase of the pole via $s_{u}^\pm= - |s_{u}^\pm|e^{\mp i\varphi_u}$. Since the quality factor must be positive, $Q_u$ is only well-defined for $-\pi/2 \leq \varphi_u \leq \pi/2$, which coincides exactly with the thermodynamic stability condition Eq.~(\ref{eq:sdispStabilityCondition}). The quality factor satisfies $Q_u \geq \frac{1}{2}$, corresponding to an underdamped simple harmonic oscillator that is critically damped for $\varphi_u = 0\Rightarrow Q_u=1/2$. Heuristically, we expect resonances to occur when the oscillatory frequency is faster than the decay rate, meaning that the displacement magnitude does not decay notably over one oscillation period, which corresponds to $Q_u > 1/\sqrt{2}$. In this regime, we find that the characteristic frequency \smash{$\omega_{u} = \text{Im}(s_{u}^+)$} provides a very good approximation to the frequency values at which the exact amplitude functions, $A_{e}(\omega)$ and $A_{o}(\omega)$, that appear in Eq.~(\ref{eq:VEgeneralSol}) become maximal (Fig.~\ref{fig:DBCresonances}a, vertical dashed line). 

\subsection{Dynamics of an odd viscoelastic solid driven by boundary stress}\label{sec:OVE SBCs}
We finally consider SBCs, for which a more complex dependence of displacement fields on material moduli gives rise to additional timescales. Specifically, we impose at the boundary of the cavity a force densities $\mathbf{\hat{f}}^{(n)}=-\boldsymbol{e}_r\cdot\boldsymbol{\sigma}^{(n)}|_R$, where
\begin{equation}\label{eq:fdbc}
\mathbf{\hat{f}}^{(n)}(\theta,s) = [R_n(s) \cos(n \theta) + S_n(s) \sin(n \theta)] \boldsymbol{e}_r.
\end{equation}
The solution for this problem on the domain $r > R$ follows from the OPN ansatz and takes the form, for $n \geq 2$,~\cite{SItext}
\begin{align}
\boldsymbol{U}^{(n)}(\boldsymbol{x},s) &= \frac{R^n}{r^n}\left[\frac{R}{r} \boldsymbol{A}^{(n)}(s) + \frac{r}{R} \boldsymbol{B}^{(n)}(s) \right]\cdot\boldsymbol{R}(n \theta)
\notag\\
&\cdot \left[ R_n(s) \boldsymbol{e}_r + S_n(s) \boldsymbol{e}_\theta \right],
\end{align}
where $\boldsymbol{A}^{(n)}(s)$ and  $\boldsymbol{B}^{(n)}(s)$ are matrices that depend on the moduli~\cite{SItext}. $\boldsymbol{A}^{(n)}(s)$ is related to the integration constants~$\boldsymbol{\alpha}^{(n)}$ that are part of the general solution Eq.~(\ref{eq:nthModeElasticSolution}). We recall that the latter describe the coupling between shear strains and shear stresses. Similarly, $\boldsymbol{B}^{(n)}(s)$ is related to the constants~$\boldsymbol{\beta}^{(n)}$ in the general solution, where they describe simultaneous contributions of dilation, rotation, and shear~\cite{SItext}.

The singularities of these matrices contain information about the stability of the material and the presence of resonances. Non-removable singularities arise in $\boldsymbol{A}^{(n)}(s)$ and $\boldsymbol{B}^{(n)}(s)$ for all $n \geq 2$ if and only if either of the matrices 
\begin{align}
    \boldsymbol{M}_{\text{s}}(s) &= (\mu + s \eta_s) \mathbb{I} + (K^o + s \eta^o) \boldsymbol{\epsilon},\label{eq:Mshear}\\
    \boldsymbol{M}_{\text{d}}(s) &= [(B + s \eta_b) \mathbb{I} - (A + s \eta_A) \boldsymbol{\epsilon}]\cdot(\mathbb{I} - \boldsymbol{\Tilde{R}}) \notag\\
    &+ [(\Gamma + s \eta_R) \mathbb{I} + (\Lambda + s \eta_\Lambda)\boldsymbol{\epsilon}]\cdot(\mathbb{I} + \boldsymbol{\Tilde{R}}),\label{eq:Md-r}
\end{align}
become singular, as $\boldsymbol{A}^{(n)}(s), \boldsymbol{B}^{(n)}(s)$ depend on their inverses, and so acquire corresponding poles~\cite{SItext}. The matrix $\boldsymbol{M}_{\text{s}}^{-1}$ is evidently associated with shear resonances: It depends only on shear moduli and any of its poles indicate singularities in $\boldsymbol{A}^{(n)}(s)$. The matrix $\boldsymbol{M}_{\text{d}}^{-1}$ is associated with dilational and rotational resonances: While it depends on all of the material moduli through~$\boldsymbol{\Tilde{R}}$, its poles also correspond to singularities in~$\boldsymbol{B}^{(n)}(s)$ and it vanishes when the moduli responding to isotropic deformations vanish. 

From $\det(\boldsymbol{M}_{\text{s}}) = 0$, we can identify a first pair of poles, $s_{s}^{\pm}$, that appears in displacement fields driven by boundary stresses. These poles are given by
\begin{equation}\label{eq:sss}
    s_{s}^{\pm} = -\frac{\mu \eta_s+ K^o \eta^o}{\eta_s^2 + (\eta^o)^2} \pm i \frac{\mu \eta^o - K^o \eta_s}{\eta_s^2 + (\eta^o)^2}.
\end{equation}
The inverse time scales given by \smash{$\text{Re}(s_{s}^\pm)$} and \smash{$\text{Im}(s_{s}^\pm)$}, corresponding to a decay rate and a characteristic oscillatory frequency, are consistent with the eigenvalue analysis of the relaxation-rate tensor $\lambda_{\alpha \beta} = \eta^{-1}_{\alpha \gamma} C_{\gamma \beta}$ by Banerjee et al.~\cite{banerjee2021active}, in which exclusively odd shear couplings were considered. Our work demonstrates that these time scales associated with the shear response remain unchanged even if the most general linear isotropic odd viscoelastic material is considered. 

Finally, we examine $\mathbf{M}_\text{d}$ given in Eq.~(\ref{eq:Md-r}), which is associated with dilation-rotation resonances and in general supports two distinct pairs of poles. For clarity, we set in the following $\Gamma = \Lambda = \eta_R = \eta_\Lambda = 0$. In this case, the viscosity tensor $\eta_{\alpha \beta}$ given in Eq.~(\ref{eq:ViscousModuliTensor}) obtains a non-trivial kernel in the dilation-rotation sub-space [components $\alpha=0,1$ in the tensor basis Eq.~(\ref{eq:ScheibnerBasis})] and is therefore singular. While the relaxation-rate tensor approach used in~\cite{banerjee2021active} breaks down for this scenario, the OPN solution provides -- via the poles of $\boldsymbol{M}_\text{d}$ -- a growth rate and oscillatory frequency associated with the odd dilation-rotation moduli~$A$ and $\eta_A$. From $\det(\boldsymbol{M}_{\text{d}}) = 0$, we find that this choice of moduli forces one of the pairs of poles to be the same as for the shear resonance, $s_s^\pm$, whilst the other pair is given by

\begin{equation}\label{eq:ssd}
s_{d}^\pm = -\frac{B \eta_b + A \eta_A}{\eta_b^2 + \eta_A^2} \pm i \frac{B \eta_A - A \eta_b}{\eta_b^2 + \eta_A^2},
\end{equation}
generalizing the passive limit found~\cite{banerjee2021active} to materials with~$A, \eta_A \neq 0$.

The magnitude of deformations at different driving frequencies of SBCs are depicted for representative material parameters in Fig.~\ref{fig:ViscoelasticResonances}a--c. For a passive KV model (\hbox{$A,K^o,\eta^A,\eta^o=0$}), we trivially recover the expected response of a monotonically decreasing displacement magnitude, irrespective of the symmetries of the stress at the cavity boundary~(Fig.~\ref{fig:ViscoelasticResonances}a). When active moduli contribute to the response, resonances associated with the poles \smash{$s_{s}^\pm$} and \smash{$s_{d}^\pm$} given in Eqs.~(\ref{eq:sss})~and~(\ref{eq:ssd}), respectively, can each appear individually~(Fig.~\ref{fig:ViscoelasticResonances}b) or simultaneously (Fig.~\ref{fig:ViscoelasticResonances}c) as the driving frequency of the SBC is varied. Since the mode $n=0$ only contains shear deformations, it only supports the shear resonance. The occurrence of these resonances is predicted by large values of the quality factors
\begin{align}
Q_{s}&:=-\frac{|s_{s}^\pm|}{2\,\text{Re}(s_{s}^\pm)}=\frac{1}{2\cos\varphi_s}\label{eq:Qs}\\
Q_{d}&:=-\frac{|s_{d}^\pm|}{2\,\text{Re}(s_{d}^\pm)}=\frac{1}{2\cos\varphi_d}\label{eq:Qd}
\end{align}
defined in analogy to the quality factor $Q_u$ given in Eq.~(\ref{eq:DispResonanceCondition}) for cavities driven by DBCs. Also for SBCs discussed here, the characteristic frequencies \smash{$\omega_{s}=\text{Im}(s_{s}^+)$} and \smash{\hbox{$\omega_{d}=\text{Im}(s_{d}^+)$}} extracted from the poles of OPN solutions provide very good approximations of the exact frequencies at which the different resonances occur (vertical dashed lines in Figs.~\ref{fig:ViscoelasticResonances}b,c).

\subsection{Unified Geometric Representation of Resonances}
A visual inspection of all the poles derived above, Eqs.~(\ref{eq:sdisppm}), (\ref{eq:sss}) and (\ref{eq:ssd}) -- each of which is associated with odd viscoelastic resonances of different physical origins -- reveals a common formal structure. Specifically, resonances associated with DBCs and with SBCs both follow schematically the form
\begin{equation}\label{eq:polgen}
s_{p}^\pm = - \frac{C^{(e)}_{p}\eta^{(e)}_{p} + C^{(o)}_{p} \eta^{(o)}_{p}}{(\eta^{(e)}_{p})^2 + (\eta^{(o)}_{p})^2} \pm i \frac{C^{(e)}_{p} \eta^{(o)}_{p} - C^{(o)}_{p} \eta^{(e)}_{p}}{(\eta^{(e)}_{p})^2 + (\eta^{(o)}_{p})^2},    
\end{equation}
where $C^{(e)}_p$ and $C^{(o)}_p$ are effective \textit{elastic} equilibrium and odd non-equilibrium moduli, respectively, associated with a given pole,~\smash{$p=u,s,d$}, that are assembled from the modulus tensor $C_{\alpha \beta}$ [Eq.~(\ref{eq:ElasticModulusTensor})] and can be read off Eqs.~(\ref{eq:sdisppm}), (\ref{eq:sss}) and (\ref{eq:ssd}). Similarly, \smash{$\eta^{(e)}_p, \eta^{(o)}_p$} are effective \textit{viscous} equilibrium and odd non-equilibrium moduli, respectively, assembled from the moduli contained in $\eta_{\alpha \beta}$ [Eq.~(\ref{eq:ViscousModuliTensor})]. Defining effective elastic and viscous moduli vectors \smash{$\boldsymbol{C}_p=(C^{(e)}_p, C^{(o)}_p)^T$} and \smash{\hbox{$\boldsymbol{\eta}_p=(\eta^{(e)}_p, \eta^{(o)}_p)^T$}}, Eq.~(\ref{eq:polgen}) -- and therefore the poles of all odd viscoelastic resonances -- have the common compact form
\begin{equation}
    s_{p}^\pm = - \frac{|\boldsymbol{C}_p|}{|\boldsymbol{\eta}_p|} e^{\mp i \varphi_p},
\end{equation}
where the angle $\varphi_p$, defined by
\begin{equation}\label{eq:phip}
\tan\varphi_{p}=\frac{\boldsymbol{C}_p\cdot\boldsymbol{\epsilon}\cdot\boldsymbol{\eta}_p}{\boldsymbol{C}_p \cdot \boldsymbol{\eta}_p}
\end{equation}
is in turn related to the quality factor [see Eqs.~(\ref{eq:DispResonanceCondition}), (\ref{eq:Qs}) and (\ref{eq:Qd})] via
\begin{equation}\label{eq:GeneralResonanceCondition}
Q_p= \frac{1}{2\cos\varphi_p}.
\end{equation}
We can use this vector representation and the angles $\varphi_p$ to generate an exhaustive illustration of the rich stability and resonance space of odd viscoelastic solids. For resonances due to boundary deformations shown in Fig.~\ref{fig:DBCresonances}, this space is illustrated in Fig.~\ref{fig:ViscoelasticResonances}d. For resonances due to stress-driven boundaries (Fig.~\ref{fig:ViscoelasticResonances}a--c) the parameters and phase spaces associated with shear and dilation-rotation resonances are individually depicted in Fig.~\ref{fig:ViscoelasticResonances}e. Their interplay is illustrated in the phase diagram Fig.~\ref{fig:ViscoelasticResonances}f. Because $Q_p>1/\sqrt{2}$ ($Q_p<1/\sqrt{2}$) provides a quality factor condition for the presence (absence) of resonances, we expect resonances for $\pi/4 < |\varphi_p| < \pi/2$, where the upper bound is a consequence of the thermodynamic stability condition Eq.~(\ref{eq:sdispStabilityCondition}). This leads to a checkerboard pattern in the phase diagram Fig.~\ref{fig:ViscoelasticResonances}f that contains regions in which resonances driven by boundary stresses can appear individually (red and blue), together (red-blue striped) or not at all (green).

The common form of the poles, Eq.~(\ref{eq:polgen}), can be understood by mapping the overdamped viscoelastic dynamics onto that of a damped harmonic oscillator. To this end, we note that near each resonance the displacement field is dominated by~$\boldsymbol{u}_p$, which can be found from the OPN method as
\begin{equation}
    \boldsymbol{u}(\boldsymbol{r}, t) \approx \boldsymbol{u}_p(\boldsymbol{r}, t) = \boldsymbol{M}_p^{-1} \boldsymbol{b}(\boldsymbol{r}, t) + (\boldsymbol{M}_p^T)^{-1} \boldsymbol{b}_T(\boldsymbol{r}, t),
\end{equation}
where $\boldsymbol{b}(\boldsymbol{r}, t), \boldsymbol{b}_T(\boldsymbol{r}, t)$ are effective forcings provided by the boundary-driving whose details do not matter for this argument, and $\boldsymbol{M}_p$ is a differential operator which, in Laplace space, is proportional to a rotation matrix. In the case of a shear resonance, we have \smash{\hbox{$\boldsymbol{M}_p = \boldsymbol{M}_s(\partial_t)$}} [see Eq.~(\ref{eq:Mshear})], and for a dilation resonance \smash{\hbox{$\boldsymbol{M}_p = \boldsymbol{M}_d(\partial_t)$}} [see Eq.~(\ref{eq:Md-r})]. Similar expressions exist for DBCs resonances~\cite{SItext}.

The dynamics of the displacement fields sourced by $\boldsymbol{b}(\boldsymbol{r}, t)$ and $\boldsymbol{b}_T(\boldsymbol{r}, t)$ are equivalent, so we keep for clarity in the following discussion only $\boldsymbol{b}(\boldsymbol{r},t)$ and describe the full system in~\cite{SItext}. $\boldsymbol{u}_p$ then satisfies an effective equation of motion 
\begin{equation}\label{eq:u_p dynamics}
    (C^{(e)}_p \mathbb{I} - C^{(o)}_p \boldsymbol{\epsilon}) \boldsymbol{u}_p + (\eta^{(e)}_p \mathbb{I} - \eta^{(o)}_p \boldsymbol{\epsilon}) \partial_t \boldsymbol{u}_p = \boldsymbol{b}.
\end{equation}
Solving Eq.~(\ref{eq:u_p dynamics}) algebraically for $\partial_t \boldsymbol{u}_p$ leads to an effective relaxation-rate tensor that can be written as
\begin{equation}
    \boldsymbol{\Lambda}_p = \frac{|\boldsymbol{C}_p|}{|\boldsymbol{\eta}_p|} \boldsymbol{R}^T(\varphi_p),
\end{equation}
where we used the definition of the phase angle $\varphi_p$ given in Eq.~(\ref{eq:phip}). The components of $\boldsymbol{u}_p$ in Eq.~(\ref{eq:u_p dynamics}) are non-reciprocally coupled, and each satisfy an equivalent decoupled second order equation
\begin{equation}\label{eq:u_p component dynamics}
    \partial^2_t u_{p,i} + \text{tr}\left(\boldsymbol{\Lambda}_p\right) \partial_t u_{p,i} + \det\left(\boldsymbol{\Lambda}_p\right) u_{p,i} = \Lambda_{ji} b'_j + \partial_t b'_i
\end{equation}
with \hbox{\smash{$\mathbf{b}'=(\eta^{(e)}_p \boldsymbol{I} - \eta^{(o)}_p \boldsymbol{\epsilon})^{-1} \cdot\mathbf{b}$}}, which corresponds to the equation of motion of a forced harmonic oscillator with damped oscillatory frequency $\boldsymbol{C}_p \cdot \boldsymbol{\epsilon} \cdot \boldsymbol{\eta}_p$ and an emergent damping coefficient \hbox{\smash{$\text{tr}(\boldsymbol{\Lambda}_p) \propto \boldsymbol{C}_p \cdot \boldsymbol{\eta}_p$}}. Hence, for \hbox{$\boldsymbol{C}_p \cdot \boldsymbol{\eta}_p > 0$}, activity enables the overdamped odd material to undergo underdamped oscillations. In the regime $\boldsymbol{C}_p \cdot \boldsymbol{\eta}_p < 0$, i.e. for effectively negative friction, the thermodynamic stability condition corresponding to Eq.~(\ref{eq:sdispStabilityCondition}) is violated and Eq.~(\ref{eq:u_p component dynamics}) predicts self-consistently an unstable system. Finally, the damping coefficient vanishes exactly when elastic and viscous moduli vectors $\boldsymbol{C}_p$ and $\boldsymbol{\eta}_p$, respectively, are orthogonal, which corresponds to an orthogonality of elastic and viscous forces in which activity enables an effectively dissipationless motion. 

\section{Conclusion}
In this work, we have introduced the odd Papkovich-Neuber ansatz for two dimensional, isotropic, odd viscoelastic materials. This generalization of the almost one century-old Papkovich-Neuber ansatz to active materials provides solutions of the force balance equation for any linear isotropic odd material in terms of harmonic scalar and vector fields. 

We have derived the general form of the deformation field for a force-free odd elastic disk to which we explicitly applied both displacement and stress boundary conditions. Access to analytical solutions on finite-sized domains allowed us to suggest experimental protocols for determining material properties, such as measurements of the shear axis rotation of an odd elastic disk, while also understanding the limitation of such measurement protocols due to inherent degeneracies in the impact of material moduli on the displacement field under different boundary conditions.

Our approach further enabled us to determine complete solutions for the dynamics of odd viscoelastic materials that are periodically driven at the boundary. These solutions provide the exact frequency-dependence of oscillation amplitudes and phase differences, revealed a set of characteristic material-intrinsic frequency scales at which different types of resonances appear, and suggested the definition of an odd quality factor with predictive power about the presence and absence of these resonances, thereby yielding a complete description of the key dynamic signatures of spatially extended isotropic odd Kelvin-Voigt materials. The generality of the OPN ansatz allows for an application of the same analysis to all linear viscoelastic systems, such as odd Maxwell models and odd Standard Linear Solids \cite{banerjee2021active, Sousl2021}.

Interestingly, odd viscoelastic materials may exhibit multiple resonances, even in the absence of inertia. These may be tuned independently of one another, and can be understood mechanistically via an equivalence between overdamped odd viscoelastic systems and damped harmonic oscillators that the OPN solution has revealed. We showed how these resonances can be interpreted geometrically, with elastic and viscous moduli defining vectors in a two dimensional moduli spaces. If elasticity and viscosity vectors are parallel, resonances and oscillations vanish and only an equilibrium-like response remains, akin to previous studies that showed how stronger non-reciprocity on the microscale does not always lead to stronger non-reciprocity on macroscopic scales~\cite{dinelli2023non,binysh2026more}. The fully quantitative understanding of these resonances we provide in this work is an important stepping stone towards the design of synthetic materials that exploit odd viscoelasticity to realize mechanical excitability at many, independently tunable, frequencies.

%\bibliography{ref}

\begin{thebibliography}{48}%
\makeatletter
\providecommand \@ifxundefined [1]{%
 \@ifx{#1\undefined}
}%
\providecommand \@ifnum [1]{%
 \ifnum #1\expandafter \@firstoftwo
 \else \expandafter \@secondoftwo
 \fi
}%
\providecommand \@ifx [1]{%
 \ifx #1\expandafter \@firstoftwo
 \else \expandafter \@secondoftwo
 \fi
}%
\providecommand \natexlab [1]{#1}%
\providecommand \enquote  [1]{``#1''}%
\providecommand \bibnamefont  [1]{#1}%
\providecommand \bibfnamefont [1]{#1}%
\providecommand \citenamefont [1]{#1}%
\providecommand \href@noop [0]{\@secondoftwo}%
\providecommand \href [0]{\begingroup \@sanitize@url \@href}%
\providecommand \@href[1]{\@@startlink{#1}\@@href}%
\providecommand \@@href[1]{\endgroup#1\@@endlink}%
\providecommand \@sanitize@url [0]{\catcode `\\12\catcode `\$12\catcode
  `\&12\catcode `\#12\catcode `\^12\catcode `\_12\catcode `\%12\relax}%
\providecommand \@@startlink[1]{}%
\providecommand \@@endlink[0]{}%
\providecommand \url  [0]{\begingroup\@sanitize@url \@url }%
\providecommand \@url [1]{\endgroup\@href {#1}{\urlprefix }}%
\providecommand \urlprefix  [0]{URL }%
\providecommand \Eprint [0]{\href }%
\providecommand \doibase [0]{https://doi.org/}%
\providecommand \selectlanguage [0]{\@gobble}%
\providecommand \bibinfo  [0]{\@secondoftwo}%
\providecommand \bibfield  [0]{\@secondoftwo}%
\providecommand \translation [1]{[#1]}%
\providecommand \BibitemOpen [0]{}%
\providecommand \bibitemStop [0]{}%
\providecommand \bibitemNoStop [0]{.\EOS\space}%
\providecommand \EOS [0]{\spacefactor3000\relax}%
\providecommand \BibitemShut  [1]{\csname bibitem#1\endcsname}%
\let\auto@bib@innerbib\@empty
%</preamble>
\bibitem [{\citenamefont {Fruchart}\ \emph {et~al.}(2023)\citenamefont
  {Fruchart}, \citenamefont {Scheibner},\ and\ \citenamefont
  {Vitelli}}]{fruchart2023odd}%
  \BibitemOpen
  \bibfield  {author} {\bibinfo {author} {\bibfnamefont {M.}~\bibnamefont
  {Fruchart}}, \bibinfo {author} {\bibfnamefont {C.}~\bibnamefont
  {Scheibner}},\ and\ \bibinfo {author} {\bibfnamefont {V.}~\bibnamefont
  {Vitelli}},\ }\href@noop {} {\bibfield  {journal} {\bibinfo  {journal} {Annu.
  Rev. Condens. Matter Phys.}\ }\textbf {\bibinfo {volume} {14}},\ \bibinfo
  {pages} {471} (\bibinfo {year} {2023})}\BibitemShut {NoStop}%
\bibitem [{\citenamefont {Soni}\ \emph {et~al.}(2019)\citenamefont {Soni},
  \citenamefont {Bililign}, \citenamefont {Magkiriadou}, \citenamefont
  {Sacanna}, \citenamefont {Bartolo}, \citenamefont {Shelley},\ and\
  \citenamefont {Irvine}}]{soni2019odd}%
  \BibitemOpen
  \bibfield  {author} {\bibinfo {author} {\bibfnamefont {V.}~\bibnamefont
  {Soni}}, \bibinfo {author} {\bibfnamefont {E.~S.}\ \bibnamefont {Bililign}},
  \bibinfo {author} {\bibfnamefont {S.}~\bibnamefont {Magkiriadou}}, \bibinfo
  {author} {\bibfnamefont {S.}~\bibnamefont {Sacanna}}, \bibinfo {author}
  {\bibfnamefont {D.}~\bibnamefont {Bartolo}}, \bibinfo {author} {\bibfnamefont
  {M.~J.}\ \bibnamefont {Shelley}},\ and\ \bibinfo {author} {\bibfnamefont
  {W.~T.~M.}\ \bibnamefont {Irvine}},\ }\href@noop {} {\bibfield  {journal}
  {\bibinfo  {journal} {Nat. Phys.}\ }\textbf {\bibinfo {volume} {15}},\
  \bibinfo {pages} {1188} (\bibinfo {year} {2019})}\BibitemShut {NoStop}%
\bibitem [{\citenamefont {Mecke}\ \emph {et~al.}(2023)\citenamefont {Mecke},
  \citenamefont {Gao}, \citenamefont {Ram{\'\i}rez~Medina}, \citenamefont
  {Aarts}, \citenamefont {Gompper},\ and\ \citenamefont {Ripoll}}]{mecke23}%
  \BibitemOpen
  \bibfield  {author} {\bibinfo {author} {\bibfnamefont {J.}~\bibnamefont
  {Mecke}}, \bibinfo {author} {\bibfnamefont {Y.}~\bibnamefont {Gao}}, \bibinfo
  {author} {\bibfnamefont {C.~A.}\ \bibnamefont {Ram{\'\i}rez~Medina}},
  \bibinfo {author} {\bibfnamefont {D.~G. A.~L.}\ \bibnamefont {Aarts}},
  \bibinfo {author} {\bibfnamefont {G.}~\bibnamefont {Gompper}},\ and\ \bibinfo
  {author} {\bibfnamefont {M.}~\bibnamefont {Ripoll}},\ }\href@noop {}
  {\bibfield  {journal} {\bibinfo  {journal} {Commun. Phys.}\ }\textbf
  {\bibinfo {volume} {6}},\ \bibinfo {pages} {324} (\bibinfo {year}
  {2023})}\BibitemShut {NoStop}%
\bibitem [{\citenamefont {Veenstra}\ \emph {et~al.}(2025)\citenamefont
  {Veenstra}, \citenamefont {Scheibner}, \citenamefont {Brandenbourger},
  \citenamefont {Binysh}, \citenamefont {Souslov}, \citenamefont {Vitelli},\
  and\ \citenamefont {Coulais}}]{veenstra2025adaptive}%
  \BibitemOpen
  \bibfield  {author} {\bibinfo {author} {\bibfnamefont {J.}~\bibnamefont
  {Veenstra}}, \bibinfo {author} {\bibfnamefont {C.}~\bibnamefont {Scheibner}},
  \bibinfo {author} {\bibfnamefont {M.}~\bibnamefont {Brandenbourger}},
  \bibinfo {author} {\bibfnamefont {J.}~\bibnamefont {Binysh}}, \bibinfo
  {author} {\bibfnamefont {A.}~\bibnamefont {Souslov}}, \bibinfo {author}
  {\bibfnamefont {V.}~\bibnamefont {Vitelli}},\ and\ \bibinfo {author}
  {\bibfnamefont {C.}~\bibnamefont {Coulais}},\ }\href@noop {} {\bibfield
  {journal} {\bibinfo  {journal} {Nature}\ }\textbf {\bibinfo {volume} {639}},\
  \bibinfo {pages} {935} (\bibinfo {year} {2025})}\BibitemShut {NoStop}%
\bibitem [{\citenamefont {Tan}\ \emph {et~al.}(2022)\citenamefont {Tan},
  \citenamefont {Mietke}, \citenamefont {Li}, \citenamefont {Chen},
  \citenamefont {Higinbotham}, \citenamefont {Foster}, \citenamefont {Gokhale},
  \citenamefont {Dunkel},\ and\ \citenamefont {Fakhri}}]{tan2022}%
  \BibitemOpen
  \bibfield  {author} {\bibinfo {author} {\bibfnamefont {T.~H.}\ \bibnamefont
  {Tan}}, \bibinfo {author} {\bibfnamefont {A.}~\bibnamefont {Mietke}},
  \bibinfo {author} {\bibfnamefont {J.}~\bibnamefont {Li}}, \bibinfo {author}
  {\bibfnamefont {Y.}~\bibnamefont {Chen}}, \bibinfo {author} {\bibfnamefont
  {H.}~\bibnamefont {Higinbotham}}, \bibinfo {author} {\bibfnamefont {P.~J.}\
  \bibnamefont {Foster}}, \bibinfo {author} {\bibfnamefont {S.}~\bibnamefont
  {Gokhale}}, \bibinfo {author} {\bibfnamefont {J.}~\bibnamefont {Dunkel}},\
  and\ \bibinfo {author} {\bibfnamefont {N.}~\bibnamefont {Fakhri}},\
  }\href@noop {} {\bibfield  {journal} {\bibinfo  {journal} {Nature}\ }\textbf
  {\bibinfo {volume} {607}},\ \bibinfo {pages} {287} (\bibinfo {year}
  {2022})}\BibitemShut {NoStop}%
\bibitem [{\citenamefont {Shankar}\ and\ \citenamefont
  {Mahadevan}(2024)}]{shankar2024active}%
  \BibitemOpen
  \bibfield  {author} {\bibinfo {author} {\bibfnamefont {S.}~\bibnamefont
  {Shankar}}\ and\ \bibinfo {author} {\bibfnamefont {L.}~\bibnamefont
  {Mahadevan}},\ }\href@noop {} {\bibfield  {journal} {\bibinfo  {journal}
  {Nat. Phys.}\ }\textbf {\bibinfo {volume} {20}},\ \bibinfo {pages} {1501}
  (\bibinfo {year} {2024})}\BibitemShut {NoStop}%
\bibitem [{\citenamefont {Gu}\ \emph {et~al.}(2025)\citenamefont {Gu},
  \citenamefont {Guiselin}, \citenamefont {Bain}, \citenamefont {Zuriguel},\
  and\ \citenamefont {Bartolo}}]{gu2025emergence}%
  \BibitemOpen
  \bibfield  {author} {\bibinfo {author} {\bibfnamefont {F.}~\bibnamefont
  {Gu}}, \bibinfo {author} {\bibfnamefont {B.}~\bibnamefont {Guiselin}},
  \bibinfo {author} {\bibfnamefont {N.}~\bibnamefont {Bain}}, \bibinfo {author}
  {\bibfnamefont {I.}~\bibnamefont {Zuriguel}},\ and\ \bibinfo {author}
  {\bibfnamefont {D.}~\bibnamefont {Bartolo}},\ }\href@noop {} {\bibfield
  {journal} {\bibinfo  {journal} {Nature}\ }\textbf {\bibinfo {volume} {638}},\
  \bibinfo {pages} {112} (\bibinfo {year} {2025})}\BibitemShut {NoStop}%
\bibitem [{\citenamefont {Essner}\ \emph {et~al.}(2005)\citenamefont {Essner},
  \citenamefont {Amack}, \citenamefont {Nyholm}, \citenamefont {Harris},\ and\
  \citenamefont {Yost}}]{essner2005kupffer}%
  \BibitemOpen
  \bibfield  {author} {\bibinfo {author} {\bibfnamefont {J.~J.}\ \bibnamefont
  {Essner}}, \bibinfo {author} {\bibfnamefont {J.~D.}\ \bibnamefont {Amack}},
  \bibinfo {author} {\bibfnamefont {M.~K.}\ \bibnamefont {Nyholm}}, \bibinfo
  {author} {\bibfnamefont {E.~B.}\ \bibnamefont {Harris}},\ and\ \bibinfo
  {author} {\bibfnamefont {H.~J.}\ \bibnamefont {Yost}},\ }\href@noop {}
  {\bibfield  {journal} {\bibinfo  {journal} {Development}\ }\textbf {\bibinfo
  {volume} {132}},\ \bibinfo {pages} {1247} (\bibinfo {year}
  {2005})}\BibitemShut {NoStop}%
\bibitem [{\citenamefont {Schweickert}\ \emph {et~al.}(2007)\citenamefont
  {Schweickert}, \citenamefont {Weber}, \citenamefont {Beyer}, \citenamefont
  {Vick}, \citenamefont {Bogusch}, \citenamefont {Feistel},\ and\ \citenamefont
  {Blum}}]{schweickert2007cilia}%
  \BibitemOpen
  \bibfield  {author} {\bibinfo {author} {\bibfnamefont {A.}~\bibnamefont
  {Schweickert}}, \bibinfo {author} {\bibfnamefont {T.}~\bibnamefont {Weber}},
  \bibinfo {author} {\bibfnamefont {T.}~\bibnamefont {Beyer}}, \bibinfo
  {author} {\bibfnamefont {P.}~\bibnamefont {Vick}}, \bibinfo {author}
  {\bibfnamefont {S.}~\bibnamefont {Bogusch}}, \bibinfo {author} {\bibfnamefont
  {K.}~\bibnamefont {Feistel}},\ and\ \bibinfo {author} {\bibfnamefont
  {M.}~\bibnamefont {Blum}},\ }\href@noop {} {\bibfield  {journal} {\bibinfo
  {journal} {Curr. Biol.}\ }\textbf {\bibinfo {volume} {17}},\ \bibinfo {pages}
  {60} (\bibinfo {year} {2007})}\BibitemShut {NoStop}%
\bibitem [{\citenamefont {Naganathan}\ \emph {et~al.}(2014)\citenamefont
  {Naganathan}, \citenamefont {F{\"u}rthauer}, \citenamefont {Nishikawa},
  \citenamefont {J{\"u}licher},\ and\ \citenamefont
  {Grill}}]{naganathan2014active}%
  \BibitemOpen
  \bibfield  {author} {\bibinfo {author} {\bibfnamefont {S.~R.}\ \bibnamefont
  {Naganathan}}, \bibinfo {author} {\bibfnamefont {S.}~\bibnamefont
  {F{\"u}rthauer}}, \bibinfo {author} {\bibfnamefont {M.}~\bibnamefont
  {Nishikawa}}, \bibinfo {author} {\bibfnamefont {F.}~\bibnamefont
  {J{\"u}licher}},\ and\ \bibinfo {author} {\bibfnamefont {S.~W.}\ \bibnamefont
  {Grill}},\ }\href@noop {} {\bibfield  {journal} {\bibinfo  {journal} {elife}\
  }\textbf {\bibinfo {volume} {3}},\ \bibinfo {pages} {e04165} (\bibinfo {year}
  {2014})}\BibitemShut {NoStop}%
\bibitem [{\citenamefont {Pfanzelter}\ \emph {et~al.}(2025)\citenamefont
  {Pfanzelter}, \citenamefont {Neipel}, \citenamefont {Lahola-Chomiak},
  \citenamefont {Tsikolia}, \citenamefont {Mietke}, \citenamefont {Gros},
  \citenamefont {J{\"u}licher},\ and\ \citenamefont
  {Grill}}]{pfanzelter2025active}%
  \BibitemOpen
  \bibfield  {author} {\bibinfo {author} {\bibfnamefont {J.}~\bibnamefont
  {Pfanzelter}}, \bibinfo {author} {\bibfnamefont {J.}~\bibnamefont {Neipel}},
  \bibinfo {author} {\bibfnamefont {A.~A.}\ \bibnamefont {Lahola-Chomiak}},
  \bibinfo {author} {\bibfnamefont {N.}~\bibnamefont {Tsikolia}}, \bibinfo
  {author} {\bibfnamefont {A.}~\bibnamefont {Mietke}}, \bibinfo {author}
  {\bibfnamefont {J.}~\bibnamefont {Gros}}, \bibinfo {author} {\bibfnamefont
  {F.}~\bibnamefont {J{\"u}licher}},\ and\ \bibinfo {author} {\bibfnamefont
  {S.~W.}\ \bibnamefont {Grill}},\ }\href@noop {} {\bibinfo {title} {An active
  torque dipole across tissue layers drives avian left-right symmetry
  breaking}} (\bibinfo {year} {2025}),\ \bibinfo {note}
  {bioRxiv:10.1101/2025.07.16.665037}\BibitemShut {NoStop}%
\bibitem [{\citenamefont {Scheibner}\ \emph {et~al.}(2020)\citenamefont
  {Scheibner}, \citenamefont {Souslov}, \citenamefont {Banerjee}, \citenamefont
  {Sur{\'o}wka}, \citenamefont {Irvine},\ and\ \citenamefont
  {Vitelli}}]{Scheibner:2020}%
  \BibitemOpen
  \bibfield  {author} {\bibinfo {author} {\bibfnamefont {C.}~\bibnamefont
  {Scheibner}}, \bibinfo {author} {\bibfnamefont {A.}~\bibnamefont {Souslov}},
  \bibinfo {author} {\bibfnamefont {D.}~\bibnamefont {Banerjee}}, \bibinfo
  {author} {\bibfnamefont {P.}~\bibnamefont {Sur{\'o}wka}}, \bibinfo {author}
  {\bibfnamefont {W.~T.~M.}\ \bibnamefont {Irvine}},\ and\ \bibinfo {author}
  {\bibfnamefont {V.}~\bibnamefont {Vitelli}},\ }\href@noop {} {\bibfield
  {journal} {\bibinfo  {journal} {Nat. Phys.}\ }\textbf {\bibinfo {volume}
  {16}},\ \bibinfo {pages} {475} (\bibinfo {year} {2020})}\BibitemShut
  {NoStop}%
\bibitem [{\citenamefont {Fodor}\ and\ \citenamefont
  {Souslov}(2021{\natexlab{a}})}]{Sousl2021}%
  \BibitemOpen
  \bibfield  {author} {\bibinfo {author} {\bibfnamefont {E.}~\bibnamefont
  {Fodor}}\ and\ \bibinfo {author} {\bibfnamefont {A.}~\bibnamefont
  {Souslov}},\ }\href@noop {} {\bibfield  {journal} {\bibinfo  {journal} {Phys.
  Rev. E}\ }\textbf {\bibinfo {volume} {104}},\ \bibinfo {pages} {L062602}
  (\bibinfo {year} {2021}{\natexlab{a}})}\BibitemShut {NoStop}%
\bibitem [{\citenamefont {Hosaka}\ \emph
  {et~al.}(2021{\natexlab{a}})\citenamefont {Hosaka}, \citenamefont {Komura},\
  and\ \citenamefont {Andelman}}]{hosaka2021nonreciprocal}%
  \BibitemOpen
  \bibfield  {author} {\bibinfo {author} {\bibfnamefont {Y.}~\bibnamefont
  {Hosaka}}, \bibinfo {author} {\bibfnamefont {S.}~\bibnamefont {Komura}},\
  and\ \bibinfo {author} {\bibfnamefont {D.}~\bibnamefont {Andelman}},\
  }\href@noop {} {\bibfield  {journal} {\bibinfo  {journal} {Phys. Rev. E}\
  }\textbf {\bibinfo {volume} {103}},\ \bibinfo {pages} {042610} (\bibinfo
  {year} {2021}{\natexlab{a}})}\BibitemShut {NoStop}%
\bibitem [{\citenamefont {Lier}\ \emph {et~al.}(2023)\citenamefont {Lier},
  \citenamefont {Duclut}, \citenamefont {Bo}, \citenamefont {Armas},
  \citenamefont {J{\"u}licher},\ and\ \citenamefont
  {Sur{\'o}wka}}]{lier2023lift}%
  \BibitemOpen
  \bibfield  {author} {\bibinfo {author} {\bibfnamefont {R.}~\bibnamefont
  {Lier}}, \bibinfo {author} {\bibfnamefont {C.}~\bibnamefont {Duclut}},
  \bibinfo {author} {\bibfnamefont {S.}~\bibnamefont {Bo}}, \bibinfo {author}
  {\bibfnamefont {J.}~\bibnamefont {Armas}}, \bibinfo {author} {\bibfnamefont
  {F.}~\bibnamefont {J{\"u}licher}},\ and\ \bibinfo {author} {\bibfnamefont
  {P.}~\bibnamefont {Sur{\'o}wka}},\ }\href@noop {} {\bibfield  {journal}
  {\bibinfo  {journal} {Phys. Rev. E}\ }\textbf {\bibinfo {volume} {108}},\
  \bibinfo {pages} {L023101} (\bibinfo {year} {2023})}\BibitemShut {NoStop}%
\bibitem [{\citenamefont {Floyd}\ \emph {et~al.}(2024)\citenamefont {Floyd},
  \citenamefont {Dinner},\ and\ \citenamefont
  {Vaikuntanathan}}]{floyd2024pattern}%
  \BibitemOpen
  \bibfield  {author} {\bibinfo {author} {\bibfnamefont {C.}~\bibnamefont
  {Floyd}}, \bibinfo {author} {\bibfnamefont {A.~R.}\ \bibnamefont {Dinner}},\
  and\ \bibinfo {author} {\bibfnamefont {S.}~\bibnamefont {Vaikuntanathan}},\
  }\href {https://doi.org/10.1103/PhysRevResearch.6.033100} {\bibfield
  {journal} {\bibinfo  {journal} {Phys. Rev. Res.}\ }\textbf {\bibinfo {volume}
  {6}},\ \bibinfo {pages} {033100} (\bibinfo {year} {2024})}\BibitemShut
  {NoStop}%
\bibitem [{\citenamefont {de~Wit}\ \emph {et~al.}(2024)\citenamefont {de~Wit},
  \citenamefont {Fruchart}, \citenamefont {Khain}, \citenamefont {Toschi},\
  and\ \citenamefont {Vitelli}}]{de2024pattern}%
  \BibitemOpen
  \bibfield  {author} {\bibinfo {author} {\bibfnamefont {X.~M.}\ \bibnamefont
  {de~Wit}}, \bibinfo {author} {\bibfnamefont {M.}~\bibnamefont {Fruchart}},
  \bibinfo {author} {\bibfnamefont {T.}~\bibnamefont {Khain}}, \bibinfo
  {author} {\bibfnamefont {F.}~\bibnamefont {Toschi}},\ and\ \bibinfo {author}
  {\bibfnamefont {V.}~\bibnamefont {Vitelli}},\ }\href@noop {} {\bibfield
  {journal} {\bibinfo  {journal} {Nature}\ }\textbf {\bibinfo {volume} {627}},\
  \bibinfo {pages} {515} (\bibinfo {year} {2024})}\BibitemShut {NoStop}%
\bibitem [{\citenamefont {Kalz}\ \emph {et~al.}(2022)\citenamefont {Kalz},
  \citenamefont {Vuijk}, \citenamefont {Abdoli}, \citenamefont {Sommer},
  \citenamefont {L\"owen},\ and\ \citenamefont {Sharma}}]{kalz2022collisions}%
  \BibitemOpen
  \bibfield  {author} {\bibinfo {author} {\bibfnamefont {E.}~\bibnamefont
  {Kalz}}, \bibinfo {author} {\bibfnamefont {H.~D.}\ \bibnamefont {Vuijk}},
  \bibinfo {author} {\bibfnamefont {I.}~\bibnamefont {Abdoli}}, \bibinfo
  {author} {\bibfnamefont {J.-U.}\ \bibnamefont {Sommer}}, \bibinfo {author}
  {\bibfnamefont {H.}~\bibnamefont {L\"owen}},\ and\ \bibinfo {author}
  {\bibfnamefont {A.}~\bibnamefont {Sharma}},\ }\href@noop {} {\bibfield
  {journal} {\bibinfo  {journal} {Phys. Rev. Lett.}\ }\textbf {\bibinfo
  {volume} {129}},\ \bibinfo {pages} {090601} (\bibinfo {year}
  {2022})}\BibitemShut {NoStop}%
\bibitem [{\citenamefont {Chao}\ \emph {et~al.}(2026)\citenamefont {Chao},
  \citenamefont {Gokhale}, \citenamefont {Lin}, \citenamefont {Hastewell},
  \citenamefont {Bacanu}, \citenamefont {Chen}, \citenamefont {Li},
  \citenamefont {Liu}, \citenamefont {Lee}, \citenamefont {Dunkel},\ and\
  \citenamefont {Fakhri}}]{chao2026selective}%
  \BibitemOpen
  \bibfield  {author} {\bibinfo {author} {\bibfnamefont {Y.-C.}\ \bibnamefont
  {Chao}}, \bibinfo {author} {\bibfnamefont {S.}~\bibnamefont {Gokhale}},
  \bibinfo {author} {\bibfnamefont {L.}~\bibnamefont {Lin}}, \bibinfo {author}
  {\bibfnamefont {A.}~\bibnamefont {Hastewell}}, \bibinfo {author}
  {\bibfnamefont {A.}~\bibnamefont {Bacanu}}, \bibinfo {author} {\bibfnamefont
  {Y.}~\bibnamefont {Chen}}, \bibinfo {author} {\bibfnamefont {J.}~\bibnamefont
  {Li}}, \bibinfo {author} {\bibfnamefont {J.}~\bibnamefont {Liu}}, \bibinfo
  {author} {\bibfnamefont {H.}~\bibnamefont {Lee}}, \bibinfo {author}
  {\bibfnamefont {J.}~\bibnamefont {Dunkel}},\ and\ \bibinfo {author}
  {\bibfnamefont {N.}~\bibnamefont {Fakhri}},\ }\href@noop {} {\bibfield
  {journal} {\bibinfo  {journal} {Nat. Phys.}\ }\textbf {\bibinfo {volume}
  {22}},\ \bibinfo {pages} {474} (\bibinfo {year} {2026})}\BibitemShut
  {NoStop}%
\bibitem [{\citenamefont {Forgacs}\ \emph {et~al.}(1998)\citenamefont
  {Forgacs}, \citenamefont {Foty}, \citenamefont {Shafrir},\ and\ \citenamefont
  {Steinberg}}]{Forgacs1998viscoelastic}%
  \BibitemOpen
  \bibfield  {author} {\bibinfo {author} {\bibfnamefont {G.}~\bibnamefont
  {Forgacs}}, \bibinfo {author} {\bibfnamefont {R.~A.}\ \bibnamefont {Foty}},
  \bibinfo {author} {\bibfnamefont {Y.}~\bibnamefont {Shafrir}},\ and\ \bibinfo
  {author} {\bibfnamefont {M.~S.}\ \bibnamefont {Steinberg}},\ }\href
  {https://doi.org/https://doi.org/10.1016/S0006-3495(98)77932-9} {\bibfield
  {journal} {\bibinfo  {journal} {Biophys. J.}\ }\textbf {\bibinfo {volume}
  {74}},\ \bibinfo {pages} {2227} (\bibinfo {year} {1998})}\BibitemShut
  {NoStop}%
\bibitem [{\citenamefont {Serwane}\ \emph {et~al.}(2017)\citenamefont
  {Serwane}, \citenamefont {Mongera}, \citenamefont {Rowghanian}, \citenamefont
  {Kealhofer}, \citenamefont {Lucio}, \citenamefont {Hockenbery},\ and\
  \citenamefont {Camp{\`a}s}}]{serwane2017}%
  \BibitemOpen
  \bibfield  {author} {\bibinfo {author} {\bibfnamefont {F.}~\bibnamefont
  {Serwane}}, \bibinfo {author} {\bibfnamefont {A.}~\bibnamefont {Mongera}},
  \bibinfo {author} {\bibfnamefont {P.}~\bibnamefont {Rowghanian}}, \bibinfo
  {author} {\bibfnamefont {D.~A.}\ \bibnamefont {Kealhofer}}, \bibinfo {author}
  {\bibfnamefont {A.~A.}\ \bibnamefont {Lucio}}, \bibinfo {author}
  {\bibfnamefont {Z.~M.}\ \bibnamefont {Hockenbery}},\ and\ \bibinfo {author}
  {\bibfnamefont {O.}~\bibnamefont {Camp{\`a}s}},\ }\href@noop {} {\bibfield
  {journal} {\bibinfo  {journal} {Nat. Methods}\ }\textbf {\bibinfo {volume}
  {14}},\ \bibinfo {pages} {181} (\bibinfo {year} {2017})}\BibitemShut
  {NoStop}%
\bibitem [{\citenamefont {Braverman}\ \emph {et~al.}(2021)\citenamefont
  {Braverman}, \citenamefont {Scheibner}, \citenamefont {VanSaders},\ and\
  \citenamefont {Vitelli}}]{braverman2020topological}%
  \BibitemOpen
  \bibfield  {author} {\bibinfo {author} {\bibfnamefont {L.}~\bibnamefont
  {Braverman}}, \bibinfo {author} {\bibfnamefont {C.}~\bibnamefont
  {Scheibner}}, \bibinfo {author} {\bibfnamefont {B.}~\bibnamefont
  {VanSaders}},\ and\ \bibinfo {author} {\bibfnamefont {V.}~\bibnamefont
  {Vitelli}},\ }\href@noop {} {\bibfield  {journal} {\bibinfo  {journal} {Phys.
  Rev. Lett.}\ }\textbf {\bibinfo {volume} {127}},\ \bibinfo {pages} {268001}
  (\bibinfo {year} {2021})}\BibitemShut {NoStop}%
\bibitem [{\citenamefont {Khain}\ \emph {et~al.}(2022)\citenamefont {Khain},
  \citenamefont {Scheibner}, \citenamefont {Fruchart},\ and\ \citenamefont
  {Vitelli}}]{khain2022stokes}%
  \BibitemOpen
  \bibfield  {author} {\bibinfo {author} {\bibfnamefont {T.}~\bibnamefont
  {Khain}}, \bibinfo {author} {\bibfnamefont {C.}~\bibnamefont {Scheibner}},
  \bibinfo {author} {\bibfnamefont {M.}~\bibnamefont {Fruchart}},\ and\
  \bibinfo {author} {\bibfnamefont {V.}~\bibnamefont {Vitelli}},\ }\href@noop
  {} {\bibfield  {journal} {\bibinfo  {journal} {J. Fluid Mech.}\ }\textbf
  {\bibinfo {volume} {934}},\ \bibinfo {pages} {A23} (\bibinfo {year}
  {2022})}\BibitemShut {NoStop}%
\bibitem [{\citenamefont {Chen}\ \emph {et~al.}(2021)\citenamefont {Chen},
  \citenamefont {Li}, \citenamefont {Scheibner}, \citenamefont {Vitelli},\ and\
  \citenamefont {Huang}}]{chen2021realization}%
  \BibitemOpen
  \bibfield  {author} {\bibinfo {author} {\bibfnamefont {Y.}~\bibnamefont
  {Chen}}, \bibinfo {author} {\bibfnamefont {X.}~\bibnamefont {Li}}, \bibinfo
  {author} {\bibfnamefont {C.}~\bibnamefont {Scheibner}}, \bibinfo {author}
  {\bibfnamefont {V.}~\bibnamefont {Vitelli}},\ and\ \bibinfo {author}
  {\bibfnamefont {G.}~\bibnamefont {Huang}},\ }\href@noop {} {\bibfield
  {journal} {\bibinfo  {journal} {Nat. Commun.}\ }\textbf {\bibinfo {volume}
  {12}},\ \bibinfo {pages} {5935} (\bibinfo {year} {2021})}\BibitemShut
  {NoStop}%
\bibitem [{\citenamefont {Banerjee}\ \emph {et~al.}(2021)\citenamefont
  {Banerjee}, \citenamefont {Vitelli}, \citenamefont {J{\"u}licher},\ and\
  \citenamefont {Sur{\'o}wka}}]{banerjee2021active}%
  \BibitemOpen
  \bibfield  {author} {\bibinfo {author} {\bibfnamefont {D.}~\bibnamefont
  {Banerjee}}, \bibinfo {author} {\bibfnamefont {V.}~\bibnamefont {Vitelli}},
  \bibinfo {author} {\bibfnamefont {F.}~\bibnamefont {J{\"u}licher}},\ and\
  \bibinfo {author} {\bibfnamefont {P.}~\bibnamefont {Sur{\'o}wka}},\
  }\href@noop {} {\bibfield  {journal} {\bibinfo  {journal} {Phys. Rev. Lett.}\
  }\textbf {\bibinfo {volume} {126}},\ \bibinfo {pages} {138001} (\bibinfo
  {year} {2021})}\BibitemShut {NoStop}%
\bibitem [{\citenamefont {Bililign}\ \emph {et~al.}(2022)\citenamefont
  {Bililign}, \citenamefont {Balboa~Usabiaga}, \citenamefont {Ganan},
  \citenamefont {Poncet}, \citenamefont {Soni}, \citenamefont {Magkiriadou},
  \citenamefont {Shelley}, \citenamefont {Bartolo},\ and\ \citenamefont
  {Irvine}}]{bililign2021chiral}%
  \BibitemOpen
  \bibfield  {author} {\bibinfo {author} {\bibfnamefont {E.~S.}\ \bibnamefont
  {Bililign}}, \bibinfo {author} {\bibfnamefont {F.}~\bibnamefont
  {Balboa~Usabiaga}}, \bibinfo {author} {\bibfnamefont {Y.~A.}\ \bibnamefont
  {Ganan}}, \bibinfo {author} {\bibfnamefont {A.}~\bibnamefont {Poncet}},
  \bibinfo {author} {\bibfnamefont {V.}~\bibnamefont {Soni}}, \bibinfo {author}
  {\bibfnamefont {S.}~\bibnamefont {Magkiriadou}}, \bibinfo {author}
  {\bibfnamefont {M.~J.}\ \bibnamefont {Shelley}}, \bibinfo {author}
  {\bibfnamefont {D.}~\bibnamefont {Bartolo}},\ and\ \bibinfo {author}
  {\bibfnamefont {W.~T.~M.}\ \bibnamefont {Irvine}},\ }\href@noop {} {\bibfield
   {journal} {\bibinfo  {journal} {Nat. Phys.}\ }\textbf {\bibinfo {volume}
  {18}},\ \bibinfo {pages} {212} (\bibinfo {year} {2022})}\BibitemShut
  {NoStop}%
\bibitem [{\citenamefont {Lee}\ \emph {et~al.}(2025)\citenamefont {Lee},
  \citenamefont {Lubensky},\ and\ \citenamefont
  {Markovich}}]{lee2025disordered}%
  \BibitemOpen
  \bibfield  {author} {\bibinfo {author} {\bibfnamefont {C.-T.}\ \bibnamefont
  {Lee}}, \bibinfo {author} {\bibfnamefont {T.~C.}\ \bibnamefont {Lubensky}},\
  and\ \bibinfo {author} {\bibfnamefont {T.}~\bibnamefont {Markovich}},\ }\href
  {https://arxiv.org/abs/2508.04468} {\bibinfo {title} {Odd elasticity in
  disordered chiral active materials}} (\bibinfo {year} {2025}),\ \Eprint
  {https://arxiv.org/abs/2508.04468} {arXiv:2508.04468} \BibitemShut {NoStop}%
\bibitem [{\citenamefont {Lee}\ and\ \citenamefont
  {Markovich}(2026)}]{lee2026nonhermitian}%
  \BibitemOpen
  \bibfield  {author} {\bibinfo {author} {\bibfnamefont {C.-T.}\ \bibnamefont
  {Lee}}\ and\ \bibinfo {author} {\bibfnamefont {T.}~\bibnamefont
  {Markovich}},\ }\href {https://arxiv.org/abs/2603.21312} {\bibinfo {title}
  {Non-hermitian chiral surface waves in disordered odd solids}} (\bibinfo
  {year} {2026}),\ \Eprint {https://arxiv.org/abs/2603.21312}
  {arXiv:2603.21312} \BibitemShut {NoStop}%
\bibitem [{\citenamefont {Duclut}\ \emph {et~al.}(2024)\citenamefont {Duclut},
  \citenamefont {Bo}, \citenamefont {Lier}, \citenamefont {Armas},
  \citenamefont {Sur{\'o}wka},\ and\ \citenamefont
  {J{\"u}licher}}]{duclut2024probe}%
  \BibitemOpen
  \bibfield  {author} {\bibinfo {author} {\bibfnamefont {C.}~\bibnamefont
  {Duclut}}, \bibinfo {author} {\bibfnamefont {S.}~\bibnamefont {Bo}}, \bibinfo
  {author} {\bibfnamefont {R.}~\bibnamefont {Lier}}, \bibinfo {author}
  {\bibfnamefont {J.}~\bibnamefont {Armas}}, \bibinfo {author} {\bibfnamefont
  {P.}~\bibnamefont {Sur{\'o}wka}},\ and\ \bibinfo {author} {\bibfnamefont
  {F.}~\bibnamefont {J{\"u}licher}},\ }\href@noop {} {\bibfield  {journal}
  {\bibinfo  {journal} {Phys. Rev. E}\ }\textbf {\bibinfo {volume} {109}},\
  \bibinfo {pages} {044126} (\bibinfo {year} {2024})}\BibitemShut {NoStop}%
\bibitem [{\citenamefont {Hosaka}\ \emph
  {et~al.}(2021{\natexlab{b}})\citenamefont {Hosaka}, \citenamefont {Komura},\
  and\ \citenamefont {Andelman}}]{hosaka2021hydrodynamic}%
  \BibitemOpen
  \bibfield  {author} {\bibinfo {author} {\bibfnamefont {Y.}~\bibnamefont
  {Hosaka}}, \bibinfo {author} {\bibfnamefont {S.}~\bibnamefont {Komura}},\
  and\ \bibinfo {author} {\bibfnamefont {D.}~\bibnamefont {Andelman}},\
  }\href@noop {} {\bibfield  {journal} {\bibinfo  {journal} {Phys. Rev. E}\
  }\textbf {\bibinfo {volume} {104}},\ \bibinfo {pages} {064613} (\bibinfo
  {year} {2021}{\natexlab{b}})}\BibitemShut {NoStop}%
\bibitem [{\citenamefont {Hosaka}\ \emph {et~al.}(2023)\citenamefont {Hosaka},
  \citenamefont {Golestanian},\ and\ \citenamefont
  {Vilfan}}]{hosaka2023lorentz}%
  \BibitemOpen
  \bibfield  {author} {\bibinfo {author} {\bibfnamefont {Y.}~\bibnamefont
  {Hosaka}}, \bibinfo {author} {\bibfnamefont {R.}~\bibnamefont
  {Golestanian}},\ and\ \bibinfo {author} {\bibfnamefont {A.}~\bibnamefont
  {Vilfan}},\ }\href@noop {} {\bibfield  {journal} {\bibinfo  {journal} {Phys.
  Rev. Lett.}\ }\textbf {\bibinfo {volume} {131}},\ \bibinfo {pages} {178303}
  (\bibinfo {year} {2023})}\BibitemShut {NoStop}%
\bibitem [{\citenamefont {Ganeshan}\ and\ \citenamefont
  {Abanov}(2017)}]{ganeshan2017}%
  \BibitemOpen
  \bibfield  {author} {\bibinfo {author} {\bibfnamefont {S.}~\bibnamefont
  {Ganeshan}}\ and\ \bibinfo {author} {\bibfnamefont {A.~G.}\ \bibnamefont
  {Abanov}},\ }\href {https://doi.org/10.1103/PhysRevFluids.2.094101}
  {\bibfield  {journal} {\bibinfo  {journal} {Phys. Rev. Fluids}\ }\textbf
  {\bibinfo {volume} {2}},\ \bibinfo {pages} {094101} (\bibinfo {year}
  {2017})}\BibitemShut {NoStop}%
\bibitem [{\citenamefont {Banerjee}\ \emph {et~al.}(2017)\citenamefont
  {Banerjee}, \citenamefont {Souslov}, \citenamefont {Abanov},\ and\
  \citenamefont {Vitelli}}]{banerjee2017odd}%
  \BibitemOpen
  \bibfield  {author} {\bibinfo {author} {\bibfnamefont {D.}~\bibnamefont
  {Banerjee}}, \bibinfo {author} {\bibfnamefont {A.}~\bibnamefont {Souslov}},
  \bibinfo {author} {\bibfnamefont {A.~G.}\ \bibnamefont {Abanov}},\ and\
  \bibinfo {author} {\bibfnamefont {V.}~\bibnamefont {Vitelli}},\ }\href@noop
  {} {\bibfield  {journal} {\bibinfo  {journal} {Nat. Commun.}\ }\textbf
  {\bibinfo {volume} {8}},\ \bibinfo {pages} {1573} (\bibinfo {year}
  {2017})}\BibitemShut {NoStop}%
\bibitem [{\citenamefont {Avron}(1998)}]{avron1998odd}%
  \BibitemOpen
  \bibfield  {author} {\bibinfo {author} {\bibfnamefont {J.~E.}\ \bibnamefont
  {Avron}},\ }\href@noop {} {\bibfield  {journal} {\bibinfo  {journal} {J.
  Stat. Phys.}\ }\textbf {\bibinfo {volume} {92}},\ \bibinfo {pages} {543}
  (\bibinfo {year} {1998})}\BibitemShut {NoStop}%
\bibitem [{\citenamefont {Lamb}(1932)}]{lamb1932}%
  \BibitemOpen
  \bibfield  {author} {\bibinfo {author} {\bibfnamefont {H.}~\bibnamefont
  {Lamb}},\ }\href@noop {} {\emph {\bibinfo {title} {Hydrodynamics}}},\
  \bibinfo {edition} {6th}\ ed.\ (\bibinfo  {publisher} {Cambridge University
  Press},\ \bibinfo {address} {Cambridge},\ \bibinfo {year} {1932})\BibitemShut
  {NoStop}%
\bibitem [{\citenamefont {Galerkin}(1930)}]{GalerkinB1930Cttg}%
  \BibitemOpen
  \bibfield  {author} {\bibinfo {author} {\bibfnamefont {B.}~\bibnamefont
  {Galerkin}},\ }\href@noop {} {\bibfield  {journal} {\bibinfo  {journal} {C.
  R. Hebd. Seances Acad. Sci.}\ }\textbf {\bibinfo {volume} {190}},\ \bibinfo
  {pages} {1047} (\bibinfo {year} {1930})}\BibitemShut {NoStop}%
\bibitem [{\citenamefont {Papkovich}(1932)}]{papkovich1932}%
  \BibitemOpen
  \bibfield  {author} {\bibinfo {author} {\bibfnamefont {P.}~\bibnamefont
  {Papkovich}},\ }\href@noop {} {\bibfield  {journal} {\bibinfo  {journal} {C.
  R. Acad. Sci. Paris}\ }\textbf {\bibinfo {volume} {195}},\ \bibinfo {pages}
  {513} (\bibinfo {year} {1932})}\BibitemShut {NoStop}%
\bibitem [{\citenamefont {Neuber}(1934)}]{neuber1934}%
  \BibitemOpen
  \bibfield  {author} {\bibinfo {author} {\bibfnamefont {H.}~\bibnamefont
  {Neuber}},\ }\href@noop {} {\bibfield  {journal} {\bibinfo  {journal} {Z.
  Angew. Math. Mech.}\ }\textbf {\bibinfo {volume} {14}},\ \bibinfo {pages}
  {203} (\bibinfo {year} {1934})}\BibitemShut {NoStop}%
\bibitem [{\citenamefont {Landau}\ and\ \citenamefont
  {Lifshitz}(1986)}]{land70}%
  \BibitemOpen
  \bibfield  {author} {\bibinfo {author} {\bibfnamefont {L.~D.}\ \bibnamefont
  {Landau}}\ and\ \bibinfo {author} {\bibfnamefont {E.~M.}\ \bibnamefont
  {Lifshitz}},\ }\href@noop {} {\emph {\bibinfo {title} {Theory of
  Elasticity}}},\ \bibinfo {edition} {3rd}\ ed.,\ \bibinfo {series} {Course of
  Theoretical Physics}, Vol.~\bibinfo {volume} {7}\ (\bibinfo  {publisher}
  {Butterworth-Heinemann},\ \bibinfo {address} {Oxford},\ \bibinfo {year}
  {1986})\BibitemShut {NoStop}%
\bibitem [{\citenamefont {Bower}(2025)}]{bower2025applied}%
  \BibitemOpen
  \bibfield  {author} {\bibinfo {author} {\bibfnamefont {A.~F.}\ \bibnamefont
  {Bower}},\ }\href@noop {} {\emph {\bibinfo {title} {Applied mechanics of
  solids}}}\ (\bibinfo  {publisher} {CRC press},\ \bibinfo {year}
  {2025})\BibitemShut {NoStop}%
\bibitem [{\citenamefont {Tran-Cong}\ and\ \citenamefont
  {Blake}(1982)}]{cong1982}%
  \BibitemOpen
  \bibfield  {author} {\bibinfo {author} {\bibfnamefont {T.}~\bibnamefont
  {Tran-Cong}}\ and\ \bibinfo {author} {\bibfnamefont {J.}~\bibnamefont
  {Blake}},\ }\href
  {https://doi.org/https://doi.org/10.1016/0022-247X(82)90045-2} {\bibfield
  {journal} {\bibinfo  {journal} {J. Math. Anal. Appl.}\ }\textbf {\bibinfo
  {volume} {90}},\ \bibinfo {pages} {72} (\bibinfo {year} {1982})}\BibitemShut
  {NoStop}%
\bibitem [{SIt()}]{SItext}%
  \BibitemOpen
  \href@noop {} {}\bibinfo {note} {See Supplemental Material for detailed
  information about anlytical derivations.}\BibitemShut {Stop}%
\bibitem [{\citenamefont {Fodor}\ and\ \citenamefont
  {Souslov}(2021{\natexlab{b}})}]{etien2021a}%
  \BibitemOpen
  \bibfield  {author} {\bibinfo {author} {\bibfnamefont {E.}~\bibnamefont
  {Fodor}}\ and\ \bibinfo {author} {\bibfnamefont {A.}~\bibnamefont
  {Souslov}},\ }\href@noop {} {\bibfield  {journal} {\bibinfo  {journal} {Phys.
  Rev. E}\ }\textbf {\bibinfo {volume} {104}},\ \bibinfo {pages} {L062602}
  (\bibinfo {year} {2021}{\natexlab{b}})}\BibitemShut {NoStop}%
\bibitem [{\citenamefont {Squires}\ and\ \citenamefont
  {Mason}(2010)}]{squires2010microrheology}%
  \BibitemOpen
  \bibfield  {author} {\bibinfo {author} {\bibfnamefont {T.~M.}\ \bibnamefont
  {Squires}}\ and\ \bibinfo {author} {\bibfnamefont {T.~G.}\ \bibnamefont
  {Mason}},\ }\href
  {https://doi.org/https://doi.org/10.1146/annurev-fluid-121108-145608}
  {\bibfield  {journal} {\bibinfo  {journal} {Annu. Rev. Fluid Mech.}\ }\textbf
  {\bibinfo {volume} {42}},\ \bibinfo {pages} {413} (\bibinfo {year}
  {2010})}\BibitemShut {NoStop}%
\bibitem [{\citenamefont {Katchalsky}\ and\ \citenamefont
  {Curran}(1965)}]{katch65}%
  \BibitemOpen
  \bibfield  {author} {\bibinfo {author} {\bibfnamefont {A.}~\bibnamefont
  {Katchalsky}}\ and\ \bibinfo {author} {\bibfnamefont {P.~F.}\ \bibnamefont
  {Curran}},\ }\href@noop {} {\emph {\bibinfo {title} {{Nonequilibrium
  Thermodynamics in Biophysics}}}}\ (\bibinfo  {publisher} {Harvard University
  Press},\ \bibinfo {year} {1965})\BibitemShut {NoStop}%
\bibitem [{\citenamefont {Taylor}(2005)}]{taylor2005classical}%
  \BibitemOpen
  \bibfield  {author} {\bibinfo {author} {\bibfnamefont {J.~R.}\ \bibnamefont
  {Taylor}},\ }\href@noop {} {\emph {\bibinfo {title} {Classical mechanics}}}\
  (\bibinfo  {publisher} {University Science Books},\ \bibinfo {address}
  {Sausalito, California},\ \bibinfo {year} {2005})\BibitemShut {NoStop}%
\bibitem [{\citenamefont {Dinelli}\ \emph {et~al.}(2023)\citenamefont
  {Dinelli}, \citenamefont {O'Byrne}, \citenamefont {Curatolo}, \citenamefont
  {Zhao}, \citenamefont {Sollich},\ and\ \citenamefont
  {Tailleur}}]{dinelli2023non}%
  \BibitemOpen
  \bibfield  {author} {\bibinfo {author} {\bibfnamefont {A.}~\bibnamefont
  {Dinelli}}, \bibinfo {author} {\bibfnamefont {J.}~\bibnamefont {O'Byrne}},
  \bibinfo {author} {\bibfnamefont {A.}~\bibnamefont {Curatolo}}, \bibinfo
  {author} {\bibfnamefont {Y.}~\bibnamefont {Zhao}}, \bibinfo {author}
  {\bibfnamefont {P.}~\bibnamefont {Sollich}},\ and\ \bibinfo {author}
  {\bibfnamefont {J.}~\bibnamefont {Tailleur}},\ }\href@noop {} {\bibfield
  {journal} {\bibinfo  {journal} {Nat. Commun.}\ }\textbf {\bibinfo {volume}
  {14}},\ \bibinfo {pages} {7035} (\bibinfo {year} {2023})}\BibitemShut
  {NoStop}%
\bibitem [{\citenamefont {Binysh}\ \emph {et~al.}(2026)\citenamefont {Binysh},
  \citenamefont {Baardink}, \citenamefont {Veenstra}, \citenamefont {Coulais},\
  and\ \citenamefont {Souslov}}]{binysh2026more}%
  \BibitemOpen
  \bibfield  {author} {\bibinfo {author} {\bibfnamefont {J.}~\bibnamefont
  {Binysh}}, \bibinfo {author} {\bibfnamefont {G.}~\bibnamefont {Baardink}},
  \bibinfo {author} {\bibfnamefont {J.}~\bibnamefont {Veenstra}}, \bibinfo
  {author} {\bibfnamefont {C.}~\bibnamefont {Coulais}},\ and\ \bibinfo {author}
  {\bibfnamefont {A.}~\bibnamefont {Souslov}},\ }\href@noop {} {\bibfield
  {journal} {\bibinfo  {journal} {Phys. Rev. X}\ }\textbf {\bibinfo {volume}
  {16}},\ \bibinfo {pages} {021012} (\bibinfo {year} {2026})}\BibitemShut
  {NoStop}%
\end{thebibliography}
%

\end{document}

\title{\Large Resonances in Overdamped Odd Materials \\ \large Supplementary Materials}

\author{Julius Kiln}
\affiliation{Rudolf Peierls Centre for Theoretical Physics, Department of Physics, University of Oxford, Parks Road, Oxford OX1 3PU, United Kingdom}
\author{Alexander Mietke}
\email{alexander.mietke@physics.ox.ac.uk}
\affiliation{Rudolf Peierls Centre for Theoretical Physics, Department of Physics, University of Oxford, Parks Road, Oxford OX1 3PU, United Kingdom}

\maketitle
\tableofcontents
\onecolumngrid
\appendix

\setcounter{figure}{0}
\renewcommand{\thefigure}{S\arabic{figure}}
\setcounter{table}{0}
\renewcommand{\thetable}{S\arabic{table}}

\section{Odd Papkovich-Neuber solution}\label{appsec:OPN}
In the main text, we have introduced the odd Papkovich-Neuber (OPN) solution
\begin{equation}\label{appeq:ElasticGeneralisedPNAnsatz}
\boldsymbol{u}(\boldsymbol{r}) = 2(a + \cos\phi) \boldsymbol{B} - \boldsymbol{R}(\phi) \cdot \nabla (\boldsymbol{r} \cdot \boldsymbol{B} + B_0),
\end{equation}
which solves the force balance equation of arbitrary isotropic odd solids in two dimensions (see Sec.~\ref{sec:OPN}, main text), 
\begin{equation}\label{appeq:CompacOEFB}
    a \nabla^2 \boldsymbol{u} + \boldsymbol{R}(\phi) \cdot \nabla (\nabla \cdot \boldsymbol{u}) = 0,
\end{equation}
if $B_0$ and $\mathbf{B}$ are scalar and vector harmonic functions, respectively. Taking the classical Papkovich-Neuber solution for passive solids (see Eq.~(\ref{eq:PassiveLinearElasticForceBalance}), main text) as inspiration, the OPN solution given in Eq.~(\ref{appeq:ElasticGeneralisedPNAnsatz}) can be found by considering an ansatz
\begin{equation}\label{appeq: Anisotropic PN Candidate}
    \boldsymbol{u} = \boldsymbol{M}_1 \cdot \boldsymbol{B} - \boldsymbol{M}_2 \cdot \nabla (\boldsymbol{r} \cdot \boldsymbol{M}_3 \cdot \boldsymbol{B} + B_0)
\end{equation}
for some matrices $\boldsymbol{M}_k$ ($k=1,2,3$). The only form these matrices can take by symmetry is $\boldsymbol{M}_k\sim a_k\mathbb{I}+b_k\mathbf{R}(\phi)$ for some constants $a_k,b_k$. Plugging the ansatz Eq.~(\ref{appeq: Anisotropic PN Candidate}) into the force balance Eq.~(\ref{appeq:CompacOEFB}), one finds that most constants are trivial or can be set to zero, which eventually yields the OPN solution Eq.~(\ref{appeq:ElasticGeneralisedPNAnsatz}).

\subsection{Poisson ratio, odd ratio, and odd angle}
The force balance equation for a passive isotropic linear elastic material is
\begin{equation}\label{appeq:PassiveLinearElasticForceBalance}
a_0 \nabla^2 \boldsymbol{u} + \nabla (\nabla \cdot \boldsymbol{u}) = 0,
\end{equation}
where $a_0$ can be expressed in terms of the conventional Poisson ratio $\nu_0$ as
\begin{equation}
a_0 = \frac{1 - \nu_0}{1 + \nu_0}.   
\end{equation}
If odd moduli are present, the force balance takes the form given in Eq.~(\ref{appeq:CompacOEFB}), which can be expanded into
\begin{equation}\label{appeq:ElasticGeneralForceBalance}
    a \nabla^2 \boldsymbol{u} + \cos(\phi) \nabla (\nabla \cdot \boldsymbol{u}) - \sin(\phi) \boldsymbol{\epsilon} \cdot \nabla (\nabla \cdot \boldsymbol{u}) = 0.
\end{equation}
 Qualitatively, the Poisson ratio may be read from Eq.~(\ref{appeq:PassiveLinearElasticForceBalance}) by comparing the coefficient of the shear term $\sim\nabla^2 \boldsymbol{u}$ and the bulk term $\sim\nabla (\nabla \cdot \boldsymbol{u})$. Proceeding similarly in Eq.~(\ref{appeq:ElasticGeneralForceBalance}) yields an implicit definition of the conventional Poisson ratio $\nu$ of an odd material in the form
\begin{equation}\label{appeq:nudef}
\frac{a}{\cos(\phi)}=\frac{1 - \nu}{1 + \nu}.
\end{equation}
Comparing the coefficient of the odd term $\sim\boldsymbol{\epsilon} \cdot \nabla (\nabla \cdot \boldsymbol{u})$ in Eq.~(\ref{appeq:ElasticGeneralForceBalance}) to the sum of the achiral coefficients yields a definition of the odd ratio~$\nu^o$ in the form
\begin{equation}\label{appeq:nuodef}
\frac{\sin(\phi)}{a + \cos(\phi)}=\nu^o.
\end{equation}
The relationships Eqs.~(\ref{appeq:nudef}) and (\ref{appeq:nuodef}) agree with definitions of the Poisson and odd ratio  in terms of the elastic moduli~\cite{Scheibner:2020, braverman2020topological}:
\begin{equation}
    \nu := \frac{\mu (B - \mu) + K^o (A - K^o)}{\mu (B + \mu) + K^o (A + K^o)}, \hspace{5pt} \nu^o := \frac{B K^o - \mu A}{\mu (B + \mu) + K^o (A + K^o)}.
\end{equation}

Using Eqs.~(\ref{appeq:nudef}) and (\ref{appeq:nuodef}), we may write the force balance equation for an isotropic odd elastic material in terms of only $\nu, \nu^o$ as 
\begin{equation}\label{appeq:Force Balance in nu,nuo}
    (1 - \nu) \nabla^2 \boldsymbol{u} + (1 + \nu)\nabla (\nabla \cdot \boldsymbol{u}) - 2 \nu^o \nabla (\nabla \cdot \boldsymbol{u}) = 0.
\end{equation}

Interestingly, Eqs.~(\ref{appeq:ElasticGeneralForceBalance}) and (\ref{appeq:Force Balance in nu,nuo}) explicitly depend on only two parameters, $(a, \phi)$ or $(\nu, \nu^o)$ respectively. Consequently, as discussed in Sec.~\ref{subsection:General Elastic Solution DBCs} of the main text, the displacement field under DBCs can only depend on two independent parameters. This is the odd analogy of the fact that deformations under DBCs for passive linear elastic materials depend only on a single parameter, namely the conventional Poisson ratio $\nu_0$.

\subsection{Displacement field solution in polar coordinates}\label{Appendix Section: General Solution Form}
In this section, we derive the form of the general OPN displacement solution in 2D polar coordinates that was given in Eqs.~(\ref{eq:ElasticSolutionSumOverModes})~and~(\ref{eq:nthModeElasticSolution}) in the main text. 

\subsubsection{Harmonic scalar and vector fields}\label{appsec:harmfields}
Separating variables in polar coordinates, scalar harmonic functions can be written as
\begin{equation}
    B_0(r, \theta) = \alpha^{\text{(def)}} \theta \log r+ \sum_{n \in \mathbb{Z}} B_0^{(n)}(r, \theta),
\end{equation}
where the $n^{th}$ scalar harmonic mode, $B_0^{(n)}$, is given by
\begin{equation}\label{appeq:Circular Scalar nth Harmonic}
    {B_0}^{(n)}(r, \theta) = \begin{cases}
        \frac{\alpha^{(n)}_1}{r^n} \cos(n \theta) + \frac{\alpha^{(n)}_2}{r^n} \sin(n \theta)  & \text{for} \hspace{4pt} n \neq 0, \\
        \alpha^{(0)}_1 \log r + \alpha^{(0)}_2 \theta + \alpha_c & \text{for} \hspace{4pt} n = 0,
    \end{cases}
\end{equation}
and $\alpha_k^{(n)}, \alpha^{(\text{def})}, \alpha_c$ are arbitrary constants ($k = 1,2; n \in \mathbb{Z}$) that can be used to solve a given boundary value problem. Similarly, general vector harmonics may be written as
\begin{equation}
    \boldsymbol{B}(r, \theta) = \boldsymbol{B}^{(\text{def})}(r, \theta) + \sum_{n \in \mathbb{Z}} \boldsymbol{B}^{(n)}(r, \theta),
\end{equation}
where the $n^{th}$ vector harmonic mode $\boldsymbol{B}^{(n)}$ is given by
\begin{equation}\label{appeq:Circular Vector nth Harmonic}
    \boldsymbol{B}^{(n)}(r,\theta) = \left( \frac{\beta^{(n)}_1}{r^{n - 1}} + \gamma^{(n)}_1 r^{n - 1} \right) \left[ \cos(n \theta) \boldsymbol{e}_r - \sin(n \theta) \boldsymbol{e}_\theta \right] + \left( \frac{\beta^{(n)}_2}{r^{n - 1}} + \gamma^{(n)}_2 r^{n - 1} \right) \left[ \sin(n \theta) \boldsymbol{e}_r + \cos(n \theta) \boldsymbol{e}_\theta \right],
\end{equation}
and $\boldsymbol{B}^{(\text{def})}$ is given by
\begin{align}
    \boldsymbol{B}^{(\text{def})}(r, \theta) &= (\beta_1^{\text{(def)}} \log r  + \delta_1^{\text{(def)}}\theta + \zeta_1^{\text{(def)}} \theta \log r )\big[\cos(\theta) \boldsymbol{e}_r - \sin(\theta) \boldsymbol{e}_\theta\big] 
    \notag \\ 
    &+ (\beta_2^{\text{(def)}} \log r  + \delta_2^{\text{(def)}} \theta + \zeta_2^{\text{(def)}} \theta \log r )\big[\sin(\theta) \boldsymbol{e}_r + \cos(\theta) \boldsymbol{e}_\theta\big],
\end{align}
and $\beta_k^{(n)}, \gamma_k^{(n)}, \beta_k^{(\text{def})}, \delta_k^{(\text{def})}, \zeta_k^{(\text{def})}$ are again arbitrary constants of integration ($k = 1,2; n \in \mathbb{Z}$).\\

If $\alpha^{(\text{def})}$ or $\boldsymbol{B}^{(\text{def})}$ are non-zero, the displacement field from the OPN ansatz becomes either multivalued (via an explicit $\theta$-dependence) or singular at the origin and infinity (via terms $\propto\log r$). These displacement fields are associated with topological defects and point force solutions, which we discuss further below in Sec.~\ref{Appendix Section: Topological Defects}. In this section, we focus on single-valued solutions that are regular at infinity or the origin and therefore drop these terms. Note that, although the scalar harmonic mode \smash{$B^{(0)}_0$} is multivalued and divergent at both $r = 0, \infty$, its gradient is single valued and decays as $r \rightarrow \infty$. The displacement field corresponding to this scalar harmonic is therefore well-behaved at infinity.\\

As with the conventional Papkovich-Neuber ansatz, multiple terms in the scalar and vector harmonics given in Eqs.~(\ref{appeq:Circular Scalar nth Harmonic}) and (\ref{appeq:Circular Vector nth Harmonic}) yield linearly dependent displacement fields. To eliminate this redundancy, we henceforth neglect $\gamma_k^{(n)}$ from Eq.~(\ref{appeq:Circular Vector nth Harmonic}), as linearly dependent solutions are already generated by terms with $\alpha^{(n)}_k$ in the scalar harmonic contributions Eq.~(\ref{appeq:Circular Scalar nth Harmonic}).\\

\subsubsection{Evaluating the odd Papkovich-Neuber solution}
For convenience, we first redefine the constants of integration $\alpha^{(n)}_k$ to absorb $\boldsymbol{R}(\phi)$, and write the $n^{th}$ mode of the displacement field, $\boldsymbol{u}_{B_0}^{(n)}$, that the OPN solution generates from the scalar harmonic field $B_0$ as
\begin{equation}\label{appeq:GeneralSolutionalphabetaPart}
\boldsymbol{u}^{(n)}_{B_0}(r, \theta) = \frac{1}{r^{n+1}} \boldsymbol{R}(n \theta) \cdot \boldsymbol{\alpha}^{(n)},
\end{equation}
where the vector $\boldsymbol{\alpha}^{(n)} = \alpha^{(n)}_1 \boldsymbol{e}_r + \alpha^{(n)}_2 \boldsymbol{e}_\theta$ collects the integration constants. Furthermore, we write $\boldsymbol{B}^{(n)}(r, \theta)$ given in Eq.~(\ref{appeq:Circular Vector nth Harmonic}) as
\begin{equation}
\boldsymbol{B}^{(n)}(r, \theta) = \frac{1}{r^{n-1}} \boldsymbol{R}^T(n \theta) \cdot \boldsymbol{\beta}^{(n)},
\end{equation}
where $\boldsymbol{\beta}^{(n)} = \beta^{(n)}_1 \boldsymbol{e}_r + \beta^{(n)}_2 \boldsymbol{e}_\theta$. The displacement field, $\boldsymbol{u}_{\boldsymbol{B}}$, that the OPN solution generates from the harmonic vector field reads
\begin{equation}\label{Appendix: Vector Harmonic part of gPN solution}
\boldsymbol{u}_{\boldsymbol{B}}(r, \theta) = 2(a + \cos\phi) \boldsymbol{B} - \boldsymbol{R}(\phi) \cdot \nabla (\boldsymbol{r} \cdot \boldsymbol{B}) = (2 a \mathbb{I} + \boldsymbol{R}^T(\phi)) \cdot \boldsymbol{B} - \boldsymbol{R}(\phi) \cdot (\nabla\otimes\boldsymbol{B}) \cdot \boldsymbol{r},
\end{equation}
where both contractions ($\cdot$) and dyadic products ($\otimes$) will be explicitly written throughout the supplementary material for clarity. Writing $\boldsymbol{B}^{(n)} = \boldsymbol{e}_r \left( \boldsymbol{e}_r \cdot \boldsymbol{B}^{(n)} \right) + \boldsymbol{e}_\theta \left( \boldsymbol{e}_\theta \cdot \boldsymbol{B}^{(n)}\right) $, we can use that
\begin{equation}\label{appeq:Theta grad trick}
\frac{\partial}{\partial \theta} \boldsymbol{v}_1 \cdot \boldsymbol{R}(n \theta) \cdot \boldsymbol{v}_2 = - n \boldsymbol{v}_1 \cdot \boldsymbol{\epsilon} \cdot \boldsymbol{R}(n \theta) \cdot \boldsymbol{v}_2,
\end{equation}
for any $\boldsymbol{v}_1, \boldsymbol{v}_2$ with constant polar coordinate components to show that
\begin{equation}
\boldsymbol{R}(\phi) \cdot (\nabla\otimes\boldsymbol{B}^{(n)}) \cdot \boldsymbol{r} = -\frac{n-1}{r^{n-1}} \boldsymbol{R}(\phi) \cdot \boldsymbol{Z} \cdot \boldsymbol{R}^T(n \theta) \cdot \boldsymbol{\beta}^{(n)}.
\end{equation}
Here, $\boldsymbol{Z} = \boldsymbol{e}_r\otimes \boldsymbol{e}_r - \boldsymbol{e}_\theta\otimes\boldsymbol{e}_\theta$ corresponds to a local reflection along the line with normal vector $\boldsymbol{e}_\theta$. Defining the matrix
\begin{equation}\label{appeq:RTildeDefinition}
    \boldsymbol{\Tilde{R}} := \boldsymbol{R}(\phi)\cdot\left[2a \mathbb{I} + \boldsymbol{R}(\phi) \right]^{-1} = \frac{1}{4a^2 + 4a \cos\phi + 1} \left( 2a \boldsymbol{R}(\phi) + \mathbb{I} \right),
\end{equation}
we transform the integration constants to $\boldsymbol{\tilde{\beta}}$ via
\begin{equation}\label{appeq beta' def}
    \boldsymbol{\beta}^{(n)} = \boldsymbol{R}(\phi) \cdot \boldsymbol{Z} \cdot \boldsymbol{\Tilde{R}} \cdot \boldsymbol{\Tilde{\beta}}^{(n)}. 
\end{equation}

Since the components in Eq.~(\ref{appeq beta' def}) are arbitrary integration constants, we drop in the following the tilde from $\boldsymbol{\Tilde{\beta}}^{(n)}$. Combining this with Eq.~(\ref{appeq:GeneralSolutionalphabetaPart}) yields the following general $n^{th}$ mode solution given in Sec.~\ref{subsection: Displacement BCs} of the main text
\begin{equation}\label{eq:nthModeElasticSolution Appendix}
    \boldsymbol{u}^{(n)}(r, \theta) = \frac{1}{r^{n+1}} \boldsymbol{R}(n \theta) \cdot \boldsymbol{\alpha}^{(n)}
+ \frac{1}{r^{n-1}} ((n-1) \boldsymbol{\Tilde{R}} + \boldsymbol{Z}) \cdot \boldsymbol{R}(n \theta) \cdot \boldsymbol{\beta}^{(n)}.
\end{equation}

By linearity of Eq.~(\ref{appeq:ElasticGeneralForceBalance}), the general displacement field $\boldsymbol{u}$ can be written as a sum of these $n^{th}$ mode solutions $\boldsymbol{u}^{(n)}$ for $n \in \mathbb{Z}$. It should be noted that changes in the sign of mode number, $n$, correspond to the same angular modes. Regularity of the solution at the origin or infinity leads to restrictions on this sign, but for annulus geometries we must consider both $\pm n$ when enforcing boundary conditions at each mode.\\

\subsection{Stress boundary conditions}\label{appsec:SBCs}
The stress tensor for an odd linearly elastic solid is
\begin{align}
    \boldsymbol{\sigma} &= \boldsymbol{\sigma}_0 + \left[ (B - \mu)(\nabla \cdot \boldsymbol{u}) - \Lambda \omega \right] \mathbb{I} + \left[ -A \left( \nabla \cdot \boldsymbol{u} \right) + \Gamma \omega \right] \boldsymbol{\epsilon} \notag\\
    &+ \mu \left[ \nabla\otimes \boldsymbol{u} + \left( \nabla\otimes\boldsymbol{u} \right)^T \right] + \frac{K^o}{2} \left[ \nabla_{\perp}\otimes \boldsymbol{u} + \nabla \otimes\boldsymbol{u}_{\perp} +\left(\nabla_{\perp}\otimes \boldsymbol{u} + \nabla \otimes\boldsymbol{u}_{\perp}\right)^T\right],\label{eq:Odd Stress Tensor Appendix}
\end{align}
where $\nabla_{\perp}=\boldsymbol{\epsilon} \cdot \nabla$, $\boldsymbol{u}_{\perp}=\boldsymbol{\epsilon} \cdot \boldsymbol{u}$, and $\smash{\boldsymbol{\sigma}_0 = -p \mathbb{I} + \tau \boldsymbol{\epsilon}}$ is a prestress with potential contributions from an isotropic pressure, $p$, and a torque density, $\tau$. A constant prestress, as discussed for example in~\cite{braverman2020topological}, does not affect the force balance equation and only becomes relevant at boundaries. If the prestress is not constant, as discussed for example in~\cite{lee2025disordered}, one can instead employ an OPN solution that incorporates external forces (see Sec.~\ref{appsec:OPN with forcing}).

\subsubsection{Axisymmetric solution ($n=0$)}
Treating the mode $n = 0$ first, we write the surface traction $\mathbf{\hat{f}}^{(0)}$ at some circular boundary ($r=R$) as
\begin{equation}
    \mathbf{\hat{f}}^{(0)}(\theta) = f^{(0)}_r \boldsymbol{e}_r + f^{(0)}_\theta \boldsymbol{e}_\theta,
\end{equation}
where $f_r^{(0)}, f_{\theta}^{(0)}$ are constants. The 0-th mode displacement field solution that is regular at $r = \infty$ (setting $- \boldsymbol{e}_r \cdot \boldsymbol{\sigma}^{(0)} = \mathbf{\hat{f}}^{(0)}$) is given by
\begin{equation}\label{appeq:u0-}
    \boldsymbol{u}^{(0)}_{-}(r, \theta) = \frac{R^2}{2 r (\mu^2 + (K^o)^2)} \begin{pmatrix}
        \mu & -K^o \\ K^o & \mu
    \end{pmatrix} \cdot (\mathbf{\hat{f}}^{(0)} - p \boldsymbol{e}_r + \tau \boldsymbol{e}_\theta).
\end{equation}

The corresponding solution that is regular at $r = 0$ (setting $\boldsymbol{e}_r \cdot \boldsymbol{\sigma}^{(0)} = \mathbf{\hat{f}}^{(0)}$) is given by
\begin{equation}\label{appeq:uo+}
    \boldsymbol{u}^{(0)}_{+}(r, \theta) = \frac{r}{2(B \Gamma - A \Lambda)} \begin{pmatrix}
        \Gamma & \Lambda \\ A & B
    \end{pmatrix} \cdot (\mathbf{\hat{f}}^{(0)} + p \boldsymbol{e}_r - \tau \boldsymbol{e}_\theta).
\end{equation}

Taking the limit $A, \Lambda, \Gamma \rightarrow 0$ (passive material without substrate coupling), we see that any torque destabilizes the material as there is no sink of angular momentum. In the absence of torques and prestress, the material will dilate with characteristic length scale $\sim \hat{f}^{(0)}/B$, as expected. Eqs~(\ref{appeq:u0-})~and~(\ref{appeq:uo+}) explicitly show how odd material parameters explicitly break Maxwell-Betti reciprocity on the macroscopic scale \cite{braverman2020topological}: The displacement field and force are related by a symmetric matrix if and only if $K^o = 0, A = \Lambda$, i.e. the modulus tensor $C_{\alpha \beta}$ is symmetric.

\subsubsection{General solution ($n\ne0$)}
To solve the stress boundary condition problem for $n\ne0$ using the general $n$-th mode displacement solution $\boldsymbol{u}^{(n)}(r, \theta)$ given in Eq.~(\ref{eq:nthModeElasticSolution Appendix}), we use the identity given in Eq.~(\ref{appeq:Theta grad trick}) to find
\begin{align}\label{appeq:dispgrad}
    \nabla \otimes \boldsymbol{u}^{(n)}(r, \theta) = &-\frac{n+1}{r^{n+2}}\left[(\boldsymbol{e}_r\otimes\boldsymbol{e}_r - \boldsymbol{e}_{\theta}\otimes\boldsymbol{e}_{\theta}) \otimes \boldsymbol{e}_r + (\boldsymbol{e}_r\otimes\boldsymbol{e}_{\theta} + \boldsymbol{e}_{\theta}\otimes\boldsymbol{e}_r) \otimes \boldsymbol{e}_{\theta} \right] \cdot \boldsymbol{R}(n \theta) \cdot\boldsymbol{\alpha}^{(n)} 
    \notag \\
    &- \frac{n-1}{r^{n}}\left[(\boldsymbol{e}_r\otimes\boldsymbol{e}_r - \boldsymbol{e}_{\theta}\otimes\boldsymbol{e}_{\theta}) \otimes (n \boldsymbol{e}_r \cdot \boldsymbol{\Tilde{R}}) + (\boldsymbol{e}_r\otimes\boldsymbol{e}_{\theta} + \boldsymbol{e}_{\theta}\otimes\boldsymbol{e}_r) \otimes (n \boldsymbol{e}_{\theta} \cdot \boldsymbol{\Tilde{R}}) \right. 
    \notag\\ 
    &  
    + \left.(\boldsymbol{e}_r \otimes \boldsymbol{e}_r + \boldsymbol{e}_{\theta} \otimes \boldsymbol{e}_{\theta}) \otimes (\boldsymbol{e}_r \cdot (\mathbb{I} - \boldsymbol{\Tilde{R}})) - (\boldsymbol{e}_r \otimes \boldsymbol{e}_{\theta} - \boldsymbol{e}_{\theta} \otimes \boldsymbol{e}_r) \otimes (\boldsymbol{e}_{\theta} \cdot (\mathbb{I} + \boldsymbol{\Tilde{R}}))
    \right] \cdot \boldsymbol{R}(n \theta)\cdot\boldsymbol{\beta}^{(n)}.
\end{align}
This is a convenient form to write the deformation gradient, since it is decomposed into second rank tensors that reflect the structure of the $s^\alpha_{ij}$ tensor basis elements given in Eq.~(\ref{eq:ScheibnerBasis}) (main text), even though they are not equivalent. However, isotropy of the modulus tensor means that a rotation of our coordinate system from polar to Cartesian has no physical effect, such that we may still read off the stress tensor components using Eq.~(\ref{appeq:dispgrad}). Assuming prestress is absent or does not contribute to the $n^{th}$ mode, we then find the surface traction that is applied by the odd elastic material to the cavity boundary reads
\begin{align}\label{appeq:GeneralForceAtBoundary}
    \boldsymbol{e}_r \cdot \boldsymbol{\sigma}^{(n)}(r, \theta) &= - \frac{2(n+1)}{r^{n+2}} \Big[\mu (\boldsymbol{e}_r\otimes\boldsymbol{e}_r + \boldsymbol{e}_{\theta}\otimes\boldsymbol{e}_{\theta}) + K^o (\boldsymbol{e}_{r}\otimes\boldsymbol{e}_{\theta} - \boldsymbol{e}_{\theta}\otimes\boldsymbol{e}_r) \Big] \cdot \boldsymbol{R}(n \theta) \cdot \boldsymbol{\alpha}^{(n)}
    \notag \\
    &- \frac{2(n-1)}{r^n} \Big[ n \mu (\boldsymbol{e}_r\otimes\boldsymbol{e}_r + \boldsymbol{e}_{\theta}\otimes\boldsymbol{e}_{\theta}) \cdot \boldsymbol{\Tilde{R}} + n K^o (\boldsymbol{e}_r\otimes\boldsymbol{e}_{\theta} - \boldsymbol{e}_{\theta}\otimes\boldsymbol{e}_r) \cdot \boldsymbol{\Tilde{R}} 
    \\ \notag
    &+ (B \boldsymbol{e}_r\otimes\boldsymbol{e}_r - A \boldsymbol{e}_{\theta}\otimes\boldsymbol{e}_r)\cdot (\mathbb{I} - \boldsymbol{\Tilde{R}}) + (\Lambda \boldsymbol{e}_r\otimes\boldsymbol{e}_{\theta} - \Gamma \boldsymbol{e}_\theta\otimes\boldsymbol{e}_\theta) \cdot (\mathbb{I} + \boldsymbol{\Tilde{R}}) \Big] \cdot \boldsymbol{R}(n \theta) \cdot \boldsymbol{\beta}^{(n)},
\end{align}
where we take only $n > 0$, so that the solution is regular at infinity. Taking $n \mapsto -n$ in these expressions yields instead the deformation gradient and boundary stress that is regular at the origin. Considering a surface traction applied to the odd elastic material at $r = R$, expanded as
\begin{equation}\label{appeq: Stress Boundary Condition on r = R}
    \boldsymbol{t}^{(n)} = - \boldsymbol{e}_r \cdot \boldsymbol{\sigma}^{(n)}|_{r=R} = [P_n \cos(n \theta) + Q_n \sin(n \theta)]\boldsymbol{e}_r + [M_n \cos(n \theta) + N_n \sin(n \theta)]\boldsymbol{e}_{\theta},
\end{equation}
we compare radial and azimuthal components of Eqs.~(\ref{appeq:GeneralForceAtBoundary}) and (\ref{appeq: Stress Boundary Condition on r = R}) to obtain
\begin{align}
        \boldsymbol{e}_r \cdot \boldsymbol{R}(n \theta) \cdot \boldsymbol{Z} \cdot \begin{pmatrix}
            P_n \\ Q_n
        \end{pmatrix} &= \boldsymbol{e}_r \cdot \boldsymbol{R}(n \theta) \cdot \Bigg\{ \frac{2(n+1)}{R^{n+2}}\left(\mu \mathbb{I} + K^o \boldsymbol{\epsilon} \right)\cdot \boldsymbol{\alpha}^{(n)} 
        \notag \\
        &+ \frac{2(n-1)}{R^n} \left[ n \mu \boldsymbol{\Tilde{R}} + n K^o \boldsymbol{\epsilon} \cdot \boldsymbol{\Tilde{R}} + B (\mathbb{I} - \boldsymbol{\Tilde{R}}) + \Lambda \boldsymbol{\epsilon} \cdot (\mathbb{I} + \boldsymbol{\Tilde{R}}) \right] \cdot \boldsymbol{\beta}^{(n)} \Bigg\},
        \label{eq:SBCid1}\\
        \boldsymbol{e}_\theta \cdot \boldsymbol{R}(n \theta) \cdot \boldsymbol{Z} \cdot \boldsymbol{\epsilon} \cdot \begin{pmatrix}
            M_n \\ N_n
        \end{pmatrix} &= \boldsymbol{e}_\theta \cdot \boldsymbol{R}(n \theta) \cdot \Bigg\{ \frac{2(n+1)}{R^{n+2}} \left( \mu \mathbb{I} + K^o \boldsymbol{\epsilon} \right) \cdot \boldsymbol{\alpha}^{(n)}
        \notag \\
        &+ \frac{2(n-1)}{R^n} \left[ n \mu \boldsymbol{\Tilde{R}} + n K^o \boldsymbol{\epsilon} \cdot \boldsymbol{\Tilde{R}} + A \boldsymbol{\epsilon} \cdot (\mathbb{I} - \boldsymbol{\Tilde{R}}) - \Gamma (\mathbb{I} + \boldsymbol{\Tilde{R}}) \right] \cdot \boldsymbol{\beta}^{(n)} \Bigg\}.\label{eq:SBCid2}
\end{align}

Using the orthogonality of $\sin(n \theta)$ and  $\cos(n \theta)$ terms, Eqs.~(\ref{eq:SBCid1}) and (\ref{eq:SBCid2}) can be converted into a single matrix equation
\begin{equation}\label{appeq: Stress BC Matrix Equation for Constants}
    \begin{pmatrix}
        P_n \\ -Q_n \\ N_n \\ M_n
    \end{pmatrix} = \frac{2}{R^n}\begin{pmatrix}
        \frac{n+1}{R^2} \boldsymbol{M}_s & (n-1)\boldsymbol{C}_{1}^{(n)} \\
        \frac{n+1}{R^2} \boldsymbol{M}_s & (n-1)\boldsymbol{C}_{2}^{(n)}
    \end{pmatrix} \cdot \begin{pmatrix}
        \alpha^{(n)}_1 \\ \alpha^{(n)}_2 \\ \beta^{(n)}_1 \\ \beta^{(n)}_2
    \end{pmatrix},
\end{equation}
where the $\boldsymbol{M}_s, \boldsymbol{C}_{1,2}^{(n)}$ are all proportional to rotation matrices, and are given by
\begin{align}
    \boldsymbol{M}_s &= (\mu \mathbb{I} + K^o \boldsymbol{\epsilon}), \notag
    \\ 
    \boldsymbol{C}_{1}^{(n)} &= \left[ n(\mu \mathbb{I} + K^o \boldsymbol{\epsilon}) \cdot \boldsymbol{\Tilde{R}} + B (\mathbb{I} - \boldsymbol{\Tilde{R}}) + \Lambda \boldsymbol{\epsilon} \cdot (\mathbb{I} + \boldsymbol{\Tilde{R}}) \right],
    \\
    \boldsymbol{C}_{2}^{(n)} &= \left[ n(\mu \mathbb{I} + K^o \boldsymbol{\epsilon}) \cdot \boldsymbol{\Tilde{R}} + A \boldsymbol{\epsilon} \cdot (\mathbb{I} - \boldsymbol{\Tilde{R}}) - \Gamma (\mathbb{I} + \boldsymbol{\Tilde{R}}) \right]. \notag
\end{align}

Inverting matrix equation~(\ref{appeq: Stress BC Matrix Equation for Constants}), one obtains
\begin{equation}\label{eq:App. alphagamma Stress BCs}
    \begin{pmatrix}
        \alpha^{(n)}_1 \\ \alpha^{(n)}_2 \\ \beta^{(n)}_1 \\ \beta^{(n)}_2
    \end{pmatrix} = 
    \frac{R^n}{2} \begin{pmatrix}
        -\frac{R^2}{n+1} \boldsymbol{M}_s^{-1} \cdot \boldsymbol{M}_d^{-1} \cdot \boldsymbol{C}_2^{(n)} & \frac{R^2}{n+1} \boldsymbol{M}_s^{-1} \cdot \boldsymbol{M}_d^{-1} \cdot \boldsymbol{C}_{1}^{(n)}\\
        \frac{1}{n-1} \boldsymbol{M}_d^{-1}& - \frac{1}{n-1} \boldsymbol{M}_d^{-1}
    \end{pmatrix} \cdot
    \begin{pmatrix}
        P_n \\ -Q_n \\ N_n \\ M_n
    \end{pmatrix},
\end{equation}
where
\begin{align}
\boldsymbol{M}_d &= \boldsymbol{C}_{1}^{(n)} - \boldsymbol{C}_{2}^{(n)} \notag\\
&= \big[ (B \mathbb{I} - A \boldsymbol{\epsilon}) \cdot (\mathbb{I} - \boldsymbol{\Tilde{R}}) + (\Gamma \mathbb{I} + \Lambda \boldsymbol{\epsilon}) \cdot (\mathbb{I} + \boldsymbol{\Tilde{R}}) \big].
\end{align}
Note that the matrices, $\boldsymbol{M}_{s,d}$, whose inverses the constants of integration, $\alpha_k^{(n)}, \beta_k^{(n)}$, depend on are both independent of the mode $n$. This leads to the independence of resonances on the mode for an odd viscoelastic material, as discussed in Sec. \ref{sec:OVE SBCs} of the main text. Substituting Eq.~(\ref{eq:App. alphagamma Stress BCs}) into the general form of the solution Eq.~(\ref{eq:nthModeElasticSolution Appendix}) and using that rotation matrices commute, we obtain a closed expression for the $n$-th mode of the displacement field resulting from stress boundary conditions characterized by constants $P_n,Q_n,N_n,M_n$ [see Eq.~(\ref{appeq: Stress Boundary Condition on r = R})]
\begin{align}
    \boldsymbol{u}^{(n)}(r,\theta) &= \frac{1}{2} \Bigg[ - \frac{R^{n+2}}{(n+1) r^{n+1}}\boldsymbol{M}_s^{-1} \cdot \boldsymbol{M}_d^{-1} \cdot \boldsymbol{C}_2^{(n)} + \frac{R^n}{(n-1) r^{n-1}}((n-1) \boldsymbol{\Tilde{R}} + \boldsymbol{Z}) \cdot \boldsymbol{M}_d^{-1}\Bigg] \cdot \boldsymbol{R}(n \theta) \cdot \begin{pmatrix}
        P_n \\ -Q_n
    \end{pmatrix} 
    \notag \\
    &+ \frac{1}{2} \Bigg[ \frac{R^{n+2}}{(n+1) r^{n+1}} \boldsymbol{M}_s^{-1} \cdot \boldsymbol{M}_d^{-1} \cdot \boldsymbol{C}_{1}^{(n)}  - \frac{R^n}{(n-1) r^{n-1}}((n-1) \boldsymbol{\Tilde{R}} + \boldsymbol{Z}) \cdot \boldsymbol{M}_d^{-1} \Bigg] \cdot \boldsymbol{R}(n \theta) \cdot \begin{pmatrix}
        N_n \\ M_n
    \end{pmatrix}.
    \label{appeq:SBCSolution}
\end{align}

\subsection{Stokes paradox}
Equations~(\ref{eq:App. alphagamma Stress BCs}) and (\ref{appeq:SBCSolution}) suggest that singularities could arise for modes $n = \pm 1$. Such singularities are associated with integration constants $\boldsymbol{\alpha}^{(-1)}, \boldsymbol{\beta}^{(1)}$ in the general displacement field solution Eq.~(\ref{eq:nthModeElasticSolution Appendix}). This is a manifestation of the Stokes paradox in 2D, where the displacement field exhibits a logarithmic divergence at infinity when a net force acts on the cavity boundary (see Sec.~\ref{Appendix Section: Topological Defects} for a discussion of these terms in the context of topological defects). As mentioned in the main text, this divergence is absent for force-free boundary conditions. The general traction force expansion in Eq.~(\ref{appeq: Stress Boundary Condition on r = R}) yields net force components
\begin{align}
F_x^{\pm}&=\mathbf{e}_x\cdot\oint\boldsymbol{t}^{(\pm1)}\,dl=R\int_0^{2\pi}[(P_{\pm1}\cos\theta\pm Q_{\pm1}\sin \theta)\cos\theta - (M_{\pm1}\cos\theta\pm N_{\pm1}\sin\theta)\sin\theta]d\theta\notag\\
&=\pi R(P_{\pm1}\mp N_{\pm1})\\
F_y^{\pm}&=\mathbf{e}_y\cdot\oint\boldsymbol{t}^{(\pm1)}\,dl=R\int_0^{2\pi}[(P_{\pm1}\cos\theta \pm Q_{\pm1}\sin \theta)\sin\theta + (M_{\pm1}\cos\theta \pm N_{\pm1}\sin\theta)\cos\theta]d\theta\notag\\
&=\pi R(Q_{\pm1}\pm M_{\pm1})
\end{align}
and is therefore force-free if $P_1=N_1$, $P_{-1}=-N_{-1}$, $Q_1=-M_1$ and $Q_{-1}=M_{-1}$. Similar integrals vanish for all other traction modes by symmetry. If these conditions hold, Eq.~(\ref{eq:App. alphagamma Stress BCs}) describes a well-defined linear system and Eq.~(\ref{appeq:SBCSolution}) yields a well-defined displacement field solution for all modes, including those with $|n|=1$.\\

As a consistency check, we may use Eq.~(\ref{appeq:GeneralForceAtBoundary}) to independently verify that the displacement field solution Eq.~(\ref{eq:nthModeElasticSolution Appendix}), from which we had deliberately excluded logarithmic and multivalued terms (see Sec.~\ref{appsec:harmfields}), yields stresses at circular boundaries that do not impose a net force. For $n = 1$, it can be directly read off from Eq.~(\ref{appeq:GeneralForceAtBoundary}) that $\boldsymbol{\beta}^{(1)}$ does not contribute to the stress. Also the total force due to $\boldsymbol{\alpha}^{(1)}$ vanishes, which follows from
\begin{equation}
    \mathbf{F}^{+}=\oint\boldsymbol{e}_r \cdot \boldsymbol{\sigma}^{(1)}|_{r=R} d l =- \frac{4}{R^2} \left(\mu \mathbb{I} + K^o \boldsymbol{\epsilon} \right) \cdot \int_0^{2 \pi} \boldsymbol{R}(2 \theta) d \theta \cdot (\alpha^{(1)}_1 \boldsymbol{e}_x + \alpha^{(1)}_2 \boldsymbol{e}_y) = 0.
\end{equation}
Similarly, for $n = -1$, the constant vector $\boldsymbol{\alpha}^{(-1)}$ does not contribute to the stress and the $\boldsymbol{\beta}^{(-1)}$ term yields a force
\begin{align}
    \mathbf{F}^{-}&=\oint\boldsymbol{e}_r \cdot \boldsymbol{\sigma}^{(-1)}|_{r=R} d l
    \notag \\
    &= 4 R^2 \int_0^{2 \pi} \left[\left(B \boldsymbol{e}_r \otimes \boldsymbol{e}_r - A \boldsymbol{e}_\theta \otimes \boldsymbol{e}_r\right)\cdot \left(\mathbb{I} - \boldsymbol{\Tilde{R}}\right)\big]  + \left(\Lambda \boldsymbol{e}_r \otimes \boldsymbol{e}_\theta - \Gamma \boldsymbol{e}_\theta \otimes \boldsymbol{e}_\theta\right) \cdot \left(\mathbb{I} + \boldsymbol{\Tilde{R}}\right) -\left(\mu \mathbb{I} + K^o \boldsymbol{\epsilon}\right) \cdot \boldsymbol{\Tilde{R}} \right] d \theta 
    \notag \\
    &\hspace{1.6cm}\cdot \left(\beta^{(-1)}_1 \boldsymbol{e}_x + \beta^{(-1)}_2 \boldsymbol{e}_y\right) 
    \notag \\
    &= 2R^2 \Big\{\big[(B-\Gamma) \mathbb{I} + (A+\Lambda) \boldsymbol{\epsilon}\big] - \big[(B + \Gamma + 2\mu) \mathbb{I} + (A - \Lambda + 2K^o) \boldsymbol{\epsilon} \big] \cdot \boldsymbol{\Tilde{R}} \Big\} \cdot\left(\beta^{(-1)}_1 \boldsymbol{e}_x + \beta^{(-1)}_2 \boldsymbol{e}_y\right)  = 0,
\end{align}
where we have used that, for $\boldsymbol{\Tilde{R}}$ defined in Eq.~(\ref{appeq:RTildeDefinition}), we have the identity
\begin{equation}\label{appeq:RtildeDefViaModuli}
    \boldsymbol{\Tilde{R}} = \big[(B-\Gamma) \mathbb{I} + (A+\Lambda) \boldsymbol{\epsilon}\big] \cdot \big[(B + \Gamma + 2\mu) \mathbb{I} + (A - \Lambda + 2K^o) \boldsymbol{\epsilon} \big]^{-1}.
\end{equation}

\subsection{Topological defects}\label{Appendix Section: Topological Defects}
In the main text and in the detailed derivation of the displacement field solution (Sec. \ref{Appendix Section: General Solution Form}), we did for brevity not include contributions to harmonic scalars and vector fields that give rise to displacement fields that are multivalued or singular at both the origin and infinity. These can be written as
\begin{equation}
    B_0 = \theta \log(r)  b_0 , \hspace{5pt} \boldsymbol{B} = \theta \boldsymbol{b}_1 + \log(r)  \boldsymbol{b}_2 + \theta \log(r)  \boldsymbol{b}_3
\end{equation}
where $\boldsymbol{b}_{1,2,3}$ are constant vectors (i.e. constant Cartesian rather than polar components) containing the integration constants, and $b_0$ is a constant scalar. The OPN displacement fields corresponding to $\boldsymbol{b}_{1,2}$ can be written as
\begin{align}
        \boldsymbol{u}_1(r, \theta) &= \theta \boldsymbol{b} - \frac{1}{2} \boldsymbol{Z} \cdot \boldsymbol{\Tilde{R}}^T \cdot \boldsymbol{\epsilon} \cdot \boldsymbol{b}, \label{appeq:theta u1}
        \\
        \boldsymbol{u}_2(r, \theta) &= \log(r)  \boldsymbol{\bar{b}} - \frac{1}{2} \boldsymbol{Z} \cdot \boldsymbol{\Tilde{R}}^T \cdot \boldsymbol{\bar{b}}. \label{appeq:log u2}
\end{align}
where we have added constant displacements to write both solutions in terms of $\boldsymbol{Z}$ and redefined the integration constants. The stresses $\boldsymbol{\sigma}_{1}$ and $\boldsymbol{\sigma}_{2}$ associated with these displacement fields go as $\sim 1/r$, such that we expect finite forces at the origin. Indeed, one can show that \hbox{\smash{$\nabla \cdot \boldsymbol{\sigma}_{1,2} = - \boldsymbol{f}_{1,2} \, \delta^{(2)}(\boldsymbol{r})$}} with
\begin{align}
    \boldsymbol{f}_1 &= \frac{4\pi}{(B + 2 \mu + \Gamma)^2 + (A + 2K^o - \Lambda)^2} \boldsymbol{\epsilon} \cdot \big[(B + 2\mu + \Gamma) \mathbb{I} + (A +2K^o - \Lambda) \boldsymbol{\epsilon} \big]
    \notag\\
    &\hspace{5.84cm} \cdot \big[ (\mu^2 + (K^o)^2 + A \Lambda - B \Gamma) \mathbb{I} + (\mu(\Lambda - A) + K^o(B + \Gamma)) \boldsymbol{\epsilon} \big] \cdot \boldsymbol{b}, \label{appeq:theta f1}
    \\
    \boldsymbol{f}_2 &= - 4 \pi \frac{(\mu + \Gamma)(B + \mu) + (K^o - \Lambda)(A + K^o)}{(B + 2 \mu + \Gamma)^2 + (A + 2K^o - \Lambda)^2} \big[(B + 2\mu + \Gamma) \mathbb{I} + (A +2K^o - \Lambda) \boldsymbol{\epsilon} \big] \cdot \boldsymbol{\bar{b}}. \label{appeq:log f2}
\end{align}

For a force-free dislocation with Burgers vector $\boldsymbol{b}$, we must take a linear combination of the solutions given in Eqs.~(\ref{appeq:theta u1}) and (\ref{appeq:log u2}) such that the forces at the origin cancel. This yields the solution -- up to a constant displacement -- found previously by Braverman et al. \cite{braverman2020topological}
\begin{equation}
    \boldsymbol{u}(r, \theta) = \theta \boldsymbol{b} + \log(r)  \boldsymbol{\bar{b}} - \frac{1}{2} \boldsymbol{Z} \cdot \boldsymbol{\Tilde{R}}^T \cdot [\boldsymbol{\epsilon} \cdot \boldsymbol{b} + \boldsymbol{\bar{b}}],
\end{equation}
with
\begin{equation}
    \boldsymbol{\bar{b}} = \frac{1}{(\mu + \Gamma)(B + \mu) + (K^o - \Lambda)(A + K^o)} \boldsymbol{\epsilon} \cdot \big[ (\mu^2 + (K^o)^2 + A \Lambda - B \Gamma) \mathbb{I} + (\mu(\Lambda - A) + K^o(B + \Gamma)) \boldsymbol{\epsilon} \big] \cdot \boldsymbol{b}.
\end{equation}

\subsection{External forcing}\label{appsec:OPN with forcing}
\subsubsection{Green's tensor}
The Green's tensor $\boldsymbol{G}(\boldsymbol{r})$ that generates a displacement field $\boldsymbol{u}(\boldsymbol{r})=\int\boldsymbol{G}(\boldsymbol{r}-\boldsymbol{r}')\cdot\boldsymbol{f}(\boldsymbol{r}')$ from any regular vector field $\boldsymbol{f}(\boldsymbol{r})$ such that $\nabla\cdot\boldsymbol{\sigma}(\boldsymbol{u}) = - \boldsymbol{f}$, can be read from Eqs.~(\ref{appeq:log u2}) and (\ref{appeq:log f2}) as
\begin{align}
    \boldsymbol{G}(\boldsymbol{r}) = - \frac{1}{4 \pi [(\mu + \Gamma)(B + \mu) + (K^o - \Lambda)(A + K^o)]} \left(\log(r)   \mathbb{I} - \frac{1}{2} \boldsymbol{Z} \cdot \boldsymbol{\Tilde{R}}^T \right) \cdot \big[(B + 2\mu + \Gamma) \mathbb{I} - (A +2K^o - \Lambda) \boldsymbol{\epsilon} \big].
\end{align}

\subsubsection{Generalized odd Papkovich-Neuber solution}
In the presence of an external force $\boldsymbol{f}$, the overdamped force balance equation becomes $\nabla \cdot \boldsymbol{\sigma} = \boldsymbol{f}$, which can be written as
\begin{equation}\label{appeq:ForcedOddElasticForceBalance}
    a \nabla^2 \boldsymbol{u} + \boldsymbol{R}(\phi) \cdot \nabla (\nabla \cdot \boldsymbol{u}) = \boldsymbol{\Tilde{f}},
\end{equation}
where
\begin{equation}
    \boldsymbol{\Tilde{f}} = \frac{1}{\sqrt{(B^2 + A^2)(\mu^2 + (K^o)^2)}}(\mu \mathbb{I} - K^o \boldsymbol{\epsilon}) \cdot \boldsymbol{f}.
\end{equation}
This is solved by the OPN ansatz
\begin{equation}\label{appeq:OPNAnsatz}
    \boldsymbol{u}(\boldsymbol{r}) = 2(a + \cos\phi) \boldsymbol{B} - \boldsymbol{R}(\phi) \cdot \nabla (\boldsymbol{r} \cdot \boldsymbol{B} + B_0),
\end{equation}
where now $B_0, \boldsymbol{B}$ satisfy Poisson's equation rather than Laplace's equation,
\begin{align}
    \nabla^2 \boldsymbol{B} &= \frac{1}{2a(a + \cos\phi)} \boldsymbol{\Tilde{f}}
    \\
    \nabla^2 B_0 &= -\frac{1}{2a(a+\cos\phi)} \boldsymbol{r} \cdot \boldsymbol{\Tilde{f}}.
\end{align}
Finding a particular solution to the force balance Eq.~(\ref{appeq:ForcedOddElasticForceBalance}) is thus reduced to finding a particular solution to the scalar and vector Poisson's equations. For a passive material ($a = a_0, \phi = 0$), this reduces to the usual forced Papkovich-Neuber solution \cite{bower2025applied}.

\subsection{Odd incompressible (Stokes) flow}\label{Appsec:IncompOPN}
An odd incompressible fluid can be described by a stress tensor~\cite{banerjee2017odd,soni2019odd}
\begin{align}\label{appeq:IncompStressTensor}
    \boldsymbol{\sigma} &= - \left(p + \eta_\Lambda \omega \right)\mathbb{I} + \eta_R \left(\omega - 2 \Omega \right)\boldsymbol{\epsilon} \notag\\
    &+ \eta_s \left(\nabla\otimes\boldsymbol{v} + \left( \nabla\otimes\boldsymbol{v} \right)^T \right) + \frac{\eta^o}{2} \left[\nabla_{\perp}\otimes\boldsymbol{v} + \nabla\otimes\boldsymbol{v}_{\perp}+\left(\nabla_{\perp}\otimes\boldsymbol{v} + \nabla\otimes\boldsymbol{v}_{\perp}\right)^T\right],
\end{align}
where $\eta_R$ is the rotational viscosity, $\Omega$ is the intrinsic rotation rate, $\omega = \epsilon_{ij} \partial_i v_j$ is the vorticity field, $\eta_s$ is the passive shear viscosity, $\eta^o$ is the odd shear viscosity, $\eta^\Lambda$ couples vorticity to an isotropic stress, and $p$ is the pressure the enforces the incompressibility condition~$\nabla\cdot\boldsymbol{v}=0$. Here, we assume that a constant intrinsic rotation rate $\Omega$ is maintained by suitable external torques and therefore do not treat the angular momentum balance explicitly. Formally, the stress of an incompressible solid takes the same form as that of a fluid in Eq.~(\ref{appeq:IncompStressTensor}), but with the velocity field $\boldsymbol{v}$ and viscous moduli replaced by the displacement field and elastic moduli. As such, the following also holds for odd incompressible linearly elastic solids.

In the force balance $\nabla \cdot \boldsymbol{\sigma} = 0$ the odd shear viscosity can be absorbed into an effective pressure \cite{avron1998odd}, yielding a Stokes-like equation and incompressibility condition
\begin{align}
    (\eta_s + \eta_R) \nabla^2 \boldsymbol{v} - \nabla \Tilde{p} &= 0,\label{appeq:StokesEquations0}
    \\
    \nabla \cdot \boldsymbol{v} &= 0,\label{appeq:StokesEquations}
\end{align}
where $\Tilde{p} = p - (\eta^o - \eta_\Lambda) \omega$ is the effective pressure, and the effective dynamic viscosity is $\eta_s + \eta_R$. The classical Papkovich-Neuber ansatz for Eqs.~(\ref{appeq:StokesEquations0})and (\ref{appeq:StokesEquations}) is \cite{cong1982}
\begin{align}
        \boldsymbol{v}(\boldsymbol{r}) &= \frac{1}{2(\eta_s + \eta_R)} [\nabla(\boldsymbol{r} \cdot \boldsymbol{B} + B_0) - 2 \boldsymbol{B}], \label{appeq: Odd Stokes PN0}
        \\
        \Tilde{p} &= \nabla \cdot \boldsymbol{B} \label{appeq: Odd Stokes PN},
        \\
        p &= \nabla \cdot \boldsymbol{B} - \frac{\eta^o - \eta_\Lambda}{\eta_s + \eta_R} \nabla \times \boldsymbol{B}, \label{appeq:pconv}
\end{align}
where $\nabla \times \boldsymbol{B} = \epsilon_{ij}\partial_i B_j$, where $\boldsymbol{B}, B_0$ must again be harmonic vector and scalar fields, respectively.\\

Applying instead the strategy used in the main text Sec. \ref{sec:OPN}, we may first write the force balance for a stress given in Eq.~(\ref{appeq:IncompStressTensor}) as
\begin{align}
        a \boldsymbol{R}^T(\phi) \cdot \nabla^2 \boldsymbol{v} - \nabla p &= 0,
        \label{appeq: Fluid Force Balance gPN Form}
\end{align}
where $\boldsymbol{R}(\phi) = \cos(\phi) \mathbb{I} - \sin(\phi) \boldsymbol{\epsilon}$ is now rotation matrix about an angle $\phi$ with
\begin{equation}
    a = \sqrt{(\eta_s + \eta_R)^2 + (\eta^o - \eta_\Lambda)^2} , \hspace{5pt} \cos(\phi) = \frac{\eta_s + \eta_R}{\sqrt{(\eta_s + \eta_R)^2 + (\eta^o - \eta_\Lambda)^2}}, \hspace{5pt} \sin(\phi) = \frac{\eta_\Lambda - \eta^o}{\sqrt{(\eta_s + \eta_R)^2 + (\eta^o - \eta_\Lambda)^2}}
\end{equation}

This admits an odd Papkovich-Neuber ansatz
\begin{align}\label{Appendix: Stokes gPN}
    \boldsymbol{v}(\boldsymbol{r}) &= \frac{1}{2a} [\boldsymbol{R}(\phi) \cdot \nabla (\boldsymbol{r} \cdot \boldsymbol{B} + B_0) - 2 \cos(\phi) \boldsymbol{B}],
    \\
    p &= \nabla \cdot \boldsymbol{B},\label{eq:OPNp}
\end{align}
which yields an incompressible flow field that solves the force balance and Eq.~(\ref{appeq: Fluid Force Balance gPN Form}) if $\boldsymbol{B}, B_0$ are harmonic vectors and scalars respectively.\\

Interestingly, the Papkovich-Neuber expressions Eqs.~(\ref{appeq: Odd Stokes PN0}) and (\ref{Appendix: Stokes gPN}) are simply related by a rotation and scaling of the harmonic vector fields each expression uses. The concrete transformation is given by
\begin{equation}\label{eq:Brot}
\boldsymbol{B}' = \frac{1}{\cos \phi} \boldsymbol{R}(\phi) \cdot \boldsymbol{B}.
\end{equation}
Evidently, $\boldsymbol{B}'$ is harmonic if and only if $\boldsymbol{B}$ is harmonic, and one can she that the pressure $p$ in the conventional ansatz, Eq.~(\ref{appeq:pconv}), follows from $p = \nabla \cdot \boldsymbol{B}'$. To discuss this relationship in more detail, we can substitute the harmonic field in Eq.~(\ref{appeq: Odd Stokes PN0}) by (the inverse of) Eq.~(\ref{eq:Brot}), which yields
\begin{align}
    \boldsymbol{v}(\boldsymbol{r}) &= \frac{1}{2 a} \big[ \nabla (\boldsymbol{r} \cdot \boldsymbol{R}^T(\phi) \cdot \boldsymbol{B}') - 2 \cos \phi \, \boldsymbol{B}' - 2\sin \phi \, \boldsymbol{\epsilon} \cdot \boldsymbol{B}' \big] \\ 
    &= \frac{1}{2a} \big[ \boldsymbol{R}(\phi) \cdot \nabla (\boldsymbol{r} \cdot \boldsymbol{B}') + \boldsymbol{A} - 2 \cos\phi \, \boldsymbol{B}'\big],\label{eq:equivPN}
\end{align}
where 
\begin{equation}
    A_i(\boldsymbol{r}) = \sin\phi \big[ \epsilon_{ij} \partial_j (r_k B'_k) + \partial_i(r_j \epsilon_{jk} B'_k) - 2 \epsilon_{ij} B'_j \big],
\end{equation}
which can be compared with Eq.~(\ref{eq:equivPN}). The vector field $\boldsymbol{A}(\boldsymbol{r})$ is both irrotational and solenoidal, i.e. $\partial_i A_i = \partial_i \epsilon_{ij} A_j = 0$, and can therefore be written as the (possibly rotated) gradient of a scalar harmonic potential, i.e. it can be written as $\boldsymbol{R}(\phi) \cdot \nabla B_0$ with $\nabla^2B_0=0$, demonstrating the relationship between the classical ansatz, Eqs.~(\ref{appeq: Odd Stokes PN0}) and (\ref{appeq: Odd Stokes PN}), and the OPN ansatz Eqs~(\ref{Appendix: Stokes gPN}) and (\ref{eq:OPNp}). Despite this close relationship, each solution has its own utility depending on boundary conditions. The conventional ansatz in Eq.~(\ref{appeq: Odd Stokes PN}) relates the velocity field to the harmonic potentials simply, making it convenient for velocity BCs. In contrast, the incompressile OPN ansatz in Eq.~(\ref{Appendix: Stokes gPN}) provides a simpler relation between pressure and the harmonic potentials, suggesting it is more appropriate for stress BCs.

\section{Odd Viscoelasticity}  
\subsection{Material parameters used in main text figures}
In Sec. \ref{sec: Odd Viscoelasticity} of the main text, we explored the resonances that arise due to intrinsic timescales in odd materials. The material parameters used in the different examples illustrated in the figures are given in Tab.~\ref{tab:ViscoelasticModuli}.

\begin{table}[H]
\includegraphics[width = 0.98\columnwidth]{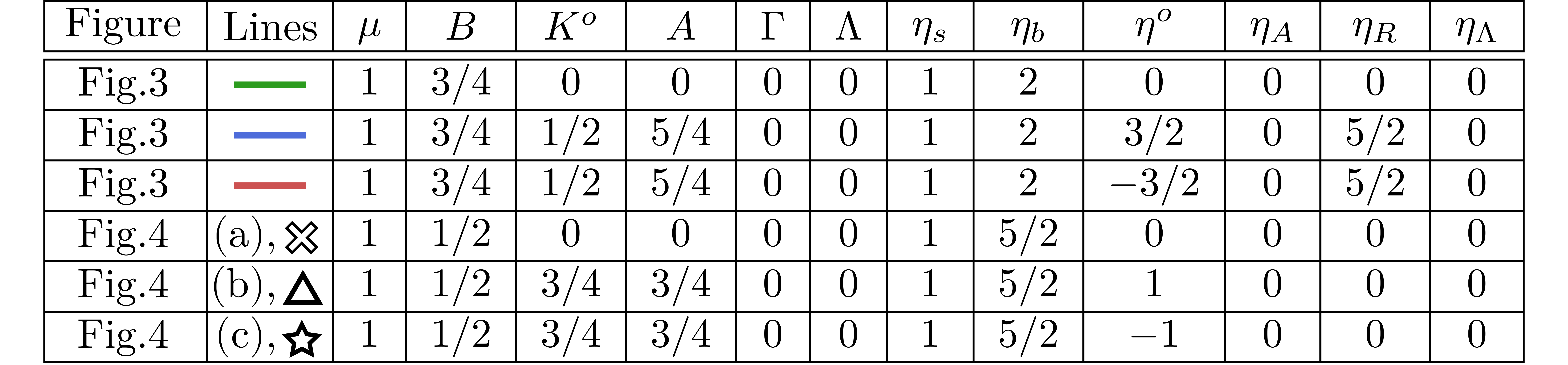}\vspace{-0.1cm}
    \caption{Material parameters used in Fig.~\ref{fig:DBCresonances}~and~Fig.~\ref{fig:ViscoelasticResonances} of the main text.}
    \label{tab:ViscoelasticModuli}
\end{table}

\subsection{Oscillatory steady states}\label{appsec:osci}
When a linear viscoelastic material is mechanically stable, it reaches an oscillatory steady-state with frequency set by the forcing frequency. This solution can be found by inverting the Laplace transform and neglecting transients, or equivalently by using a Fourier ansatz. Considering DBCs as in main text Sec.~\ref{sec:OVE DBCs}, we impose a boundary driving $\boldsymbol{\hat{u}}(\theta,t):=\boldsymbol{u}(R,\theta,t)$ with
\begin{equation}\label{appeq:Viscoelastic DBCs}
\boldsymbol{\hat{u}}^{(n)}(\theta, t) = \text{Re} \big[ u_0 e^{i \omega t} \cos(n \theta) \boldsymbol{e}_r \big].
\end{equation}
Steady-state displacement oscillations can be found by solving the force balance equation for an odd elastic material, with complex effective elasticity tensor $C^{\text{eff}}_{\alpha \beta} = C_{\alpha \beta} + i \omega \eta_{\alpha \beta}$. The solution reads
\begin{equation}
\boldsymbol{u}(r, \theta, t) = \frac{\text{Re} \left(u_0 e^{i \omega t} \right)}{2} \frac{R^n}{r^n} \left( \frac{r}{R} \boldsymbol{Z} + \frac{R}{r} \mathbb{I} \right)\cdot\boldsymbol{R}(n \theta)\cdot\boldsymbol{e}_r 
+\frac{(n-1)}{2}\frac{R^n}{r^n}\left(\frac{r}{R} - \frac{R}{r} \right) \boldsymbol{R}(n \theta) \cdot \text{Re}\big[ u_0 e^{i \omega t} \boldsymbol{\Tilde{R}}(i \omega) \big] \cdot \boldsymbol{e}_r.
\end{equation}
Taking $u_0 \in \mathbb{R}$, since any complex phase may be absorbed as a phase shift in the displacement field, the solution may be written as (Eq.~(\ref{eq:VEgeneralSol}) provided in the main text)
\begin{align}
\boldsymbol{u}(r, \theta, t) = u_0\frac{\cos(\omega t)}{2} &\frac{R^n}{r^n} \left( \frac{r}{R} \boldsymbol{Z} + \frac{R}{r} \mathbb{I} \right)\cdot\boldsymbol{R}(n \theta)\cdot\boldsymbol{e}_r\notag\\
+u_0\frac{(n-1)}{2} &\frac{R^n}{r^n}\left(\frac{r}{R} - \frac{R}{r} \right) \boldsymbol{R}(n \theta) \cdot \left[A_e(\omega) \cos(\omega t + \varphi_e(\omega)) \boldsymbol{e}_r + A_o(\omega) \cos(\omega t + \varphi_\text{o}(\omega)) \boldsymbol{e}_\theta \right],
\end{align}
where $A_e(\omega), A_o(\omega)$ are (positive) real amplitudes, and $\varphi_e(\omega), \varphi_o(\omega)$ are real phases given by
\begin{equation}
    A_e(\omega) = |\Tilde{R}_e(i \omega)|, \hspace{5pt}
    A_o(\omega) = |\Tilde{R}_o(i \omega)| , \hspace{5pt}
    \tan(\varphi_e) = \frac{\text{Im}\big[ \Tilde{R}_e(i \omega) \big]}{\text{Re}\big[ \Tilde{R}_e(i \omega) \big]} , \hspace{5pt}
    \tan(\varphi_o) = \frac{\text{Im}\big[ \Tilde{R}_o(i \omega) \big]}{\text{Re}\big[ \Tilde{R}_o(i \omega) \big]},
\end{equation}
with
\begin{align}
    \Tilde{R}_e(i \omega) &= \frac{(B + i \omega \eta_b)(B + 2 \mu + i\omega(\eta_b + 2\eta_s)) + (A + i \omega \eta_A)(A + 2 K^o + i\omega(\eta_A + 2\eta^o))}{(B + 2\mu + i \omega(\eta_b + 2 \eta_s))^2 + (A + 2K^o + i \omega (\eta_A + 2 \eta^o))^2}, \\ \notag
    \Tilde{R}_o(i \omega) &= 2 \frac{(B + i \omega \eta_b)(K^o + i \omega \eta^o) - (\mu + i \omega \eta_s)(A + i \omega \eta_A)}{(B + 2\mu + i \omega(\eta_b + 2 \eta_s))^2 + (A + 2K^o + i \omega (\eta_A + 2 \eta^o))^2},
\end{align}
are the diagonal and antisymmetric parts of $\boldsymbol{\Tilde{R}}(i \omega)$.

\subsection{Relaxation dynamics}
In the main text Sec. \ref{sec: Odd Viscoelasticity}, we focused on solutions to the homogeneous part of the force balance equation in Laplace space
\begin{equation}\label{appeq:ViscoelasticForceBalanceLaplace}
    \partial_i C^{\text{eff}}_{ijkl} \partial_k U_l(\boldsymbol{x},s) = \partial_i \eta_{ijkl} \partial_k [\boldsymbol{u}_0(\boldsymbol{x})]_l,
\end{equation}
where $s$ is our Laplace space variable, $\boldsymbol{U}(\boldsymbol{x},s)$ is the Laplace transform of $\boldsymbol{u}(\boldsymbol{x},t)$, $C^{\text{eff}}_{ijkl} = C_{ijkl} + s \eta_{ijkl}$ is the effective elasticity tensor, and $\boldsymbol{u}_0 (\boldsymbol{x}) = \boldsymbol{u}(\boldsymbol{x}, t = 0)$ is the initial displacement field. In the following, we solve for a particular solution for $\boldsymbol{U}(s, \boldsymbol{x})$ of Eq.~(\ref{appeq:ViscoelasticForceBalanceLaplace}). Writing Eq.~(\ref{appeq:ViscoelasticForceBalanceLaplace}) in tensor form (with moduli renormalized by $\Gamma, \Lambda, \eta_R, \eta_\Lambda$ as in Eq.~(\ref{eq:OddElasticForceBalanceModuliVersion}) of main text), one obtains
\begin{equation}\label{appeq:ViscoelasticForceBalanceLaplaceTensor}
\nabla^2 [(\mu + \eta_s s)\mathbb{I} + (K^o + \eta^o s)\boldsymbol{\epsilon}]\cdot \boldsymbol{U} + [(B + \eta_b s) \mathbb{I} + (A + \eta_A s) \boldsymbol{\epsilon}] \cdot \nabla (\nabla \cdot \boldsymbol{U}) = 
\nabla^2(\eta_s \mathbb{I} + \eta^o \boldsymbol{\epsilon})\cdot \boldsymbol{u}_0 + (\eta_b \mathbb{I} + \eta_A \boldsymbol{\epsilon})\cdot \nabla (\nabla \cdot \boldsymbol{u}_0).
\end{equation}
First, we consider a Helmholtz decompositions of $\boldsymbol{U}$ and $\boldsymbol{u}_0$,
\begin{equation}
    \boldsymbol{U} = \boldsymbol{U}^\text{sol} + \boldsymbol{U}^\text{irr} , \hspace{5pt} \boldsymbol{u}_0 = \boldsymbol{u}_0^\text{sol} + \boldsymbol{u}_0^\text{irr},
\end{equation}
where $\nabla \cdot \boldsymbol{U}^\text{sol} = 0, \nabla \times \boldsymbol{U} = \nabla \cdot \boldsymbol{\epsilon} \cdot \boldsymbol{U}^\text{irr} = 0$, and similarly for $\boldsymbol{u}_0$. Using that 
\begin{equation}
    \nabla^2 \boldsymbol{v} = \nabla (\nabla \cdot \boldsymbol{v}) - \nabla \times \nabla \times \boldsymbol{v},
\end{equation}
where $\nabla \times \nabla \times \boldsymbol{v} := \boldsymbol{\epsilon} \cdot \nabla (\nabla \cdot \boldsymbol{\epsilon} \cdot \boldsymbol{v})$ for any 2D vector field $\boldsymbol{v}$, Eq.~(\ref{appeq:ViscoelasticForceBalanceLaplaceTensor}) becomes

\begin{align}\label{appeq: Laplace inhom force balance Helmhotz decomp}
    &\nabla \nabla \cdot [(\mu + B + \eta_s s + \eta_b s) \boldsymbol{U}^\text{irr}_\text{inh} + (K^o + \eta^o s) (\boldsymbol{U}^\text{sol}_\text{inh})_\bot ]
    + \nabla \times \nabla \times\left[(\mu + \eta_s s) \boldsymbol{U}^\text{sol}_\text{inh} + (K^o + A + \eta^o s + \eta_A s) (\boldsymbol{U}^\text{irr}_\text{inh})_\bot\right] 
    \notag \\ 
    =&\ \nabla \nabla \cdot \left[(\eta_s + \eta_b) \boldsymbol{u}_0^\text{irr} + \eta^o (\boldsymbol{u}_0^\text{sol})_\bot ) - \nabla \times \nabla \times (\eta_s \boldsymbol{u}_0^\text{sol} + (\eta^o + \eta_A) (\boldsymbol{u}_0^\text{irr})_\bot\right],
\end{align}
where $\boldsymbol{v}_{\bot} = \epsilon \cdot \boldsymbol{v}$ is the orthogonal conjugate of $\boldsymbol{v}$. We then make the ansatz
\begin{align}\label{appeq: Laplace inh ansatz}
    \boldsymbol{U}^\text{sol}_\text{inh}(s, \boldsymbol{x}) &= M_1(s) \boldsymbol{u}_0^\text{sol}(\boldsymbol{x}) + M_3(s) (\boldsymbol{u}_0^\text{irr}(\boldsymbol{x}))_\bot, 
    \\ \notag
    \boldsymbol{U}^\text{irr}_\text{inh}(s, \boldsymbol{x}) &= M_2(s) \boldsymbol{u}_0^\text{irr}(\boldsymbol{x}) + M_4(s) (\boldsymbol{u}_0^\text{sol}(\boldsymbol{x}))_\bot.
\end{align}
Noting that $(\boldsymbol{v}_\bot)_\bot = -\boldsymbol{v}$, the force balance Eq.~(\ref{appeq: Laplace inhom force balance Helmhotz decomp}) is solved identically by substituting in ansatz Eq.~(\ref{appeq: Laplace inh ansatz}), when

\begin{align}
        M_1(s) &= \frac{\Bar{B} \eta_s + \Bar{A} \eta^o + s(\eta_s \Bar{\eta}_B + \eta^o \Bar{\eta}_A)}{\mu \Bar{B} + K^o \Bar{A} + s(\eta_s \Bar{B} + \mu \Bar{\eta}_B + \eta^o \Bar{A} + K^o \Bar{\eta}_A) + s^2 (\eta_s \Bar{\eta}_B + \eta^o \Bar{\eta}_A)}
        \\ \notag        
        M_2(s) &=  \frac{K^o \Bar{\eta}_A + \mu \Bar{\eta}_B + s(\eta^o \Bar{\eta}_A + \eta_s \Bar{\eta}_B)}{\mu \Bar{B} + K^o \Bar{A} + s(\eta_s \Bar{B} + \mu \Bar{\eta}_B + \eta^o \Bar{A} + K^o \Bar{\eta}_A) + s^2 (\eta_s \Bar{\eta}_B + \eta^o \Bar{\eta}_A)}
        \\ \notag
        M_3(s) &= \frac{\Bar{B} \Bar{\eta}_A - \Bar{A} \Bar{\eta}_B}{\mu \Bar{B} + K^o \Bar{A} + s(\eta_s \Bar{B} + \mu \Bar{\eta}_B + \eta^o \Bar{A} + K^o \Bar{\eta}_A) + s^2 (\eta_s \Bar{\eta}_B + \eta^o \Bar{\eta}_A)}\notag\\
        M_4(s) &= \frac{\mu \eta^o - K^o \eta_s}{\mu \Bar{B} + K^o \Bar{A} + s(\eta_s \Bar{B} + \mu \Bar{\eta}_B + \eta^o \Bar{A} + K^o \Bar{\eta}_A) + s^2 (\eta_s \Bar{\eta}_B + \eta^o \Bar{\eta}_A)},
        \notag
\end{align}
where $\Bar{B} = B + \mu, \Bar{A} = A + K^o, \Bar{\eta}_b = \eta_b + \eta_s, \Bar{\eta}_A = \eta_A + \eta^o$.

Inverting the Laplace transform yields a particular solution, $\boldsymbol{u}_\text{inh}(\boldsymbol{x},t)$, given by
\begin{equation}
    \boldsymbol{u}_\text{inh}(\boldsymbol{x}, t) = m_1(t) \boldsymbol{u}_0^\text{sol}(\boldsymbol{x}) + m_2(t) \boldsymbol{u}_0^\text{irr}(\boldsymbol{x}) + m_3(t) [\boldsymbol{u}_0^\text{irr}(\boldsymbol{x})]_\bot + m_4(t)[\boldsymbol{u}_0^\text{sol}(\boldsymbol{x})]_\bot.
\end{equation}

As for resonances in the main text, the oscillatory frequencies and decay rates of $m_{1,2,3,4}(t)$ are given by the real and imaginary parts of the poles of $M_{1,2,3}(s)$. These poles are the same for all $M_{1,2,3,4}(s)$, and are given by $s = s_\text{inh}^\pm$ where
\begin{equation}
s_\text{inh}^\pm = - \frac{\eta_s \Bar{B} + \mu \Bar{\eta}_B + \eta^o \Bar{A} + K^o \Bar{\eta}_A \pm \sqrt{(B \eta_s - \mu \eta_b + A \eta^o - K^o \eta_A)^2 - 4 (\Bar{B} \Bar{\eta}_A - \Bar{A} \Bar{\eta}_B)(\mu \eta^o - K^o \eta_s)}}{2(\eta_s \Bar{\eta}_B + \eta^o \Bar{\eta}_A)}.
\end{equation}

Stability requires $\text{Re}\left(s_\text{inh}^\pm\right)< 0$. In the limit of passive viscosity, the reduces to the stability condition found by Scheibner et al.\cite{Scheibner:2020},
\begin{equation}
    \mu (B + \mu) + K^o (A + K^o) > 0.
\end{equation}

The poles $s_{\text{inh}}^\pm$ acquire non-zero imaginary part when
\begin{equation}
    (B \eta_s - \mu \eta_b + A \eta^o - K^o \eta_A)^2 < 4 (\Bar{B} \Bar{\eta}_A - \Bar{A} \Bar{\eta}_B)(\mu \eta^o - K^o \eta_s).
\end{equation}
In the overdamped regimes, this extends the result of Scheibner et al. \cite{Scheibner:2020} for the onset of waves in a viscoelastic material by allowing odd viscosity and additional elastic moduli $\Gamma, \Lambda$.

\subsection{Overdamped odd viscoelasticity as damped harmonic oscillation}
 Resonances occur when the forcing frequency of the solution is near the location of a pole in Laplace space. For an odd Kelvin-Voigt model, such poles arise only from matrix inverses. Under SBCs, the only non-removable singularities are due to $\boldsymbol{M}_s^{-1}$ and $\boldsymbol{M}_d^{-1}$, whilst for DBCs they arise solely from $\boldsymbol{M}_u^{-1}$, where

\begin{align}
    \boldsymbol{M}_u(s) &= \big[(B + \Gamma + 2\mu + (\eta_b + \eta_R +2 \eta)s) \mathbb{I} + (A - \Lambda + 2K^o + (\eta_A - \eta_\Lambda +2 \eta^o)s) \boldsymbol{\epsilon} \big] \notag \\
    \boldsymbol{M}_{\text{s}}(s) &= (\mu + s \eta_s) \mathbb{I} + (K^o + s \eta^o) \boldsymbol{\epsilon},\label{appeq:Mshear}\\
    \boldsymbol{M}_{\text{d}}(s) &= [(B + s \eta_b) \mathbb{I} - (A + s \eta_A) \boldsymbol{\epsilon}]\cdot(\mathbb{I} - \boldsymbol{\Tilde{R}}) + [(\Gamma + s \eta_R) \mathbb{I} + (\Lambda + s \eta_\Lambda)\boldsymbol{\epsilon}]\cdot(\mathbb{I} + \boldsymbol{\Tilde{R}}) \notag\\
    &= 2\boldsymbol{M}^{-1}_u(s)\cdot\big\{ [(B + s \eta_b) \mathbb{I} - (A + s \eta_A) \boldsymbol{\epsilon}]\cdot[(\mu + s \eta_s + \Gamma + s \eta_R)\mathbb{I} + (K^o + s \eta^o - \Lambda - s \eta_\Lambda) \boldsymbol{\epsilon}] \notag\\
    &+ [(\Gamma + s \eta_R) \mathbb{I} + (\Lambda + s \eta_\Lambda)\boldsymbol{\epsilon}]\cdot[(\mu + s \eta_s + B + s \eta_b)\mathbb{I} + (K^o + s \eta^o + A + s \eta_A) \boldsymbol{\epsilon}] \big\}.\label{appeq:Md-r}
\end{align}

Since all of $\boldsymbol{M}_{u,s,d}(s)$ are independent of the angular mode, $n$, resonances are as well. Each of $\boldsymbol{M}_{u,s}(s)$ become singular at one pair of complex conjugate poles $s_{u,s}^\pm$. However, as noted in the main text Sec. \ref{sec:OVE SBCs}, $\boldsymbol{M}_d(s)$ becomes singular at two distinct pairs of complex conjugate poles. Closed form expressions for these poles are in general too complicated to write down explicitly. However, taking $\Gamma = \Lambda = \eta_R = \eta_\Lambda = 0$ as done in the main text, allows for a factorization of $\boldsymbol{M}_d(s)$ as
\begin{equation}
    \boldsymbol{M}_d(s) = 2 \boldsymbol{M}^{-1}_u(s)\cdot\boldsymbol{M}_s(s)\cdot[(B + s \eta_b) \mathbb{I} - (A + s \eta_A) \boldsymbol{\epsilon}],
\end{equation}
revealing $s^\pm_d$ as a new pair of poles, as well as recovering $s^\pm_s$. However, a similar decomposition of $\boldsymbol{M}_d(s)$ into products of matrices remains possible even under completely general isotropic moduli. This can be shown by considering $P(s) \mathbb{I} + Q(s) \boldsymbol{\epsilon}$ -- where $P(s), Q(s)$ are quadratic polynomials in $s$ with real coefficients -- as $C(s) = P(s) + i Q(s)$, i.e. a quadratic polynomial with complex coefficients, by the identification of rotations and complex numbers, $1 \leftrightarrow \mathbb{I}, i \leftrightarrow \boldsymbol{\epsilon}$. By the fundamental theorem of algebra, this can be decomposed into the product of two linear polynomials with complex coefficients, which can be transformed back into matrices to obtain
\begin{equation}
    \boldsymbol{M}_d(s) = 2 \boldsymbol{M}^{-1}_u(s)\cdot\boldsymbol{M}_{d1}(s)\cdot\boldsymbol{M}_{d2}(s),
\end{equation}
where $\boldsymbol{M}_{d1}(s)$ and $\boldsymbol{M}_{d2}(s)$ are proportional to rotation matrices, and linear in $s$ with real coefficients. Near where one of $\boldsymbol{M}_p$ ($p = u,s,d1,d2$) becomes singular, subject to the relevant BCs, the displacement field is dominated by 
\begin{equation}
    \boldsymbol{u}(r, \theta,t) \approx \boldsymbol{u}_p(r, \theta,t) + \boldsymbol{u}^{(T)}_p(r, \theta,t) = \boldsymbol{M}^{-1}_p\cdot\boldsymbol{b}(r, \theta,t) + \left(\boldsymbol{M}^T_p\right)^{-1}\cdot\boldsymbol{b}_T(r, \theta,t),
\end{equation}
where $\boldsymbol{b}(\boldsymbol{r},t), \boldsymbol{b}_T(\boldsymbol{r},t)$ contain the boundary conditions. For DBCs we may take $\boldsymbol{b}_T=0$, however under SBCs $\boldsymbol{Z}$ appears in dominant parts of the displacement field, $\boldsymbol{u}$. Since $\boldsymbol{Z}$ geometrically describes a reflection, rotation matrices are transposed when exchanging order of multiplication with $\boldsymbol{Z}$, hence we require $\boldsymbol{b}_T \neq 0$ in general. The dominant contribution to the displacement field from $\boldsymbol{b}(r,\theta,t)$ evolves as
\begin{equation}\label{appeq:up dynamics as couples oscillators}
    \boldsymbol{M}_p(\partial_t)\cdot\boldsymbol{u}_p = (C^{(e)}_p \mathbb{I} - C^{(o)}_p \boldsymbol{\epsilon})\cdot\boldsymbol{u}_p + (\eta^{(e)}_p \mathbb{I} - \eta^{(o)}_p \boldsymbol{\epsilon})\cdot\partial_t \boldsymbol{u}_p = \boldsymbol{b}.
\end{equation}
This can be written in terms of the effective relaxation-rate tensor $\boldsymbol{\Lambda}_p$,
\begin{equation}
    \boldsymbol{\Lambda}_p \cdot \boldsymbol{u}_p + \partial_t \boldsymbol{u}_p = \boldsymbol{b}',
\end{equation}
with $\mathbf{b}'=(\eta^{(e)}_p \boldsymbol{I} - \eta^{(o)}_p \boldsymbol{\epsilon})^{-1} \cdot\mathbf{b}$, and
\begin{equation}
    \boldsymbol{\Lambda}_p = \frac{|\boldsymbol{C}_p|}{|\boldsymbol{\eta}_p|} \boldsymbol{R}^T(\varphi_p),
\end{equation}
where $\varphi_p$ is the angle between vectors $\boldsymbol{C}_p, \boldsymbol{C}_p$. Since $\boldsymbol{\Lambda}_p$ is proportional to a rotation matrix, we have that
\begin{equation}\label{appeq:decoupled_up_dynamics}
    \boldsymbol{\Lambda}_p^T \cdot \boldsymbol{b}' + \partial_t \boldsymbol{b}' = \partial^2_t \boldsymbol{u}_p + \left(\boldsymbol{\Lambda}_p + \boldsymbol{\Lambda}_p^T \right) \cdot \partial_t \boldsymbol{u}_p + \boldsymbol{\Lambda}_p^T \cdot \boldsymbol{\Lambda}_p \cdot \boldsymbol{u}_p = \partial^2_t \boldsymbol{u}_p + \text{tr}\left(\boldsymbol{\Lambda}_p\right) \boldsymbol{u}_p + \det \left(\boldsymbol{\Lambda}_p \right) \boldsymbol{u}_p.
\end{equation}
This yields two decoupled second order ODEs for the components of $\boldsymbol{u}_p$, which are each equivalent to that for a damped harmonic oscillator.

The dynamics for $\boldsymbol{u}_p$ are governed by effective relaxation rate tensor $\boldsymbol{\Lambda}_p^T$ instead, and so the components of $\boldsymbol{u}_p$ are non-reciprocally coupled with opposite sign to those for $\boldsymbol{u}_p$. Since the decoupled dynamics for $\boldsymbol{u}_p$ in Eq.~(\ref{appeq:decoupled_up_dynamics}) depend only on the trace and determinant of $\boldsymbol{\Lambda}_p$ -- which are left invariant under taking the transpose -- $\boldsymbol{u}_p$ obeys dynamics equivalent to those for $\boldsymbol{u}_p$. It should be noted that whilst for $p=u,s$, each of $C^{(e)}_p, C^{(o)}_p, \eta^{(e)}_p, \eta^{(o)}_p$ are linear in material moduli, for $p=d1,d2$, they are in general rational functions of material moduli. However, symmetry implies that $C^{(e)},\eta^{(e)}$ are both parity even, whilst $C^{(o)},\eta^{(o)}$ are both parity odd.

\section{Anisotropic odd materials}\label{Appsec:AnisotropicOPN}
Dropping the requirement of isotropy, the most general elastic modulus tensor can be written as
\begin{equation}\label{appeq: Anisotropic Modulus Tensor}
    C_{\alpha \beta} = 2 \begin{pmatrix}
        B & \Lambda & t_1 & p_1 \\
        A & \Gamma & t_2 & p_2 \\
        d_1 & r_1 & \mu + s_1 & s_2 + K^o \\
        d_2 & r_2 & s_2 - K^o & \mu - s_1
    \end{pmatrix}.
\end{equation}
In the absence of external forces, the overdamped force balance equation is
\begin{equation}\label{appeq: Anisotropic Force Balance}
    \nabla \cdot \boldsymbol{\sigma} = \frac{1}{2} C_{\alpha \beta} [(\boldsymbol{s}^\alpha)^T \cdot (\nabla\otimes\nabla) \cdot \boldsymbol{s}^\beta] \cdot \boldsymbol{u} = 0.
\end{equation}
To reduce this to a form that closer reflects the structure of the force balance in the isotropic case, Eq.~(\ref{Appendix Section: General Solution Form}), we may use the (anti)commutation rules 
\begin{align}
        [\nabla\otimes\nabla, \boldsymbol{s}^0] &= 0, \notag
        \\
        [\nabla\otimes\nabla, \boldsymbol{s}^2] &= 2 \boldsymbol{s}^1 \partial_x \partial_y, \notag
        \\
        \{\nabla\otimes\nabla, \boldsymbol{s}^1\} &= \boldsymbol{s}^1 \nabla^2,
        \\
        \{\nabla\otimes\nabla, \boldsymbol{s}^3\} &= 2 \mathbb{I} \partial_x \partial_y + \boldsymbol{s}^3 \nabla^2. \notag
\end{align}

The force balance Eq.~(\ref{appeq: Anisotropic Force Balance}) then becomes
\begin{align}\label{appeq: Anisotropic Expanded Force Balance}
    0 = &\nabla^2 [(\mu + \Gamma - s_1) \mathbb{I} + (K^o - \Lambda + s_2)\boldsymbol{\epsilon}  + (r_2 + p_2) \boldsymbol{s}^2 + (p_1 - r_1) \boldsymbol{s}^3] \cdot \boldsymbol{u} + 
    \notag\\ &+
    [(B - \Gamma + 2 s_1) \mathbb{I} + (A + \Lambda - 2 s_2) \boldsymbol{\epsilon} + (t_1 - p_2 - r_2 + d_1)\boldsymbol{s}^2 + (d_2 + r_1 - p_1 - t_2) \boldsymbol{s}^3]\cdot \nabla (\nabla \cdot \boldsymbol{u}) + 
    \\ \notag &+
    2 \partial_x \partial_y [(p_1 + t_2) \mathbb{I} + (p_2 - t_1)\boldsymbol{\epsilon} + 2 s_2 \boldsymbol{s}^2 - 2 s_1 \boldsymbol{s}^3] \cdot \boldsymbol{u}.
\end{align}
There are two important new features to this force balance equation: (1) an anisotropic Laplacian term, $2 \partial_x \partial_y = \nabla \cdot \boldsymbol{s}^2 \cdot \nabla$; and (2) the nematic tensors $\boldsymbol{s}^2, \boldsymbol{s}^3$ appear explicitly.\\

The Papkovich-Neuber ansatz relies on the fact that the only non-trivial terms which remain after substituting the displacement field Eq.~(\ref{appeq: Anisotropic PN Candidate}) into the force balance Eq.~(\ref{appeq: Anisotropic Expanded Force Balance}) contain Laplacians acting on $\boldsymbol{B}, B_0$. However, anisotropic Laplacians in the force balance Eq.~(\ref{appeq: Anisotropic Expanded Force Balance}) cannot in general be removed, and so no anisotropic Papkovich-Neuber ansatz exists. In 3D, odd elasticy requires anisotropy, and hence the odd Papkovich-Neuber ansatz does not exist in 3D in general.\\

For an isotropic elastic material, only rotation matrices enter the force balance equation, and these preserve the harmonic structure, since
\begin{equation}
\nabla \cdot \boldsymbol{R}(\phi) \cdot \nabla = \nabla \cdot (\cos(\phi) \mathbb{I} + \sin(\phi) \boldsymbol{s}^1) \cdot \nabla = \cos(\phi) \nabla^2
\end{equation}

Interestingly, there is still a class of modulus tensors with non-zero anisotropic terms where the force balance Eq.~(\ref{appeq: Anisotropic Expanded Force Balance}) still admits a rotation matrix structure. These are given by
\begin{equation}
    C_{\alpha \beta} = 2 \begin{pmatrix}
        B & \Lambda & t & p \\
        A & \Gamma & -p & t \\
        -t & p & \mu & K^o \\
        -p & -t & - K^o & \mu
    \end{pmatrix}
\end{equation}
For such a material, the anisotropic moduli $t, p$ cancel out in the force balance Eq.~(\ref{appeq: Anisotropic Expanded Force Balance}). This is analogous to a gauge transformation for the stress tensor, given by
\begin{equation}\label{appeq: Gauge Transformation}
    \boldsymbol{\sigma} \mapsto \boldsymbol{\sigma} + \nabla \times \boldsymbol{A}
\end{equation}
where $\nabla \times \boldsymbol{A} = \boldsymbol{\epsilon} \cdot \nabla \boldsymbol{A}$, and $\boldsymbol{A} = p \boldsymbol{s}^2 \cdot\boldsymbol{u} + t \boldsymbol{\epsilon} \cdot \boldsymbol{s}^2 \cdot \boldsymbol{u}$. This redundancy is to be expected, since there are two linearly independent ways to anisotropically construct a vector $\boldsymbol{A}$ from the displacement field $\boldsymbol{u}$ in 2D. Although this form of anisotropy does not enter the force balance equations in the bulk, it will affect the solution through SBCs.

%\bibliography{ref}
%